\documentclass[10pt,journal,compsoc]{IEEEtran}

\pdftrailerid{} % Remove ID
\usepackage{hyperref}
\hypersetup{
    unicode=false,          % non-Latin characters in Acrobat’s bookmarks
    pdftoolbar=true,        % show Acrobat’s toolbar?
    pdfmenubar=true,        % show Acrobat’s menu?
    pdffitwindow=false,     % window fit to page when opened
    pdftitle={Robust Dynamic CPU Resource Provisioning in Virtualized Servers},    % title
    pdfauthor={Evagoras~Makridis,} {Kyriakos Deliparaschos,} {Evangelia Kalyvianaki,} {Argyrios Zolotas,} {Themistoklis~Charalambous},     % author
    pdfsubject={Cloud computing},   % subject of the document
    pdfcreator={Dr. K. Deliparaschos},   % creator of the document
    pdfproducer={},  % producer of the document
    pdfkeywords={Resource provisioning,} {virtualized servers,} {CPU allocation,} {CPU usage,} {RUBiS,} {Robust prediction,} {Hinfinity filter,} {MCC-KF,} {Kalman filter}, % list of keywords
    pdfnewwindow=true,      % links in new window
    colorlinks=false,       % false: boxed links; true: colored links
    linkcolor=red,          % color of internal links
    citecolor=green,        % color of links to bibliography
    filecolor=magenta,      % color of file links
    urlcolor=cyan           % color of external links
}

\usepackage[latin1]{inputenc}
\usepackage{amssymb,amsmath,amsfonts}
\usepackage{algorithm,algorithmic}  
\usepackage[algo2e,ruled, vlined]{algorithm2e} 
\usepackage{cite}
\usepackage{float}
\usepackage{subfigure}
\usepackage{xspace}
\usepackage[usenames,dvipsnames]{color}
\usepackage{epsfig, hyperref}
\usepackage{verbatim}
\usepackage{url,changebar,bm,xspace,dsfont}
\usepackage[T1]{fontenc}
\usepackage{soul} % Strikethrough text
\usepackage{multicol}
\usepackage{url,changebar,bm,xspace,dsfont}
\usepackage[table,xcdraw]{xcolor}
\usepackage{graphicx}
\usepackage{caption}
\usepackage{fixltx2e}
\usepackage{dblfloatfix}
\usepackage{hyperref}

\hypersetup{
    pdfproducer={},  % producer of the document
}

\newlength\myindent
\floatstyle{ruled}
\newfloat{model}{H}{mod}
\floatname{model}{\footnotesize Model}
\newfloat{notatio}{H}{not}
\floatname{notatio}{\footnotesize Notation}

\newenvironment{list4}{
  \begin{list}{$\bullet$}{%
      \setlength{\itemsep}{0.05cm}
      \setlength{\labelsep}{0.2cm}
      \setlength{\labelwidth}{0.3cm}
      \setlength{\parsep}{0in} 
      \setlength{\parskip}{0in}
      \setlength{\topsep}{0in} 
      \setlength{\partopsep}{0in}
      \setlength{\leftmargin}{0.17in}}}
      {\end{list}}

\let\mathbb=\mathds % I much prefer the dsfont over the bbfont
\def\var{\mathrm{var}}
\def\cov{\mathrm{cov}} 

\def\mRT{\mathop{\mathrm{mRT}}}
\newcommand{\hinf}{$\mathcal{H}_{\infty}$\xspace}

\newcommand{\E}{\ensuremath{\mathbb{E}}}

\newtheorem{remark}{Remark}

\begin{document}

\title{Robust Dynamic CPU Resource Provisioning in Virtualized Servers}

\author{Evagoras~Makridis, Kyriakos~Deliparaschos, Evangelia~Kalyvianaki, Argyrios~Zolotas,~\IEEEmembership{Senior~Member,~IEEE}, and Themistoklis~Charalambous,~\IEEEmembership{Member,~IEEE}\IEEEcompsocitemizethanks{%
\IEEEcompsocthanksitem E. Makridis and T. Charalambous are with the Department of Electrical Engineering and Automation, Aalto University.\par E-mail: name.surname@aalto.fi%
\IEEEcompsocthanksitem K. Deliparaschos is with the Department of Electrical Engineering, Computer Engineering and Informatics, Cyprus University of Technology.\par E-mail: k.deliparaschos@cut.ac.cy%
\IEEEcompsocthanksitem E. Kalyvianaki is with the Department of Computer Science and Technology, University of Cambridge, United Kingdom.\par E-mail: ek264@cam.ac.uk%
\IEEEcompsocthanksitem A. Zolotas is with the School of Engineering, University of Lincoln, United Kingdom.\par E-mail: azolotas@lincoln.ac.uk}}%

%{Shell \MakeLowercase{\textit{et al.}}: Bare Demo of IEEEtran.cls for Computer Society Journals}

\IEEEtitleabstractindextext{%
\begin{abstract}
We present robust dynamic resource allocation mechanisms to allocate application resources meeting Service Level Objectives (SLOs) agreed between cloud providers and customers. In fact, two filter-based robust controllers, i.e. \hinf filter and Maximum Correntropy Criterion Kalman filter (MCC-KF), are proposed.  The controllers are self-adaptive, with process noise variances and covariances calculated using previous measurements within a time window. In the allocation process, a bounded client mean response time ($\mRT$) is maintained. Both controllers are deployed and evaluated on an experimental testbed hosting the RUBiS (Rice University Bidding System) auction benchmark web site. The proposed controllers offer improved performance under abrupt workload changes, shown via rigorous comparison with current state-of-the-art.  On our experimental setup, the Single-Input-Single-Output (SISO) controllers can operate on the same server where the resource allocation is performed; while Multi-Input-Multi-Output (MIMO) controllers are on a separate server where all the data are collected for decision making. SISO controllers take decisions not dependent to other system states (servers), albeit MIMO controllers are characterized by increased communication overhead and potential delays. While SISO controllers offer improved performance over MIMO ones, the latter enable a more informed decision making framework for resource allocation problem of multi-tier applications.
\end{abstract}

}

\maketitle

\IEEEdisplaynontitleabstractindextext
\IEEEpeerreviewmaketitle

\IEEEraisesectionheading{\section{Introduction}\label{sec:introduction}}
\IEEEPARstart{C}{loud} computing has revolutionized the way applications are delivered over the Internet. A typical Cloud offers resources on demand, such as CPU cycles, memory and storage space to applications typically using a pay-as-you-go model. Clouds often employ multiple data centers geographically distributed to allow application deployment in various locations around the world for reduced response times.  A data center contains tens of thousands of server machines located in a single warehouse to reduce operational and capital costs. Within a single data center modern applications are typically deployed over multiple servers to cope with their resource requirements. The task of allocating resources to applications is referred to as \emph{resource management}. In particular, resource management aims at allocating cloud resources in a way to align with application performance requirements and to reduce the operational cost of the hosted data center. However, as applications often exhibit highly variable and unpredicted workload demands, resource management remains a challenge; see, for example, a comprehensive review on resource management in \cite{MUSTAFA:2015}. 

A common practice to resource management has been to over-provision applications with resources to cope even with their most demanding but rare workloads. Although simple, this practice has led to substantial under-utilization of data centers (see, e.g., \cite{5934691})  since practitioners are devoting disjoint groups of server machines to a single application. At the same time, the advent of virtualization enables a highly configurable environment for application deployment. A server machine can be partitioned into mutliple \emph{Virtual Machines (VMs)} each providing an isolated server environment capable of hosting a single application or parts of it in a secure and resource assured manner. The allocation of resources to VM can be changed at runtime to dynamically match the virtualized application workload demands. Virtualization enables \emph{server consolidation} where a single physical server can run multiple VMs while sharing its resources and running different applications within the VMs. In addition, studies have shown that reducing the frequency of VM migrations and server switches can be very beneficial for energy saving \cite{nguyen2017virtual}. Ultimately, server consolidation increases data center utilization and thus reduces energy consumption and operational costs. 

The main challenge of server consolidation is how to dynamically adjust the allocation of VM resources so as to match the virtualized application demands, meet their Service Level Objectives (SLOs) and achieve increased server utilization. Towards this end, different autonomic resource management methods have been proposed to dynamically allocate resources across virtualized applications with diverse workload and highly fluctuating workload demands. Autonomic resource management in a virtualized environment using control-based techniques has recently gained significant attention; see \cite{Zhang:2016}, \cite{Ullah:2018} and \cite{al2018elasticity} for a survey. One of the most common approaches to control the application performance is by controlling its CPU utilization within the VM; see, e.g., \cite{Kalyvianaki:2014} and references therein.

\subsection{Contributions}

Cloud service providers encounter abrupt varying loads that deteriorate cloud elasticity on handling peak demands and potential unpredictable system faults and failures \cite{Marinescu:2013}. For this reason, we formulate the problem of CPU resource provisioning using robust control techniques, where the controllers aim at anticipating abrupt workload changes in order to maintain a certain headroom of the allocation above the utilization so that a certain SLO is satisfied. In particular, we use robust filters to predict the random CPU utilizations of a two-tier virtualized server application and, therefore, provide the CPU allocations needed dynamically to satisfy a certain upper bound on the $\mRT$. 

The contributions of this paper are as follows:
\begin{list4}
		\item An adaptive \hinf filter (SISO and MIMO) which minimizes the worst-case estimation error hence improving robustness in the state estimation problem. \hinf tracks the CPU resource utilization and adapts the state estimation based on previous observations and noises. The \hinf filter is designed and evaluated using our experimental setup.
		\item An adaptive MCC-KF (SISO and MIMO), i.e., an extended version of the Kalman filter, that utilizes higher-order statistics to predict state(s) and track the CPU resource utilization. MCC-KF is designed to adapt the state estimation based on previous observations and noises similar to that of a \hinf filter. Its performance is evaluated via our experimental setup using real-data CPU resource demands.
		\item A generic algorithm for dynamic CPU allocation is presented in order to illustrate how control theoretic approaches can address the aspect of resource provisioning in virtualized servers. 
\end{list4}

\subsection{Organization}

The rest of the paper is organized as follows. Section~\ref{notation} presents the notation used throughout the paper. In Section~\ref{metrics}, we introduce the client mean request response times and the performance metric used to evaluate the performance of the system. Section~\ref{model} discusses the model adopted for capturing the dynamics of the CPU utilization. Section~\ref{controller} presents the robust controllers developed, while the experimental setup is described in Section~\ref{setup}. The performance of the proposed controllers is evaluated and compared with other state-of-the-art solutions in Section~\ref{performance}. Related work, to the topic presented here, is discussed in Section~\ref{sec:relatedwork}. Finally, Section~\ref{conclusions} presents conclusions and discusses directions of future research.

% ===============================================
% PRELIMINARIES
% ===============================================
\section{Notation}\label{notation}

Note that $\mathbb{R}$ and $\mathbb{R}_{+}$ represent the real and the nonnegative real numbers, respectively.  Vectors, matrices and sets, are denoted by lowercase, uppercase and calligraphic uppercase letters, respectively. $A^{T}$ and $A^{-1}$ denote the transpose and inverse of matrix $A$, respectively. The identity matrix is represented by $I$. Also, $\hat{x}_{k|k-1}$ and $\hat{x}_{k|k}$ denote the \emph{a priori} and \emph{a posteriori} estimates of random value/vector ${x}_{k}$ for time instant $k$. $P_{k}$ denotes the matrix $P$ at time instant $k$.  $\mathbb{E}\{\cdot \}$ represents the expectation of its argument. Given any vector norm $\|\cdot\|$, a weighted vector norm can be written as $\|x\|_Q \triangleq \|Qx\|$, where $Q$ is an arbitrary nonsingular matrix.

% ===============================================
%
%
% Performance metrics
%
%
% ===============================================
\section{Performance metric}\label{metrics}

One of the most widely used metrics for measuring server performance is the \emph{client mean request response times} ($\mRT$). It is difficult to predict the values of the $\mRT$ of server applications across operating regions, and different applications and workloads. However, it is known to have certain characteristics~\cite{theory}. In particular, its values can be divided into three regions:
\begin{itemize}
	\item[(a)] when the application has abundant resources and, therefore, all requests are served as they arrive and the response times are kept low;
	\item[(b)] when the utilization approaches 100\% (e.g. around 70-80\% on average) the $\mRT$ increases above the low values from the previous region, due to the fact that there are instances in which the requests increase abruptly, approaching 90-100\%;
	\item[(c)] when resources are scarce and very close to 100\%, since requests compete for limited resources, they wait in the input queues for long and, as a result, their response times increase dramatically to relatively high values.
\end{itemize}

\begin{figure}[h]
	\begin{center}
		\includegraphics[width=1\columnwidth]{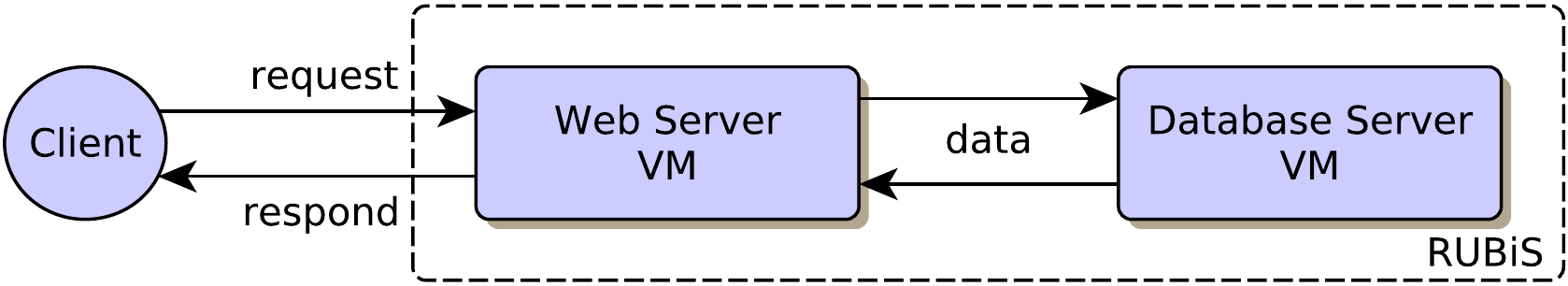}
		\caption{Request-to-response path.}
		\label{figure2}
	\end{center}
\end{figure}

In this work, the response time of every type of request was captured calculating the time difference between the request and its response, as Fig.~\ref{figure2} shows. All requests were issued to our RUBiS cluster and specifically to the Web Server, through the Client Emulator that was deployed on a separate physical machine. When all requests were completed, a mean value of the response times of the requests within a time interval of $1$s was calculated in order to have an estimate of the $\mRT$ over time. Note that in the results for the experiments presented in Section~\ref{performance}, the $\mRT$ is smoothed over the sampling/control interval.

To maintain a good  server performance, the operators try to keep the CPU utilization below 100\% of the machine capacity by a certain value, which is usually called \emph{headroom}. Headroom values are chosen such that they form the boundary between the second and the third $\mRT$ regions. At such values the server is well provisioned and response times are kept low. If the utilization exceeds the boundary due to increased workload demands, operators should increase the server resources.

\begin{figure}[!h]
	\centering
	\includegraphics[width=1\columnwidth]{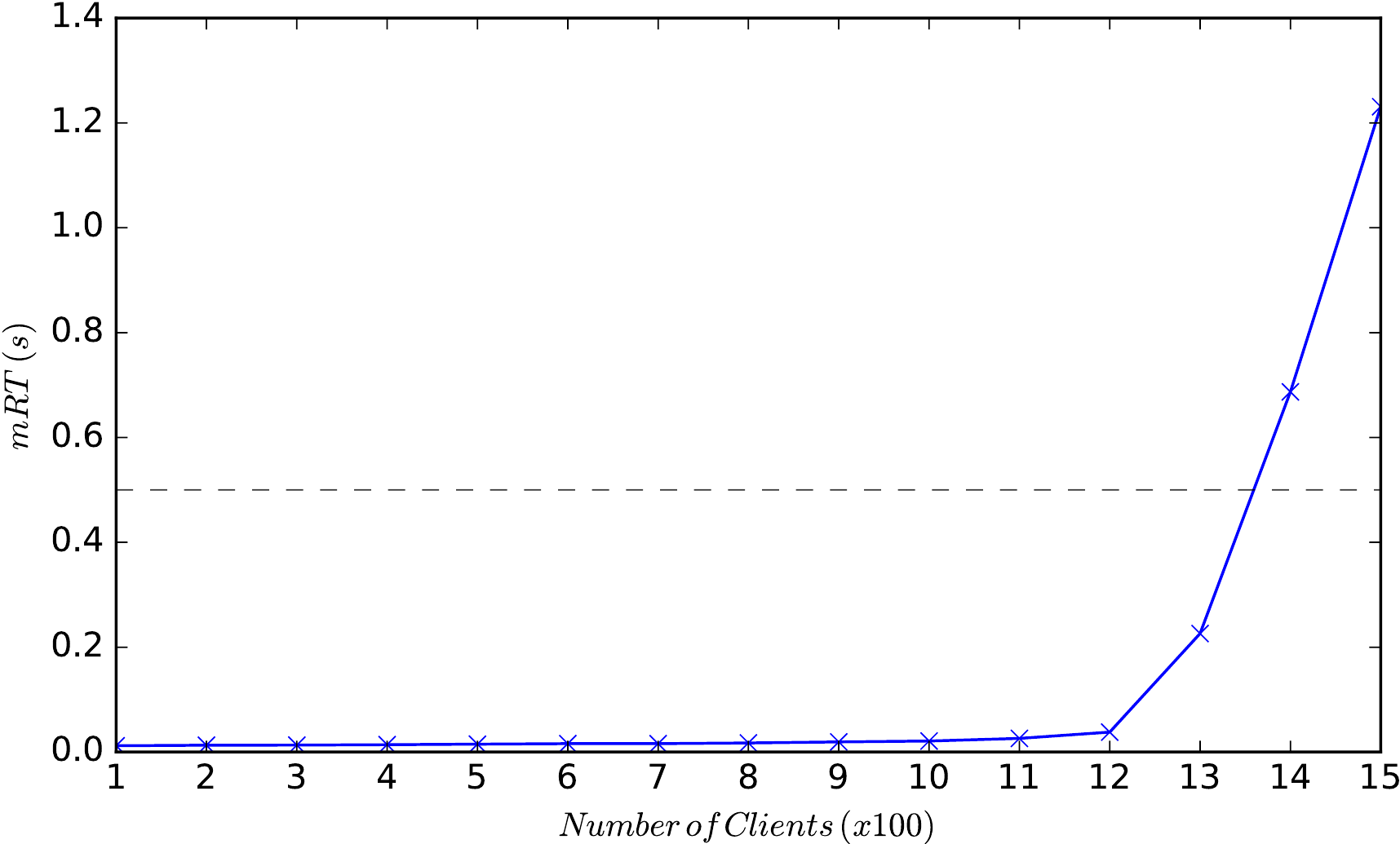}
	\caption{Mean Response Times ($\mRT$) for different workloads}
	\label{figure3}
\end{figure}

Firstly, we measure the server's performance when 100\% of resources is provisioned, without any controller adjusting the allocation of resources, in order to extract what is the required headroom. In this work, we consider a Browsing Mix workload type, in order to specify the server's performance while the number of clients varies. Fig.~\ref{figure3} shows the mean response times ($\mRT$) with number of clients increasing in steps of $100$ until $\mRT$ crosses the $0.5$s level. Clearly, the $\mRT$ increases rapidly when the number of clients exceeds $1350$ and the SLO is violated.

Initially, with increasing number of clients the $\mRT$ stays low,  albeit when the number of clients exceeds $1200$ the $\mRT$ increases above the low values. Note that, the QoS threshold of $0.5$s is exceeded when the number of clients, simultaneously issuing requests to the server, is approximately $1350$.

\begin{figure}[!h]
	\centering
	\includegraphics[width=1\columnwidth]{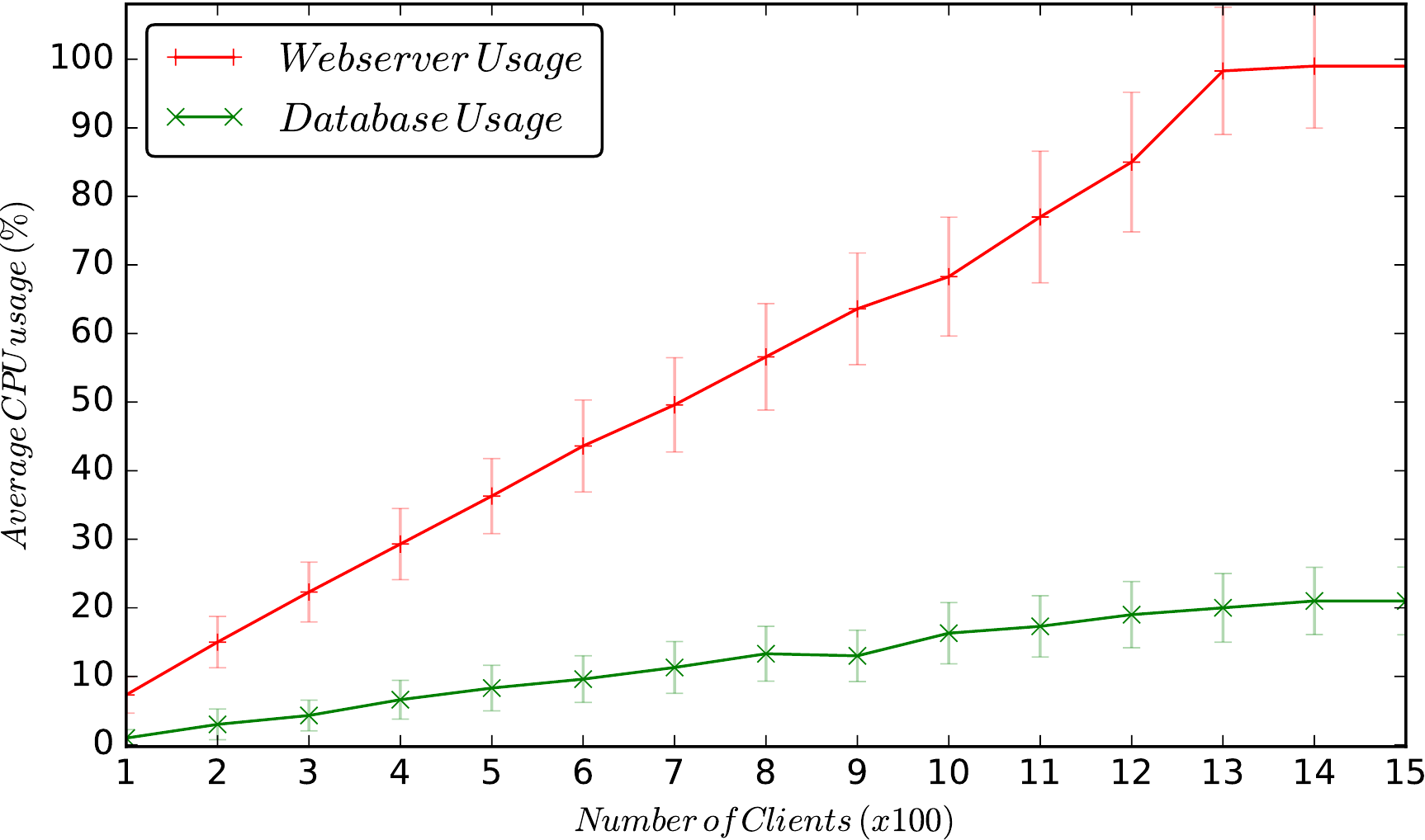}
	\caption{Average CPU usages per component for different workloads\vspace{-0.2cm}.}
	\label{figure4}
\end{figure}

Fig.~\ref{figure4} shows the average CPU usage per component while the number of clients increases. As shown in this figure, the database server demand is lower than the web server's one with the same number of clients. The error bars in Fig.~\ref{figure4} show one standard deviation above and below the mean CPU usage. When, the number of clients exceeds $1350$, the web server's CPU usage becomes the bottleneck and even though the database server does not use 100\% of its resources, it remains (almost) constant. Hence, it is important to establish the required resources for all the involved components comprising the requests.

% ===============================================
%
%
% Model
%
%
% ===============================================
\section{System model}\label{model}

% ===============================================
% SISO
% ===============================================
\subsection{SISO system}\label{SISO}

The time-varying CPU utilization per component is modeled as a random walk given by the following  linear stochastic difference equation as introduced in \cite{selfAdaptive,2010:CDC_themisEva,Kalyvianaki:2014,ecc:2016}:

\begin{equation}\label{stoch1}
	x_{k+1}=x_{k}+w_k,
\end{equation}

where $x_k\in [0,1]$ is the CPU utilization, i.e., the percentage of the total CPU capacity actually used by the application component during time-interval $k$. The independent random process $w_k$ is the \emph{process noise} which models the utilization between successive intervals caused by workload changes, e.g., requests being added to or removed from the server; it is often assumed to be normally distributed \cite{selfAdaptive, Kalyvianaki:2014}, but it can also be a distribution of finite support \cite{2010:CDC_themisEva}. The total CPU utilization of a VM which is actually \emph{observed} by the Xen Hypervisor, $y_k \in [0,1]$, is given by
\begin{align}\label{stoch2}
	y_k = x_k+{v}_k,
\end{align}
where the independent random variable $v_k$ is the utilization \emph{measurement noise} which models the utilization difference between the measured and the actual utilization; $v_k$, as it is the case with $w_k$, is often assumed to be normally distributed \cite{selfAdaptive, Kalyvianaki:2014}, but it can also be a distribution of finite support \cite{2010:CDC_themisEva}. Note that $y_k$ models the observed utilization in addition to any usage noise coming from other sources, such as the operating system, to support the application. 

% ===============================================
% MIMO
% ===============================================
\subsection{MIMO system}\label{MIMO}

For the MIMO system, the dynamics of all the components (VMs) can be written compactly as
\begin{subequations}
	\label{eq:system}
	\begin{align}
		{x}_{k+1} & =  A{x}_{k}+w_{k},   \\
		{y}_{k}   & = C {x}_{k} + v_{k},
	\end{align}
\end{subequations}
where $x_{k}\in[0,1]^{n_x}$ is the system's state vector representing the actual total CPU capacity precentages used by the application components during time-interval $k$. The \emph{process} and \emph{measurement} noise vectors, $w_{k}\in\mathbb{R}^{n_{x}}$ and $v_{k}\in\mathbb{R}^{n_{y}}$, are stochastic disturbances with zero mean and finite second-order matrices $W_k$ and $V_k$, respectively. The observed state $x_{k}$ of the system by the Xen Hypervisor is $y_{k}\in \mathbb{R}^{n_{y}}$. Matrix $A$ shows the interdependencies between different VMs and matrix $C$ captures what is actually the Xen Hypervisor observing. In the case where the CPU utilizations at the VMs are independent, matrices $A$ and $C$ are given by
\begin{gather}
 	A = 	\begin{bmatrix}
		1 & 0 \\ 
		0 & 1 
		\end{bmatrix}, \indent
 	C =	\begin{bmatrix}
		1 & 0 \\ 
		0 & 1
		\end{bmatrix}.	
\end{gather}

% ===============================================
% CPU allocation
% ===============================================
\subsection{CPU allocation}\label{CPUallocation}

By ${a}_k\in \mathbb{R}_{+}$ we denote the CPU capacity of a physical machine allocated to the VM, i.e., the maximum amount of resources a VM can use. The purpose of a designed controller is to control the allocation of the VM running a server application while observing its utilization in the VM, maintaining good server performance in the presence of workload changes. This is achieved by adjusting the allocation to values above the utilization. For each time-interval $k$, the \emph{desired} relationship between the two quantities is given by:
\begin{equation}\label{stoch3}
	a_k = \max\left\{a_{\min}, \min \{(1+h) x_k, a_{\max}\} \right\},
\end{equation}
where $h\in (0,1)$ represents the headroom (i.e., how much extra resources are provided above the actual CPU utilization), $a_{\min}$ is the minimum CPU allocated at any given time (if allocation goes very small, then even small usage may lead to high $\mRT$), and $a_{\max}$ is the maximum CPU that can be allocated. To maintain good server performance, the allocation $a_k$ should adapt to the utilization $x_k$.

Let $\mathcal{Y}_{k}$ represent the set of all observations up to time $k$. Let the \emph{a posteriori} and \emph{a priori} state estimates be denoted by $\hat{x}_{k|k} = \E\left\{ x_{k} | \mathcal{Y}_{k}\right\}$ and $\hat{x}_{k+1|k} = \E\left\{ x_{k+1} | \mathcal{Y}_{k}\right\}$, respectively; hence, $\hat{x}_{k+1|k}$ is the predicted CPU utilization for time-interval $k+1$. In order to approach the desired CPU allocation, as given in \eqref{stoch3}, the CPU allocation mechanism uses the prediction of the usage and is thus given by
\begin{equation}\label{stoch4}
	a_{k+1} =\max\left\{a_{\min}, \min \{(1+h) \hat{x}_{k+1|k}, a_{\max}\} \right\}.
\end{equation}

% ===============================================
% Computation of variances/covariances
% ===============================================
\subsection{Computation of variances/covariances}\label{statistics}
To estimate the variance of the process noise at time step $k$, $W_k$, using real-data for each component, we use a sliding window approach in which the variance of the data belonging in a sliding window of size $T$ steps at each time step $k$ is computed.  Initially, the variance is chosen based on some prior information. $T$ steps after the process is initiated, and $T$ CPU usages have been stored, the variance is estimated. 

While the mean of a \emph{random-walk-without-a-drift} is still zero, the covariance is non-stationary. For example, for the SISO case,
\begin{align*}
\var\{x_k\} &= \var\{w_{k-1} + w_{k-2 } + ... \} \\
&=  \var\{w_{k-1}\} +  \var\{w_{k-2}\} +\ldots + \var\{w_0\} \\
&= W_{k-1} + W_{k-2} + \ldots W_{0}.
\end{align*}
By taking the difference between two observations, i.e., $z_k\triangleq y_{k}-y_{k-1}$, we get:
\begin{align*}
z_k&=x_{k}-x_{k-1}+v_{k}-v_{k-1} \\
&= w_{k-1}+v_{k}-v_{k-1}.
\end{align*}
The variance of the difference between observations is thus
\begin{align*}
\var\{z_k\}&=\var\{ w_{k-1}+v_{k}-v_{k-1}\} \\
&= W_{k-1}+V_{k}-V_{k-1}.
\end{align*}
While the experiment is running, the \emph{last} $T$ CPU usages is stored and used for updating the variance at each step $k$. Computing the variance based on the difference between observations, we get: 
\begin{align*}
&\var\{z_{k-T+1}+ \ldots + z_k\}=\var\{z_{k-T+1}\}+\ldots+\var\{z_k\} \\
& \quad = W_{k-T}+V_{k-T+1}-V_{k-T}+ \ldots + W_{k-1}+V_{k}-V_{k-1}.
\end{align*}
The measurement noise variance $V_k$ was set to a small value because we observed that since the CPU usage is captured every $1$s, and therefore, the measurement is relatively accurate. In other words, our measurements of the CPU usage are relatively very close to the real ones. This fact let us pin the measurement noise variance to a fixed value (herein $V_k=1$). 
As a result, the variance of the difference breaks down to
\begin{align}\label{eq:var1}
\var\{z_{k-T+1}+ \ldots + z_k\} = W_{k-T} +\ldots+ W_{k-1}.
\end{align}
Hence, using \eqref{eq:var1} and assuming that the variance does not change (much) over a time horizon $T$, the estimate of the variance at time $k$, denoted by $\widehat{W}_k^{\mathrm{SISO}}$, is given by:

\begin{align}\label{eq:variance}
\widehat{W}_k^{\mathrm{SISO}} &{=} \frac{1}{T} \var\{z_{k-T+1}+\ldots + z_k \}  \nonumber                                  \\
& = \frac{1}{T}\left\{ \frac{\sum_{t=k-T+1}^{k} z_{t}^2}{T} - \left(\frac{\sum_{t=k-T+1}^{k} z_{t}}{T}\right)^2 \right\}.
\end{align}

The process noise covariance of the components is calculated using a similar methodology \emph{mutatis mutandis} as the variances. At this point, we have to capture each component's CPU usage somewhere centrally (e.g., on the MIMO controller node) in order to compute the covariances using the approach of sliding window, as before. The estimate of the covariance $\widehat{W}_k^{\mathrm{MIMO}}$ for the two components of our system is given by:
\begin{align}\label{eq:covariance}
\widehat{W}_k^{\mathrm{MIMO}} & \stackrel{(a)}{=} \cov\{z_{1,t}, \ldots, z_{1,k} \textbf{ , } z_{2,t}, \ldots, z_{2,k} \}  \nonumber            \\
& = \left(\frac{\sum_{t=k-T+1}^{k} \left(z_{1,t}-\mu_{z_1}\right)  \left(z_{2,t}-\mu_{z_2}\right)}{T}\right) ,
\end{align}
where $\mu_{z_1}$ and $\mu_{z_2}$ denote the mean CPU usages for the Web Server and Database Server components, respectively, for a window of size $T$, and are given by
\begin{align*}
\mu_{z_1} = \left(\frac{\sum_{t=k-T+1}^{k} z_{1,t}}{T}\right) \ \ \text{and} \ \ \mu_{z_2} = \left(\frac{\sum_{t=k-T+1}^{k} z_{2,t}}{T}\right),
\end{align*}
while $z_{1,t}$ and $z_{2,t}$ denote the differences between observed CPU utilizations at time instant $t$ and $t-1$ of the first and the second component of the application, respectively.

\begin{remark}
Note that the size of the sliding window, $T$, is chosen to be large enough so that it captures the variance of the random variable, but also it is small enough so that it can also track the change in variance due to changes in the dynamics of the requests. Numerical investigation helps in choosing the sliding window $T$; see Section~\ref{performance}.
\end{remark}

Note that each VM can be controlled either \emph{locally} or via a \emph{remote} physical machine. Using the locally controlled VM as a SISO system, the estimate of the VM's variance can be obtained, but the noise covariances with respect to other applications cannot be obtained. Using a remotely controlled VM to host the MIMO controller, the noise covariances of the whole system can be estimated via \eqref{eq:covariance}.

% ===============================================
%
%
% CONTROLLER DESIGN
%
%
% ===============================================
\section{Controller design}\label{controller}

This work emphasizes robust dynamic resource provisioning that accounts for model uncertainties and non-Gaussian noise. Two robust controllers are proposed in order to predict and hence allocate the CPU resources in a realistic scenario for each VM that constitutes the RUBiS application. More specifically:
\begin{itemize}
	\item \textbf{\hinf filter:} This controller minimizes the worst-case estimation error of the CPU allocation and provides robust state estimation. It can be modeled either as a SISO filter to control a single VM or as MIMO filter to control all VMs of a multi-tier application.
	\item \textbf{MCC-KF:} This controller is an enhanced Kalman filter version that utilizes the Maximum Correntropy Criterion for the state estimation of the CPU resources. Note the MCC-KF measures the similarity of two random variables  using information of high-order signal statistics, essentially handling cases of non-Gaussian noises (which are not directly handled by the standard Kalman filter. As in the \hinf filter case, this controller can be also modeled as a SISO system (controlling a single VM of a multi-tier application), or as a MIMO system (controlling all VMs).
\end{itemize}

% ===============================================
% H_INFINITY FILTER
% ===============================================
\subsection{\hinf Filter}

$\mathcal{H}_{\infty}$ filters, called \emph{minimax} filters, minimize the worst-case estimation error hence facilitates better robustness for the state estimation problem. In this work, we adopt a game theoretic approach to $\mathcal{H}_{\infty}$ filters proposed in \cite{Ban:1992} and thoroughly described in \cite[Chapter~11]{2006:hinf}.

The cost function for our problem formulation is given by:
\begin{align}\label{cost}
	J=\frac{\sum_{k=0}^{N-1}\| x_{k}-\hat{x}_{k|k} \|_{2}^{2}}{\| x_0-\hat{x}_{0|0}\|_{P_{0|0}^{-1}}^{2}+\sum_{k=0}^{N-1}\left(\| {w}_{k}\|_{W_{k}^{-1}}^{2}+\| {v}_{k}\|_{V_{k}^{-1}}^{2}\right)}
\end{align}
where $P_{0|0} \in \mathbb{R}^{N \times N}$, $W_k \in \mathbb{R}^{N \times N}$ and $V_k\in  \mathbb{R}^{N \times N}$ are symmetric, positive definite matrices defined by the problem specifications, i.e., $P_{0|0}$ is the initial error covariance matrix, $W_k$ and  $V_k$ are the process and measurement covariance matrices for time interval $k$, respectively; $\hat{x}_{k|k}$ is the estimate of the CPU allocation. The direct minimization of $J$ in \eqref{cost} is not tractable and, therefore, a performance bound is chosen, i.e., $J<1/\theta$, $\theta>0$, and attempt to find an estimation strategy (controller, in this case) that satisfies the bound. In our problem, the target is to  keep the $\mRT$ below a certain threshold (e.g., less than $0.5$s).  Therefore, $\theta$ is tuned such that  the desired $\mRT$ is less than a  certain user-specified threshold, i.e., so that the designed controller satisfies the desired target. The choice of $\theta$ will be investigated in Section~\ref{performance}. Considering \eqref{cost}, the steady-state \hinf filter bounds the following cost function:
\begin{align}\label{cost1}
	J=\lim_{N \rightarrow \infty}\frac{\sum_{k=0}^{N-1}\| x_{k}-\hat{x}_{k|k} \|_{2}^{2}}{\sum_{k=0}^{N-1}\left(\| {w}_{k}\|_{W_{k}^{-1}}^{2}+\| {v}_{k}\|_{V_{k}^{-1}}^{2}\right)}.
\end{align}
Let $G_{\hat{x}{e}}$ be the system that has ${e}=[{w} \ \ {v} ]^T$ as its input and $\hat{x}$ as its output. Since the \hinf filter makes cost \eqref{cost1} less than $1/\theta$ for all ${w}_{k}$ and ${v}_{k}$, then according to~\cite[Equation~(11.109)]{2006:hinf}:
\begin{align}\label{system_hinf}
	\|G_{\hat{x}{e}}\|_{\infty}^{2}=\sup_{\zeta}\frac{\| x-\hat{x} \|_{2}^{2}}{\| {w}\|_{W^{-1}}^{2}+\| {v}\|_{V^{-1}}^{2}} \leq \frac{1}{\theta},
\end{align}
where $\zeta$ is the phase of $\| {w}\|_{W^{-1}}^{2}+\| {v}\|_{V^{-1}}^{2}$ comprised by the sampling time of the system and the frequency of the signals. Since we want the $\mRT$ to be less than a certain value (usually around $1$ second), we have to keep the CPU usage to less than a threshold set by our $\mRT$ model. Therefore, using \eqref{system_hinf} we want:
\begin{align}\label{theta_phi}
	\sup_{\zeta}\frac{\| D \|_{2}^{2}}{\| {w}\|_{W^{-1}}^{2}+\| {v}\|_{V^{-1}}^{2}} \leq \frac{1}{\theta},
\end{align}
which is equivalent to:
\begin{align}\label{thetais}
	\theta \leq \inf_{\zeta}\frac{\| {w}\|_{W^{-1}}^{2}+\| {v}\|_{V^{-1}}^{2}}{\| D \|_{2}^{2}}.
\end{align}
where $D$ is a diagonal matrix with the allowable error for each component along the diagonal.

Let the \emph{a posteriori} (updated) and \emph{a priori} (predicted) error covariances be given by
\begin{align*}
	P_{k|k}   & = \E\left\{ (x_k-\hat{x}_{k|k}) (x_k-\hat{x}_{k|k})^T | \mathcal{Y}_k\right\} ,      \\
	P_{k+1|k} & = \E\left\{  (x_k-\hat{x}_{k+1|k})(x_k-\hat{x}_{k+1|k})^T | \mathcal{Y}_k \right\} .
\end{align*}
The necessary condition to ensure that $P_{k|k}$ remains positive definite and the system retains stability for the \hinf filter is that:
\begin{align}\label{condition}
	I-\theta P_{k|k-1} +C^{T}V_{k}^{-1}CP_{k|k-1} \succ 0.
\end{align}
To design the controller we consider inequalities~\eqref{thetais} and~\eqref{condition}.

The equations for the \hinf filter are summarized below \cite{2006:hinf}. For the \emph{prediction phase}:
\begin{subequations}
	\begin{align}
		\hat{x} _{k|k-1} & = A\hat{x} _{k-1|k-1}, \label{state_prior_hinf}        \\
		P_{k|k-1}        & = A P_{k-1|k-1}{A^T} + {W_k}. \label{covar_prior_hinf}
	\end{align}
	For the cost function \eqref{cost}, the \emph{update phase} of the $\mathcal{H}_{\infty}$ filter is given by:
	\begin{align}\label{controller_equations}
		K_{k}         & =P_{k|k-1}[I-\theta P_{k|k-1}+C^{T}V_{k}^{-1}CP_{k|k-1}]^{-1}C^{T}V_{k}^{-1} \\
		\hat{x}_{k|k} & = \hat{x}_{k|k-1} +K_{k}({y}_{k}-C \hat{x}_{k|k-1} )                         \\
		P_{k|k}       & =  P_{k|k-1}[I-\theta P_{k|k-1}+C^{T}V_{k}^{-1}CP_{k|k-1}]^{-1} \label{P}
	\end{align}
\end{subequations}
where $K_{k}$ is the gain matrix.

When using a SISO controller for our SISO model given by \eqref{stoch1}-\eqref{stoch2}, then the \emph{prediction phase} is given by
\begin{subequations}
	\begin{align}
		\hat{x} _{k|k-1} & = \hat{x} _{k-1|k-1}, \label{state_prior_hinf_SISO}   \\
		P_{k|k-1}        & =  P_{k-1|k-1} + {W_k}, \label{covar_prior_hinf_SISO}
	\end{align}
	and the \emph{update phase} is given by
	\begin{align}
		{K}_k         & = \frac{{P}_{k|k-1}}{V_k\left (1- \theta{P}_{k|k-1}+{P}_{k|k-1} V_k^{-1}  \right)}, \\
		\hat{x}_{k|k} & = \hat{x}_{k|k-1} + {K}_k \left( y_k -  \hat{x}_{k|k-1} \right),                    \\
		{P}_{k|k}     & = \frac{P_{k|k-1}}{1- \theta{P}_{k|k-1}+{P}_{k|k-1} V_k^{-1}}.
	\end{align}
\end{subequations}

The Kalman filter gain is  less than the $\mathcal{H}_{\infty}$ filter gain for $\theta > 0$,  meaning that the $\mathcal{H}_{\infty}$ filter relies more on the  measurement and less on the system model. As $\theta\rightarrow 0$, the $\mathcal{H}_{\infty}$ filter gain and Kalman filter gain coincide \cite{2006:hinf}. For a comparison between Kalman and \hinf filters see \cite{2012:Comparison}.

% ===============================================
% MCC Kalman Filter
% ===============================================
\subsection{Maximum Correntropy Criterion Kalman Filter}

In this section, a new Kalman filter approach is deployed that uses the Maximum Correntropy Criterion (MCC) for state estimation, referred in literature as MCC Kalman filter (MCC-KF) in \cite{CHEN:2017} and \cite{izanloo_kalman_2016}. The correntropy criterion measures the similarity of two random variables using information from high-order signal statistics \cite{liu_correntropy:_2006,liu_correntropy:_2007,he_maximum_2011,singh_using_2009}.
Since the Kalman filter uses only second-order signal information is not optimal if the process and measurement noises are non-Gaussian noise disturbances, such as shot noise or mixture of Gaussian noise.

\noindent The equations for the MCC-KF are summarized below \cite{izanloo_kalman_2016}. For the \emph{prediction phase}:
\begin{subequations}
	\begin{align}
		\hat{x} _{k|k-1} & = A\hat{x} _{k-1|k-1}, \label{state_prior}        \\
		P_{k|k-1}        & = A P_{k-1|k-1}{A^T} + {W_k}, \label{covar_prior}
	\end{align}
	and for the \emph{update phase}:
	\begin{align}
		{L_k}          & = \frac{{{G_\sigma }\left( {\parallel {y_k} - C{{\hat{x} }_{k|k-1}}{\parallel _{V_k^{ - 1}}}} \right)}}{{{G_\sigma }\left( {\parallel  \hat{x} _{k|k-1}  - A{{\hat{x} }_{k - 1|k-1}}{\parallel _{P_{k|k - 1}^{ - 1}}}} \right)}}, \label{mcc_innov_covar_posterior} \\
		{K_k}          & = {(P_{k|k - 1}^{ - 1} + {L_k}{C^T}V_k^{ - 1}C)^{ - 1}}{L_k}{C^T}V_k^{ - 1}, \label{gain}                                                                                                                                                                           \\
		\hat{x} _{k|k} & = \hat{x} _{k|k-1} + K_k (y_k - C \hat{x} _{k|k-1} ), \label{state_posterior}                                                                                                                                                                                       \\
		{P_{k|k}}      & = (I - {K_k}C){P_{k|k - 1}}{(I - {K_k}C)^T} + {K_k}{V_k}K_k^T, \label{covar_posterior}
	\end{align}
\end{subequations}
where $G_\sigma$ is the Gaussian kernel, i.e.,
\begin{align*}
	G_{\sigma}( \parallel x_i-y_i \parallel )=\exp\left(-\frac{\parallel x_i-y_i \parallel ^2}{2\sigma^2}\right)
\end{align*}
with kernel size $\sigma$\footnote{The kernel bandwidth $\sigma$ serves as a parameter weighting the second- and higher-order moments; for a very large $\sigma$ (compared to the dynamic range of the data), the correntropy will be dominated by the second-order moment \cite{CHEN:2017}.}. Note that $L_k$ is called the \emph{minimized correntropy estimation cost function} and $K_k$ is the Kalman gain (as in the \hinf filter).

When using a SISO controller in our model, given by \eqref{stoch1} and \eqref{stoch2}, the original MCC-KF equations \eqref{state_prior}-\eqref{covar_posterior} are simplified to:
\begin{subequations}
	\begin{align}
		\hat{x} _{k|k-1} & = \hat{x} _{k-1|k-1}, \label{state_prior_siso}                                                                                                                                                                                                                         \\
		P_{k|k-1}        & = P_{k-1|k-1} + {W_k}, \label{covar_prior_siso}                                                                                                                                                                                                                        \\
		{L_k}            & = \frac{{{G_\sigma }\left( {\parallel {y_k} - {{\hat{x} }_{k|k-1}}{\parallel _{V_k^{ - 1}}}} \right)}}{{{G_\sigma }\left( {\parallel  \hat{x} _{k|k-1}  - {{\hat{x} }_{k - 1|k-1}}{\parallel _{P_{k|k - 1}^{ - 1}}}} \right)}}, \label{mcc_innov_covar_posterior_siso} \\
		{K_k}            & = \frac{L_k}{(P_{k|k - 1}^{ - 1} + {L_k}V_k^{-1})V_k}, \label{gain_siso}                                                                                                                                                                                               \\
		\hat{x} _{k|k}   & = \hat{x} _{k|k-1} + K_k (y_k - \hat{x} _{k|k-1} ), \label{state_posterior_siso}                                                                                                                                                                                       \\
		{P_{k|k}}        & = (1 - {K_k})^2{P_{k|k - 1}} + {K_k}^2{V_k}, \label{covar_posterior_siso}
	\end{align}
\end{subequations}
As it can be observed in \eqref{state_prior_siso}-\eqref{covar_posterior_siso}, MCC-KF has the same structure as the Kalman filter, but in addition, it uses high-order statistics to improve state estimation.

% ===============================================
%
%
% RESOURCE PROVISIONING ALGORITHM
%
%
% ===============================================
\subsection{Resource Provisioning Algorithm}\label{sec:algorithm}

Irrespective of which filter is being used, a generic algorithm for allocating the CPU is given in Algorithm~\ref{algorithm_resourceprovisioning}. A thorough discussion on possible filtering approaches is presented in Section~\ref{sec:relatedwork}.

\begin{algorithm}
	\begin{algorithmic}[1]
		\caption{Dynamic Resource Provisioning\label{algorithm_resourceprovisioning}.}
		\STATE \textbf{Input:} $a_{\min}$, $a_{\max}$, $h$,  $T$, $\theta$ (for the \hinf filter), $\sigma$ (for the MCC-KF), 
		\STATE \textbf{Initialization:} $W_0$, $V_0$ ($V_k=1~\forall k$), $P_{0|0}$
		\FOR{each time step $k$}
		\STATE \textbf{Data:} $y_k$
		\STATE \textbf{variances/covariances}
		\begin{ALC@g}
			\STATE Compute $\widehat{W}_k$ for SISO and MIMO controllers according to \eqref{eq:variance} and \eqref{eq:covariance}, respectively
		\end{ALC@g}
		\STATE \textbf{filter}
		\begin{ALC@g}
			\STATE \textbf{Update phase:}
			\begin{ALC@g}
				\STATE{Compute $L_k$ (for the MCC-KF)}, $K_k$, $\hat{x} _{k|k}$, $P_{k|k}$
			\end{ALC@g}
			\STATE \textbf{Prediction phase:}
			\begin{ALC@g}
				\STATE{Compute $\hat{x} _{k+1|k}$, $P_{k+1|k}$}
			\end{ALC@g}
		\end{ALC@g}
		\STATE \textbf{CPU allocation:}
		\begin{ALC@g}
			\STATE Compute $a_{k+1}$ using \eqref{stoch4}
		\end{ALC@g}
		\ENDFOR
		\STATE \textbf{Output:} CPU allocation $a_{k+1}$.
	\end{algorithmic}
\end{algorithm}

The Algorithm~\ref{algorithm_resourceprovisioning} describes the steps of our approach for dynamically provisioning the CPU resources of any cloud application which is hosted on virtualized servers and by using any estimation technique proposed herein.
\begin{list4}
\item \textbf{Step 1 (Input):} Firstly, the minimum ($a_{min}$) and maximum ($a_{max}$) allocations for application's components, the sliding window width $T$ for the computation of the variances and covariances as well as the tunable parameters $\theta$ and $\sigma$ for the \hinf and MCC-K filters, respectively, are needed as inputs to the system. 
\item \textbf{Step 2 (Initialization):} Initial values for the process and measurement noise matrices and for the initial error covariance matrix must be declared in advance.
\item \textbf{At each time step $k$}, the observed utilization, using the Xen Hypervisor, is set as the control input signal.
\begin{list4}
\item \textbf{Step 3 (Variance/Covariance Computation:)} At this step, the algorithm calculates the variances and/or the covariances of $T$ passed utilizations using the approach in Section~\ref{statistics} in order to estimate the process covariance error $\widehat{W}_k$ at each time instance $k$.
\item \textbf{Step 4 (Filtering):} Using the statistics from the previous step, the filter updates the state while it computes the $L_k$, $K_k$, $\hat{x} _{k|k}$ and $P_{k|k}$ for each time instance $k$. Right after, the filter predicts the next state of the system with computing the $\hat{x} _{k+1|k}$, $P_{k+1|k}$. With this process, the $a_{k+1}$ is computed and it can be exported as the new allocation for the next step $k+1$.
\item \textbf{Step 5 (Output):} The new predicted allocation $a_{k+1}$ is adapted in the appropriate VM using the Xen scheduler.
\end{list4}
\end{list4}

%\clearpage
% ===============================================
%
%
% EXPERIMENTAL SETUP
%
%
% ===============================================
\section{Experimental Setup}\label{setup}

The main target is to continuously provision each virtualized application with enough CPU resources to adequately serve its incoming requests from a variable workload. The purpose of the Dom0 component is to monitor the CPU usage by the percentage of CPU cycles of each VM running on the Hypervisor. The controller utilizes the CPU measurements in order to predict the CPU usage for the next time interval and hence determine the CPU allocation, which is then fed back to to the Hypervisor to set the new allocation.

\begin{figure}[h]
	\begin{center}
		\includegraphics[width=1\columnwidth]{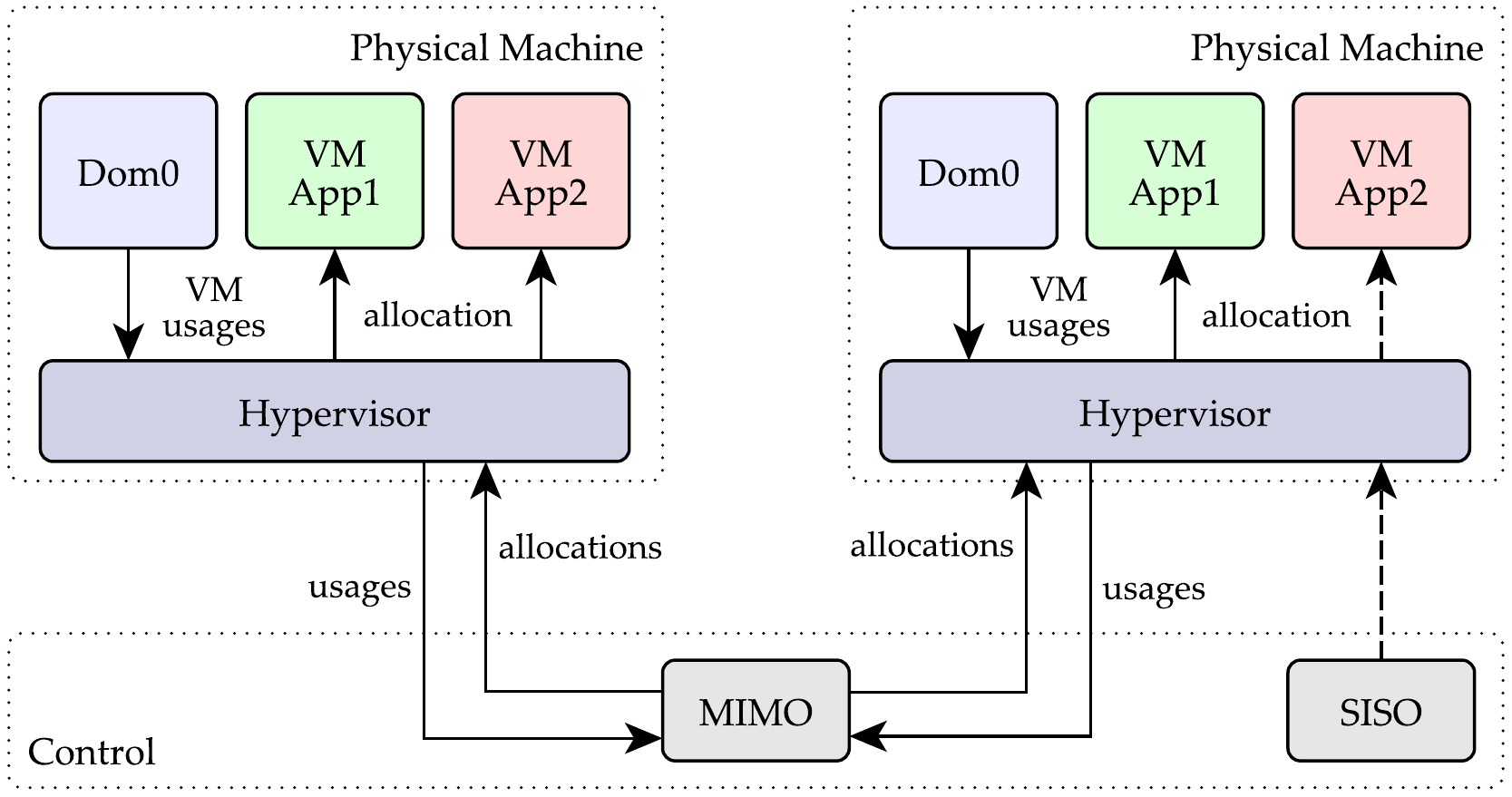}
		\caption{Resource allocation manager architecture.}
		\label{fig:architecture}
	\end{center}
\end{figure}

% ===============================================
% Experimental System Infrastructure 
% ===============================================
\subsection{Experimental Setup Base}
Our experimental setup is divided in two different infrastructure installations. One for the SISO controller model and one for the MIMO controller model. To evaluate the performance of each control system, we set up a small-scale data center in order to host the RUBiS auction site as the cloud application. The data center consists of two physical blade servers with Intel Xeon 5140 and 1.0GB of RAM, running Debian 8 Jessie Linux distribution with 3.16.0-4-amd64 kernel and Xen 4.4.1 Virtualization technology. Note that Xen Hypervisor was also has been widely used for experimental evaluations in the literature; for example, \cite{selfAdaptive}, \cite{zhu2012resource} and \cite{Makridis:2017}. These physical machines are used for hosting the VMs of the two-tier RUBiS benchmark application. Each physical machine, namely PM1 and PM2, hosts a VM running on Debian Jessie 8 Linux with Apache 2.4.10 web server and  MySQL 5.5.55 database respectively. Note that, this infrastructure does not reflect the performance of modern servers, but it adequately serves our purpose of studying the performance of the controllers over RUBiS workload. For each physical machine we created a couple of configurations on the Xen Credit Scheduler overriding the default \emph{time-slice} and \emph{rate-limit} values. The default values for the Credit Scheduler are $30$ms for the time-slice and $1$ms for the rate-limit. We set rate-limit unchanged at it's default value and time-slice at $10$ms, since we determined experimentally that reducing the time-slice value increased the performance of the RUBiS application.

% ===============================================
% RUBiS
% ===============================================
\subsection{Benchmark Application - RUBiS}
Rice University Bidding System (RUBiS), an auction site benchmark, implements the core functionality of an auction site, i.e., selling, browsing and bidding. It is modeled after ebay.com and involves a client-browser emulator, a web server, an application server and a database. It was originally used to evaluate application design patterns and application servers performance scalability. In our work, RUBiS is hosted on a two-tier application model (web server, database), while the Client Emulator generates the workload for the RUBiS application. Several RUBiS implementations exist using Java Servlets, PHP, and Enterprise Java Bean (EJB) technologies. It has provisions for selling, browsing and bidding items, allowing for different sessions for different type of users in the form of visitor, buyer and seller to be implemented. RUBiS auction site defines 26 type of actions that can be performed through client's Web browser. In our work, the clients were modeled using the Client Emulator which is mentioned below. The MySQL database contains 7 tables which stores: bids, buy now, categories, comments, items, regions and users. With cloud computing increasingly attracting the attention of researchers, RUBiS became the classic real-data benchmark for resource management problems \cite{selfAdaptive, resourceProvisioning, Makridis:2017, adaptiveControlV, automatedControl}.

% ===============================================
% Client Emulator
% ===============================================
\subsection{Client Emulator}
The Client Emulator is hosted on a third physical machine (PM3) which is dedicated for generating the RUBiS auction site workload. A Java code is responsible for generating the workload and creating user sessions to send HTTP requests to the RUBiS auction site for the purpose of emulating the client's behavior. The original version of Client Emulator provides visual statistics for throughput, response times and other information for the sessions. However, in our experiments we modified the original Client Emulator's source code in order to capture the response time of each completed request and as a next step to calculate the $\mRT$ each second or time interval. For more information about the workload generation see \cite{RubisSpecificationImplementation}.

% ===============================================
% Resource Allocation Control
% ===============================================
\subsection{Resource Allocation Control}
All controllers presented in this work were added on the base project code called ViResA\footnote{ViResA (Virtualized [server] Resource Allocation) is a base project code hosted in Atlassian Bitbucket (https://bitbucket.org) as a private Git repository. For download requests, please contact authors.}, first developed for the synthetic data generation and the performance evaluation of the controllers in \cite{ecc:2016} and later for the real-data performance evaluation of the \hinf and MCC-KF SISO controllers in \cite{Makridis:2017}. The performance of the dynamic CPU allocation is evaluated using the $\mRT$, which is measured at the Client-side of our prototype RUBiS server application, as the performance metric. The goal of the control system is to adapt the CPU resource allocations of a single VM or group of VMs in exchange for saving resources for other applications that are hosted on the same physical machine. The resource allocation can be managed using the SISO or the MIMO model of ViResA application respectively.

There are two parameters by which a server's performance can be affected: \emph{(i)} the number of clients that send requests simultaneously to the server and \emph{(ii)} the workload type. The workload scheme used in our work will be discussed further in the next section.

% ===============================================
% Single Tier Control
% ===============================================
\subsubsection{Single Tier Control}\label{siso_tier_control}
An overview of the SISO system's architecture is shown in Fig.~\ref{siso-system} below. The web server component and the database server component communicate with each other via HTTP requests, as shown in Fig.~\ref{figure2}
Each VM of this setup is controlled in parallel via the SISO model (e.g., Kalman SISO, \hinf SISO and MCC-KF SISO) while keeping them isolated from each other. However, changes in the demand affect both components in a relative way because each request follows a path from client to Web Server to Database Server and back, as shown in Fig.~\ref{figure2}. Hence, the SISO system is not able to calculate the correlation between the two components. The CPU usage measurements are recorded every $1$s using the Xen Hypervisor through the Domain-0, which also hosts the controllers for each component. With this setup, there are not any network delays for the allocation control which is adapted normally every $1$ sample ($5$s).
\begin{figure}[!h]
	\centering
	\includegraphics[width=0.75\columnwidth]{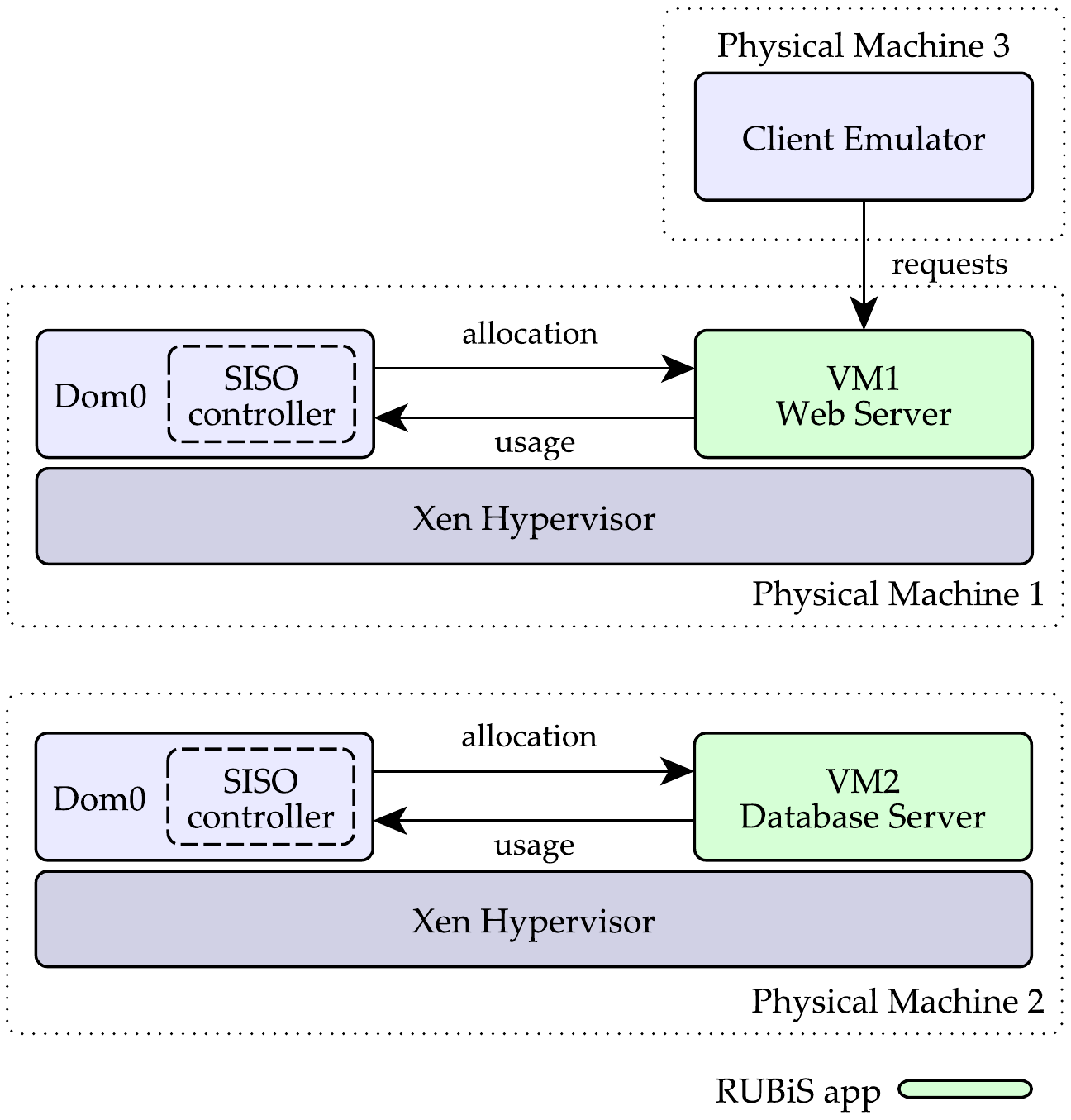}
	\caption{SISO system architecture.}\label{siso-system}
\end{figure}

% ===============================================
% Multiple Tier Control
% ===============================================
\subsubsection{Multiple Tier Control}
An overview of the MIMO system's architecture is shown in Fig.~\ref{mimo-system} below.
On this setup, each VM is controlled remotely (i.e., with an external physical machine (PM4) over network) via the MIMO model (e.g., Kalman MIMO, \hinf MIMO and MCC-KF MIMO). Similar to the case of the SISO model, the workload of the application affects both components in a correlated manner. Nevertheless in the MIMO system model, the resource couplings between the two components are calculated using the covariance computation method mentioned in Section~\ref{statistics}. The CPU usage measurements are recorded with the same method as in the SISO case, however the controllers are deployed on the same external physical machine for calculating the covariances between the two components.

\begin{figure}[!h]
	\centering
	\includegraphics[width=0.90\columnwidth]{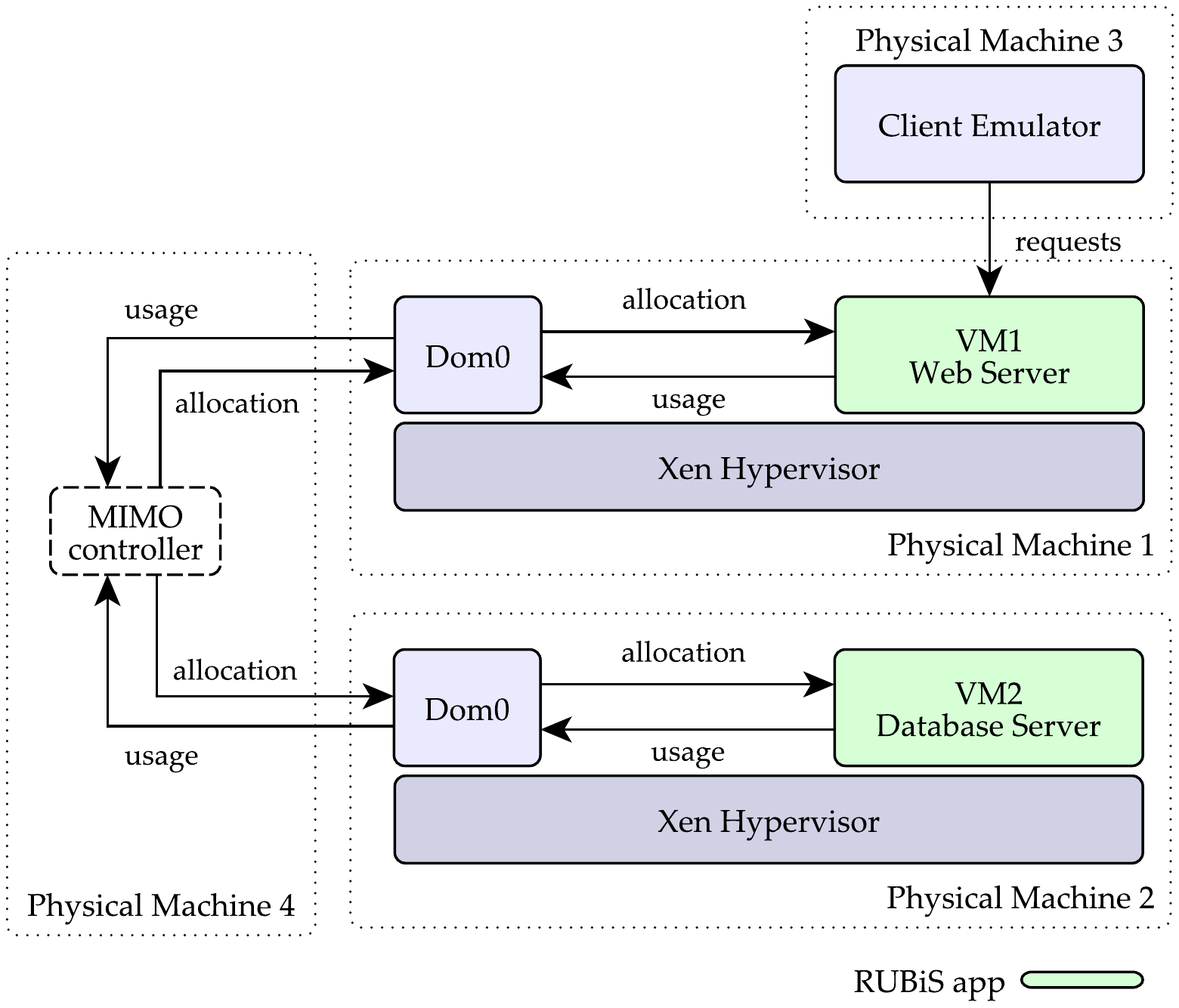}
	\caption{MIMO system architecture.}\label{mimo-system}
\end{figure}

% ===============================================
%
%
% PERFORMANCE EVALUATION
%
%
% ===============================================
\section{Performance Evaluation}\label{performance}

% ===============================================
% Evaluation Metrics
% ===============================================
\subsection{Evaluation Metrics}
For each experiment that will be presented in this section, several evaluation metrics have been calculated and captured, in order to get a full picture besides of the basic performance metric (i.e., $\mRT$).
	\begin{list4}
		\item \textbf{Completed Requests (CR):} denotes the total number of completed requests for all users during the full time duration of each experiment.
        \item \textbf{Average VM1 CPU usage:} denotes the average CPU usage of the Web Server component (VM1) for the full time duration of each experiment.
        \item \textbf{Average VM2 CPU usage:} denotes the average CPU usage of the Database Server component (VM2) for the full time duration of each experiment.
		\item \textbf{SLO Obedience (SLOO):} denotes the ratio of the requests with $\mRT$ below the QoS threshold (e.g., $\mRT$ $\leq$ $0.5$s), over the total number of completed requests (CR).
		\item \textbf{Average $\mRT$ (A$\mRT$):} denotes the average $\mRT$ for the full time duration of each experiment.
	\end{list4}

% ===============================================
% Sliding Window
% ===============================================
\subsection{Sliding window}
As discussed in Section~\ref{statistics}, the size of sliding window, $T$, is chosen carefully such as to be adequately large for capturing the variance of the random variable, but at the same time small enough for tracking the change in variance due to changes in the dynamics of the requests.
For choosing the sliding window size $T$, we setup a number of experiments keeping different values of $T$ each time in order to observe the behavior of the variance for each value of $T$. In the following experiment we recorded the CPU usage of the web server component under a workload applied on RUBiS application. The workload initiates with $500$ clients sending requests to RUBiS application. At sampling point $20$, another $500$ clients are inserted to the workload for almost another $30$ samples (one sample corresponds to $5$s). Then, at sampling point $100$ the number of clients sending requests to RUBiS rise up from $500$ to $1500$ clients for $10$ samples. Finally, at sampling point $150$, $1200$ clients send requests to RUBiS for $30$ more samples. The workload of this experiment provides a thorough CPU usage variability thus making the dynamics of the system detectable with different values of window sizes ($T$).

\begin{figure*}[!ht]
\centering
\mbox{\vbox{
\subfigure[\label{statistics_subfig_a}]{\includegraphics[width=.32\textwidth]{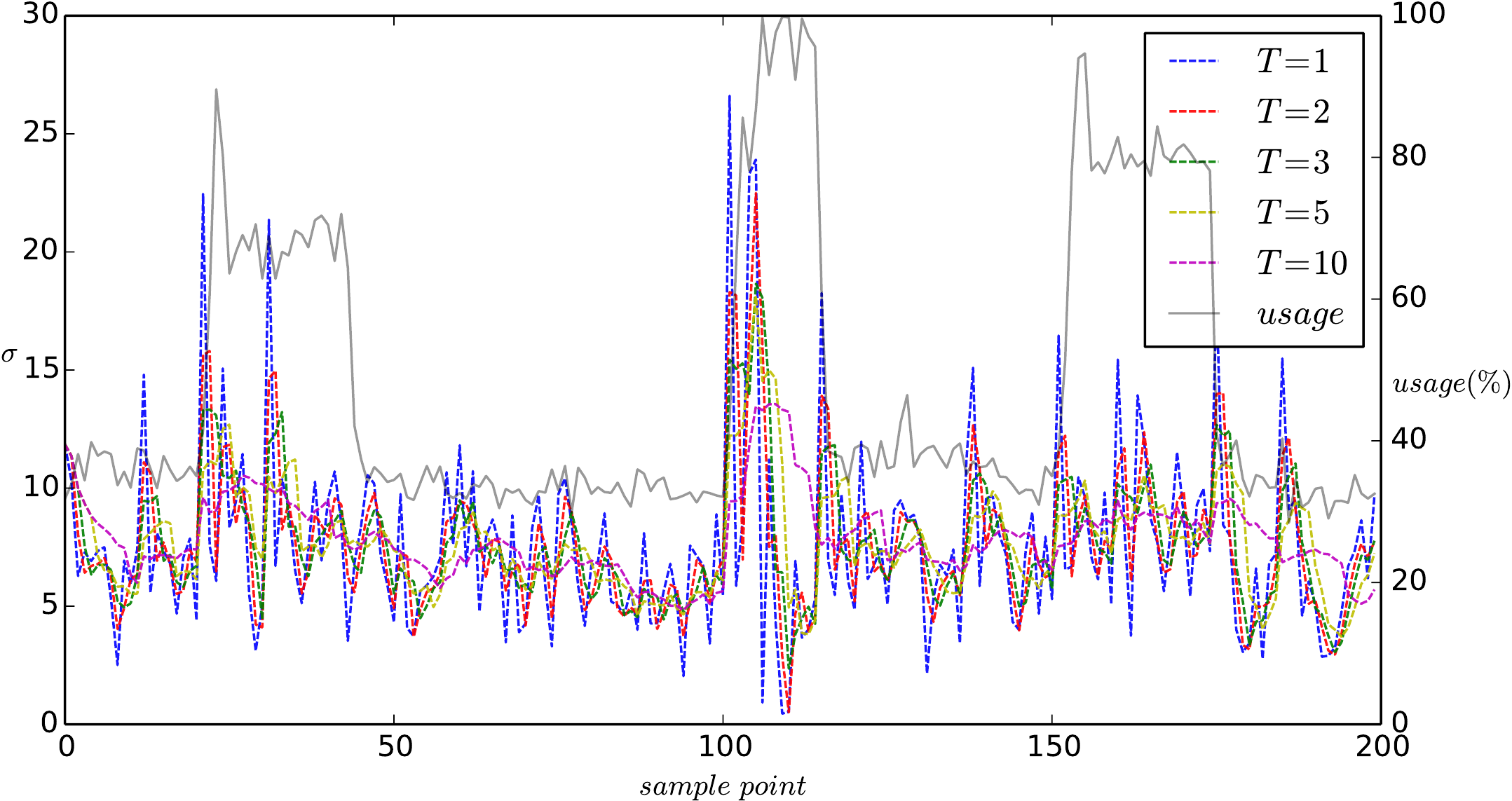}}\quad
\subfigure[\label{statistics_subfig_b}]{\includegraphics[width=.32\textwidth]{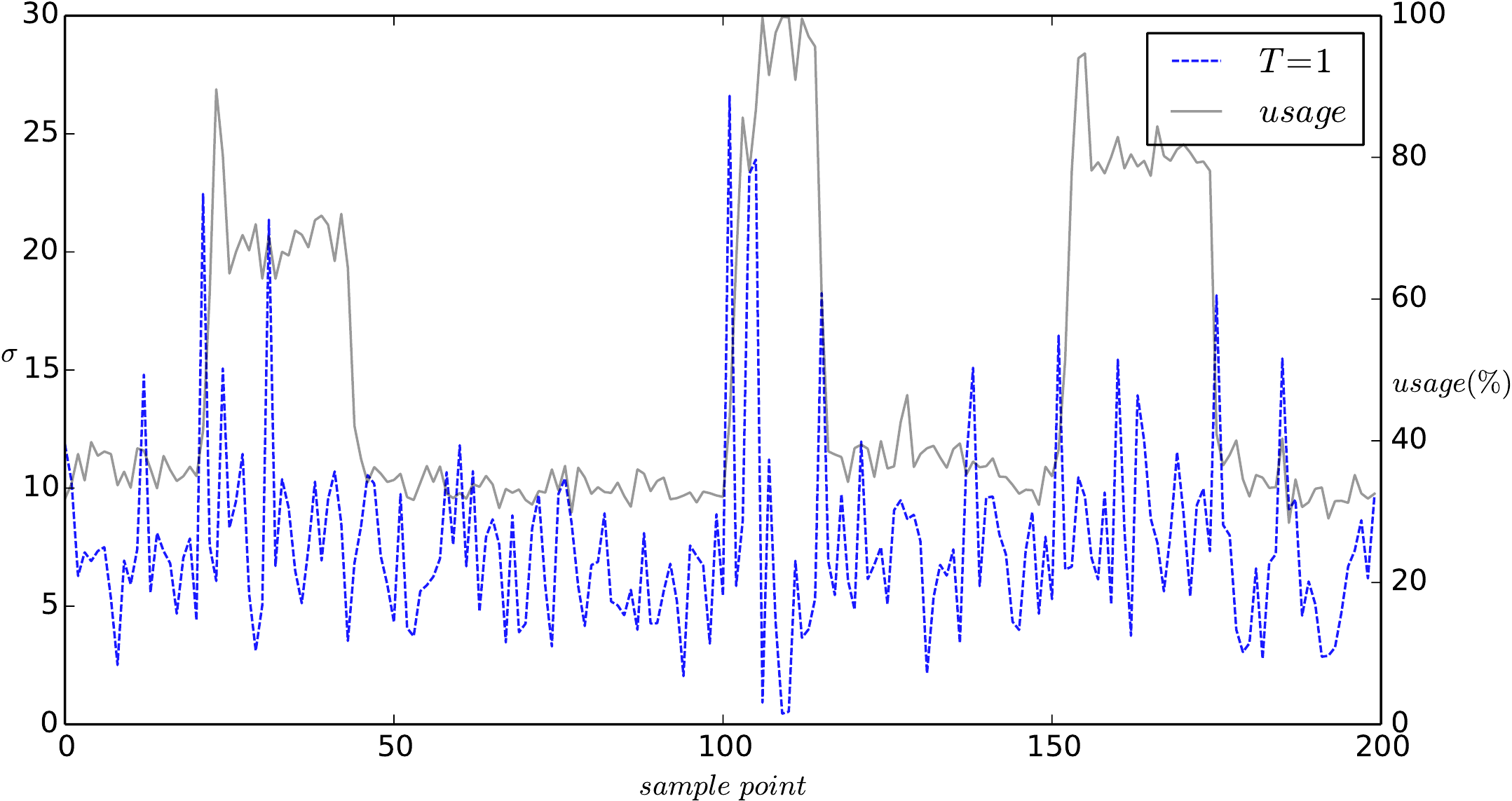}}\quad
\subfigure[\label{statistics_subfig_c}]{\includegraphics[width=.32\textwidth]{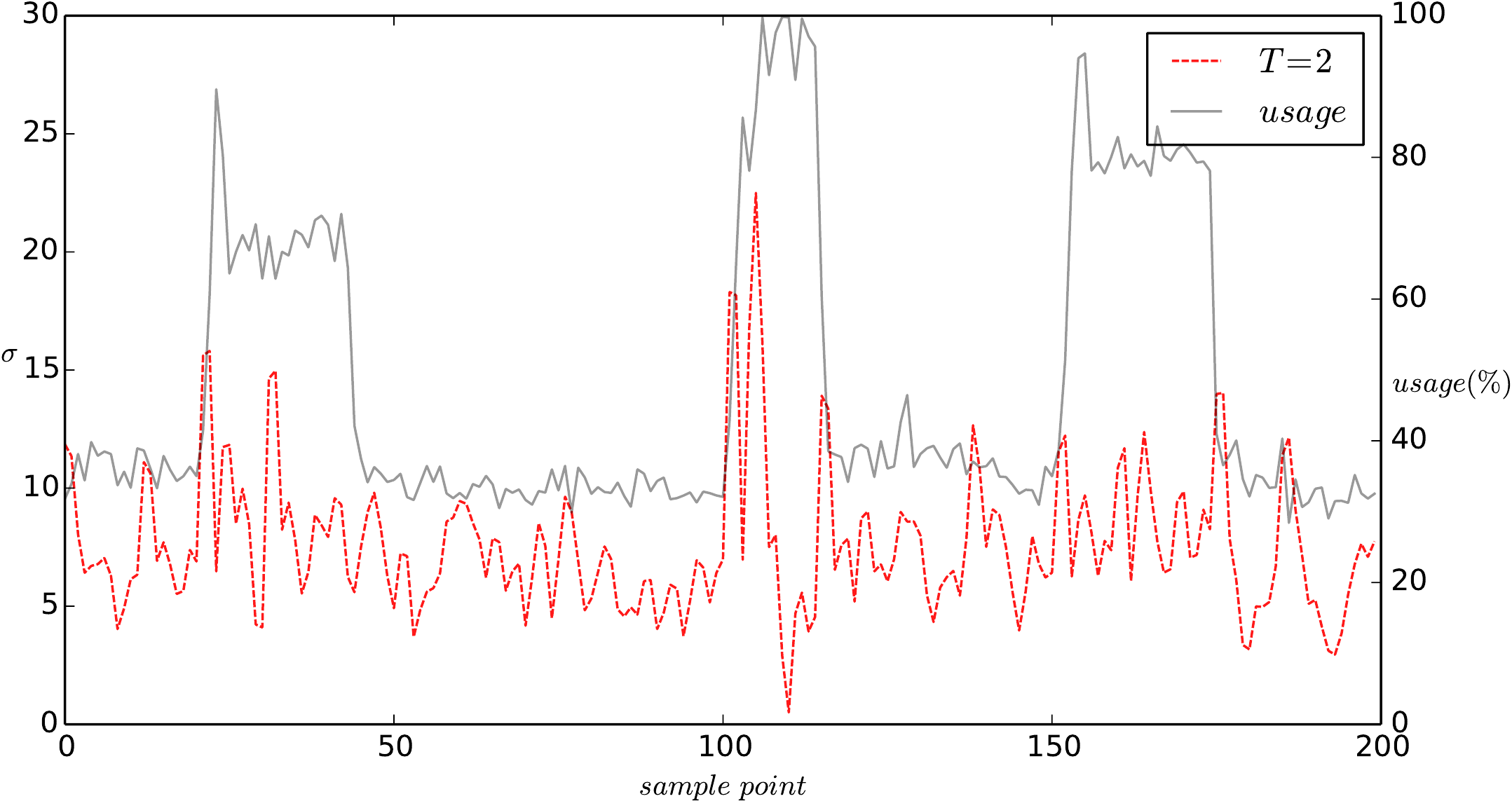}}\quad%\newline\hspace*{-0.3cm}
\subfigure[\label{statistics_subfig_d}]{\includegraphics[width=.32\textwidth]{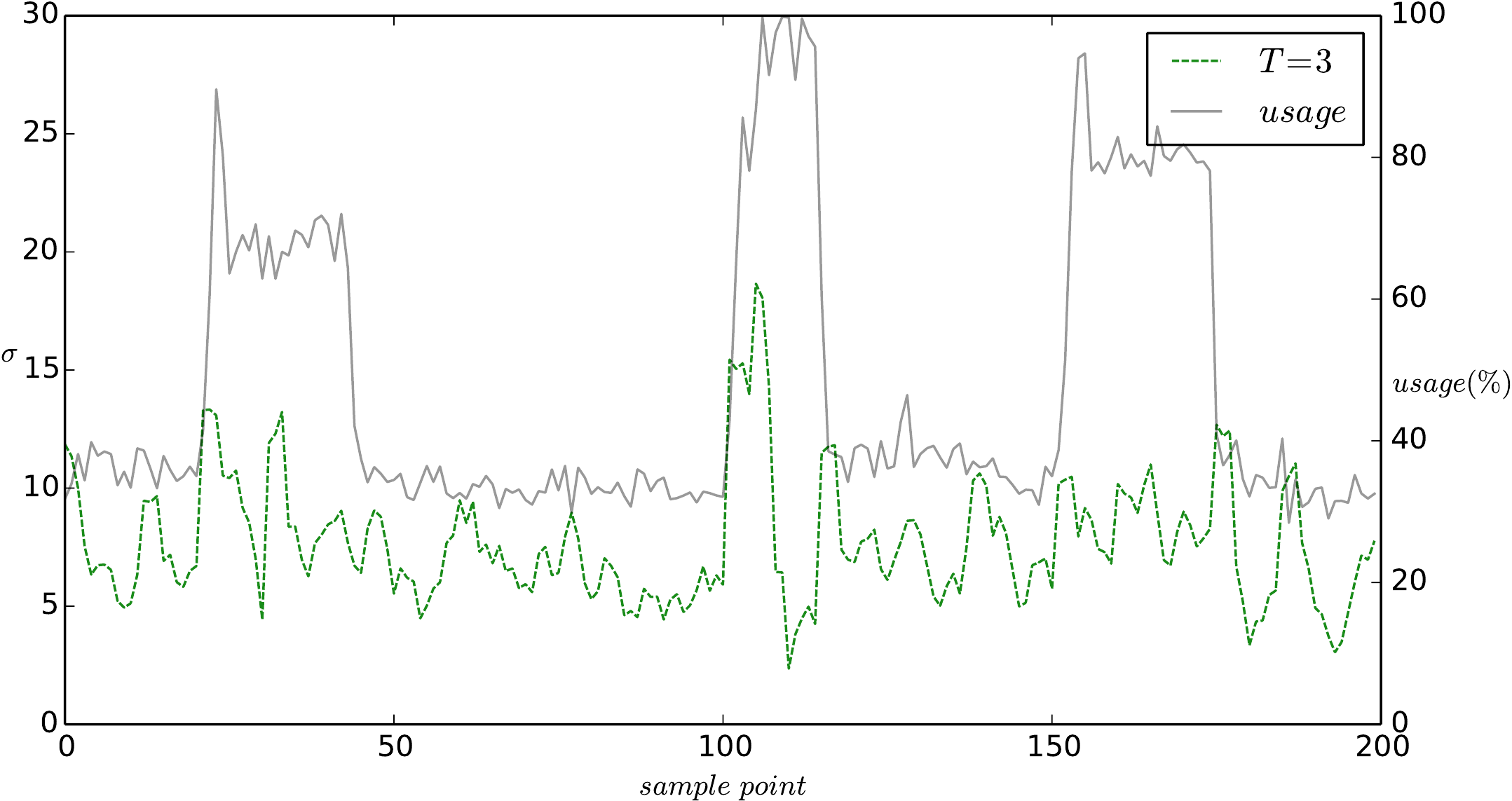}}\quad
\subfigure[\label{statistics_subfig_e}]{\includegraphics[width=.32\textwidth]{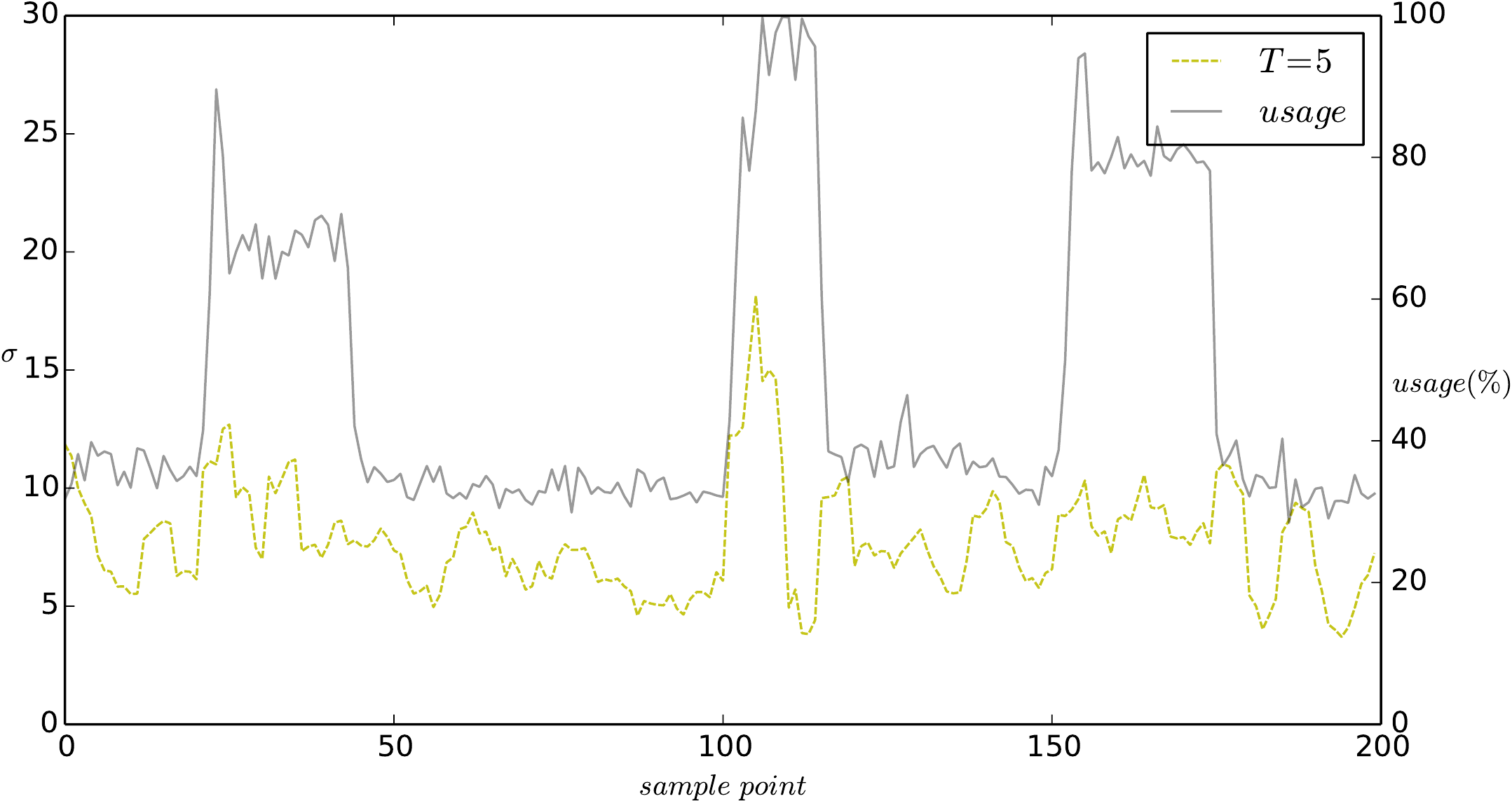}}\quad
\subfigure[\label{statistics_subfig_f}]{\includegraphics[width=.32\textwidth]{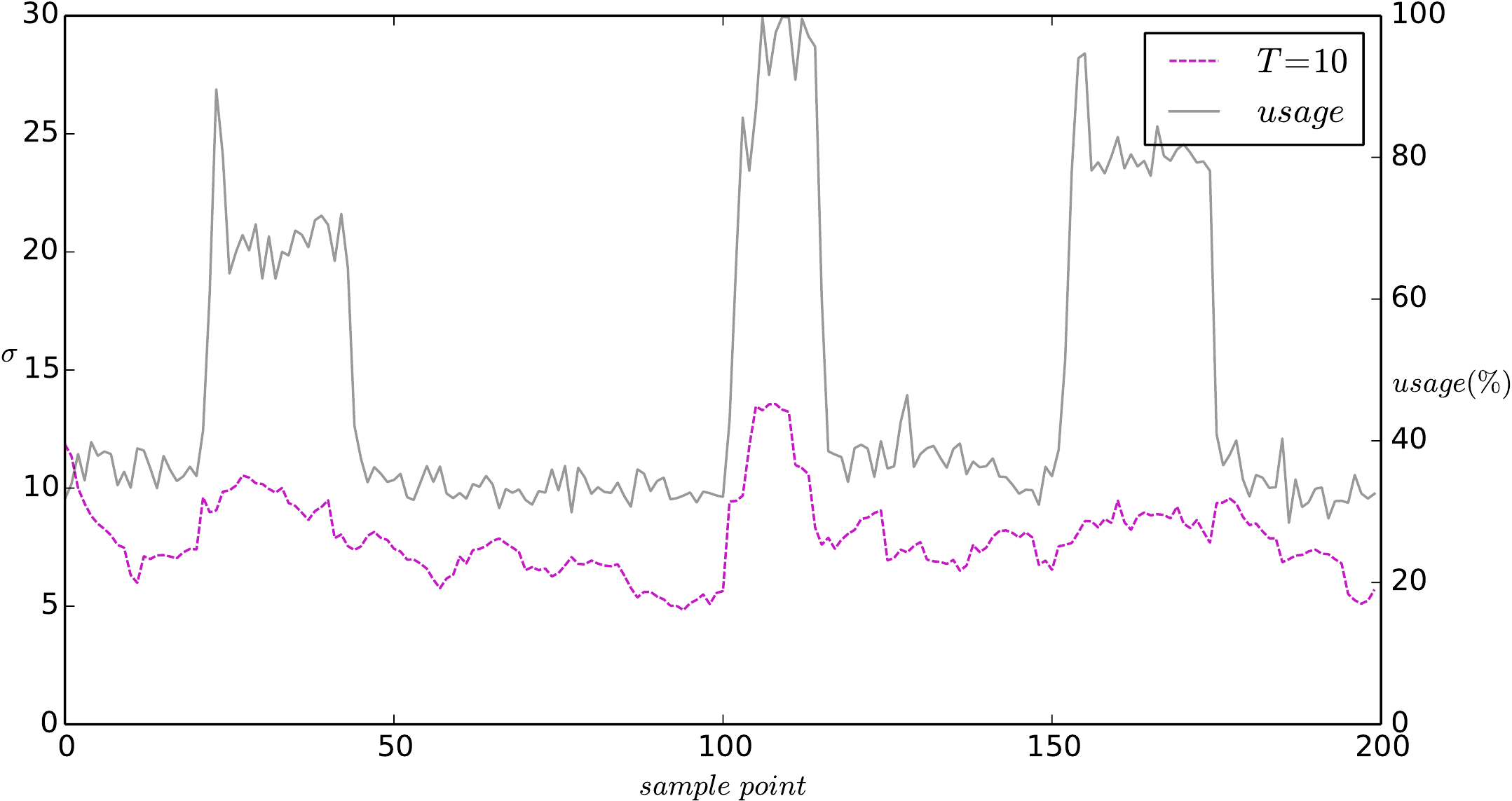}}\quad}}
\caption{Standard deviation ($\sigma$) of the CPU usage signal for different window sizes (T) for the same workload scheme. Fig.~\ref{statistics_subfig_a}: Standard deviation ($\sigma$) of the CPU usage all different values of window sizes ($T$). Fig.~\ref{statistics_subfig_b}-\ref{statistics_subfig_f}: Standard deviation ($\sigma$) of the CPU usage using window size $T=1$, $T=2$, $T=3$, $T=5$ and $T=10$ sampling points respectively.}
\end{figure*}

From Fig.~\ref{statistics_subfig_b} and ~\ref{statistics_subfig_c} it is easy to see that for $T=1$ and $T=2$ the change in variance is too noisy and sensitive to mild workload changes which can be observed by the abruptness and the sharpness of the standard deviation. For $T=10$ (Fig.~\ref{statistics_subfig_f}), we can see that the variance is not following the variability of the workload and if it does to a certain extend, this is delayed considerably. The best responses of the variance are given for $T=3$ and $T=5$ (Fig.~\ref{statistics_subfig_d} and \ref{statistics_subfig_f}), in which cases it is less sensitive to mild changes and it captures large and abrupt variabilities. 
\begin{figure}[!ht]
\includegraphics[width=.5\textwidth]{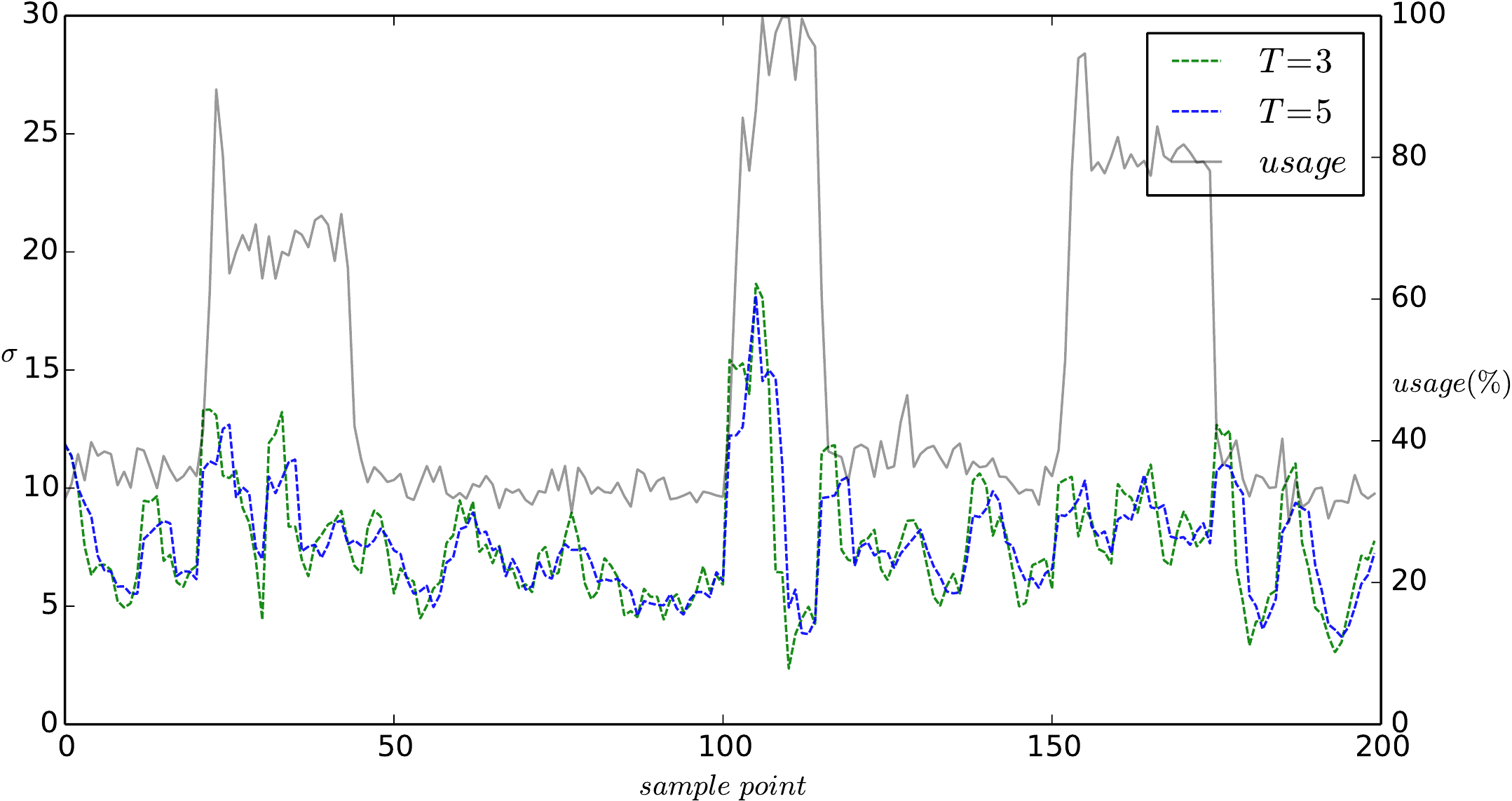}
\caption{Standard deviation ($\sigma$) of the CPU usage signal for $T=3$ and $T=5$ sampling point window size.}\label{statistics_final}
\end{figure}
Comparing the $2$ sliding windows ($T=3$ and $T=5$) in Fig.~\ref{statistics_final}, we conclude that there is little difference (maybe the variance for $T=5$ there is some more delay and less variability than that for $T=3$) and any of them should work more or less fine for our experiments. To reduce the communication and computational overhead, for the experiments in this work we choose the window size to be $5$ sampling points ($T=5$) as it is large enough to capture the variance of the random variable and small enough to track the change in variance due to the changes in the dynamics of the requests.

% ===============================================
% Headroom
% ===============================================
\subsection{Headroom}

\begin{figure}
\centering
\minipage{1\textwidth}
\subfigure[\label{subfig6a}]{\includegraphics[width=.48\textwidth]{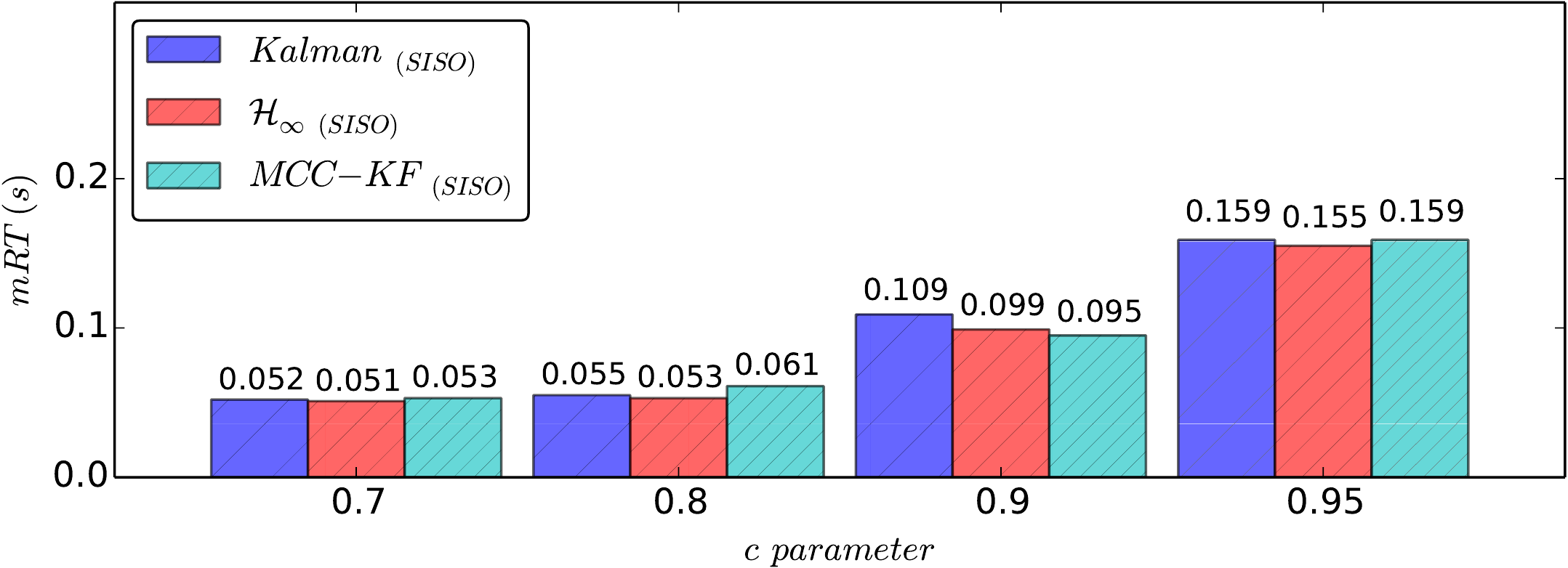}}
\endminipage\hfill
\minipage{1\textwidth}
\subfigure[\label{subfig6b}]{\includegraphics[width=.48\textwidth]{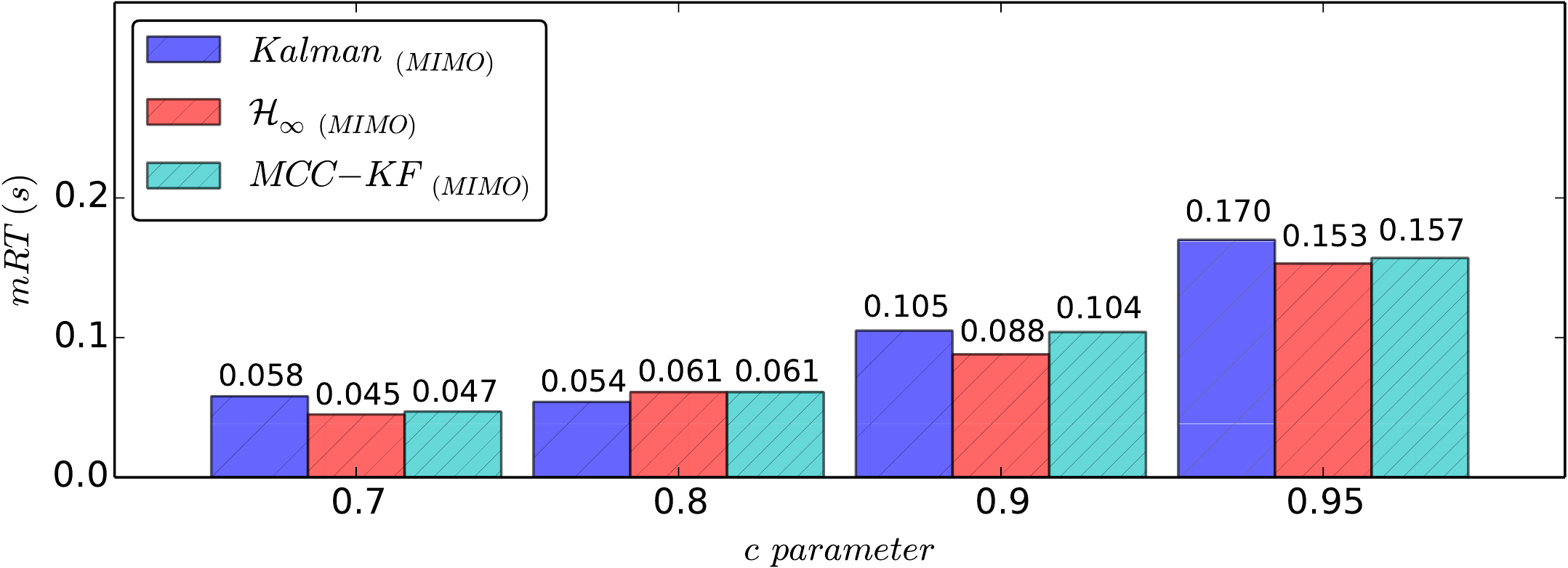}}
\endminipage
\caption{$\mRT$ with respect to $c$ parameter - (a) SISO controllers, (b)  MIMO controllers applied with control interval at $5$s. (Note $\mRT$ increases as $c$ tends to unity)}\label{figure6}
\end{figure}

Let parameter $c$ denote the desired CPU utilization to CPU allocation ratio, i.e., $c=1/(1+h)$, where $h$ is the headroom as used in and defined right after \eqref{stoch3}. The $\mRT$ with respect to parameter $c$ for Kalman filter - SISO, $\mathcal{H}_\infty$ - SISO filter and MCC-KF - SISO, is shown in the Fig.~\ref{subfig6a} and for Kalman filter - MIMO $\mathcal{H}_\infty$ - MIMO filter and MCC-KF - MIMO in the lower Fig.~\ref{subfig6b} while the control action is applied every $5$s. In this evaluation, we set a stable workload of $1000$ clients sending requests simultaneously to the RUBiS auction site. Each measurement is derived from experiments where $c$ has values of $0.7$, $0.8$, $0.9$ and $0.95$ which are enough to present the behavior of $\mRT$ as parameter $c$ grows. While $c$ approaches $1$, more resources are saved for other applications to run, but the $\mRT$ of requests is increasing and thus the performance of the RUBiS benchmark is decreasing. This happens because the headroom approaches $0$ which means that there are few resources left for RUBiS to use.
As it can be observed in Fig.~\ref{subfig6a} and Fig.~\ref{subfig6b}, both SISO and MIMO controllers can allocate resources without a big increase of $\mRT$ when the parameter $c$ is less than $0.8$. However, when the $c$ parameter increases above $0.8$, the $\mRT$ starts to grow exponentially. Note that the current experiment was conducted with a static number of clients and thus the $\mRT$ values are relatively low. Nevertheless, a larger static number of clients sending requests to the RUBiS does not always lead to a higher $\mRT$ value than a dynamic smaller number of clients. Workloads with relatively smooth dynamics can use a larger $c$ parameter without an effective influence on the $\mRT$. However, abrupt workloads with high frequency dynamics should use smaller $c$ parameter in order to let the system to adapt on abrupt CPU usage changes.

Using these results, we select parameter $c$  to be $0.8$ for the future experiments as the workload will be more dynamic and aggressive. With the selection of such a large value of parameter $c$, enough resources would be saved for other applications to run on the same physical machine.

% ===============================================
% Workload
% ===============================================
\subsection{Workload}\label{subsection:workload}
The conducted experiments run in total for $250$s, an interval adequate enough to evaluate the system's performance. Specifically, two different workload patterns are generated for the experiments, namely Workload 1 (WL1) and Workload 2 (WL2). Both workloads are initiated with $700$ clients sending requests to the RUBiS application. At sampling points $10$ and $30$ another $500$ clients are inserted to each workload for about $15$ samples. The RUBiS Client Emulator deployed on PM3, sends HTTP requests to the RUBiS application during the entire time of the experiments. The workload type for RUBiS application is set to Browsing Mix (BR), where each client waits for a \emph{think time} \footnote{Time between two sequential requests of an emulated client. Thorough discussion and experiments are presented in \cite{bahga2011synthetic}} following a negative exponential distribution with a mean of $7$ seconds (WL1) or a custom think time (WL2) which is included in the default RUBiS workload files, in order to send the next request.

\subsection{Experiment configuration}\label{subsection:exp_configuration}
Each experiment that follows (see Fig.~\ref{figure7}-~\ref{figure12}) was conducted for a total duration of $250$s and with $c$ parameter set to $0.8$. All CPU measurements for the utilization and the allocations are exported from each component through the Xen Hypervisor. The CPU usage measurements are recorded every $1$s and after the completion of one sample (i.e., $5$s), the mean value of the previous interval is forwarded to the controllers to take action. Using this sampling approach, the control action is applied every $5$s in such a way that the high frequency variations of the workload are smoothed and better responses to workload increases are achieved \cite{kalyvianaki2009resource}. For the control schemes, the initial value of the error covariance matrix, $P_0$, the variance of the process noise, $W$, and the variance of the measurement noise, $V$ are set to $10$, $4$, and $1$ respectively. The values of the process and measurement noises are updated on-line whenever the interval $k$ uses the sliding window approach, mentioned in Section~\ref{statistics}. The sliding window width $T$ is set to $5$ samples, which correspond to $25$s.

\subsection{Parameter tuning}\label{subsection:parameter_tuning}
\begin{figure}[t]
\centering
\minipage{1\textwidth}
\subfigure[\label{subfig_theta}]{\includegraphics[width=.48\textwidth]{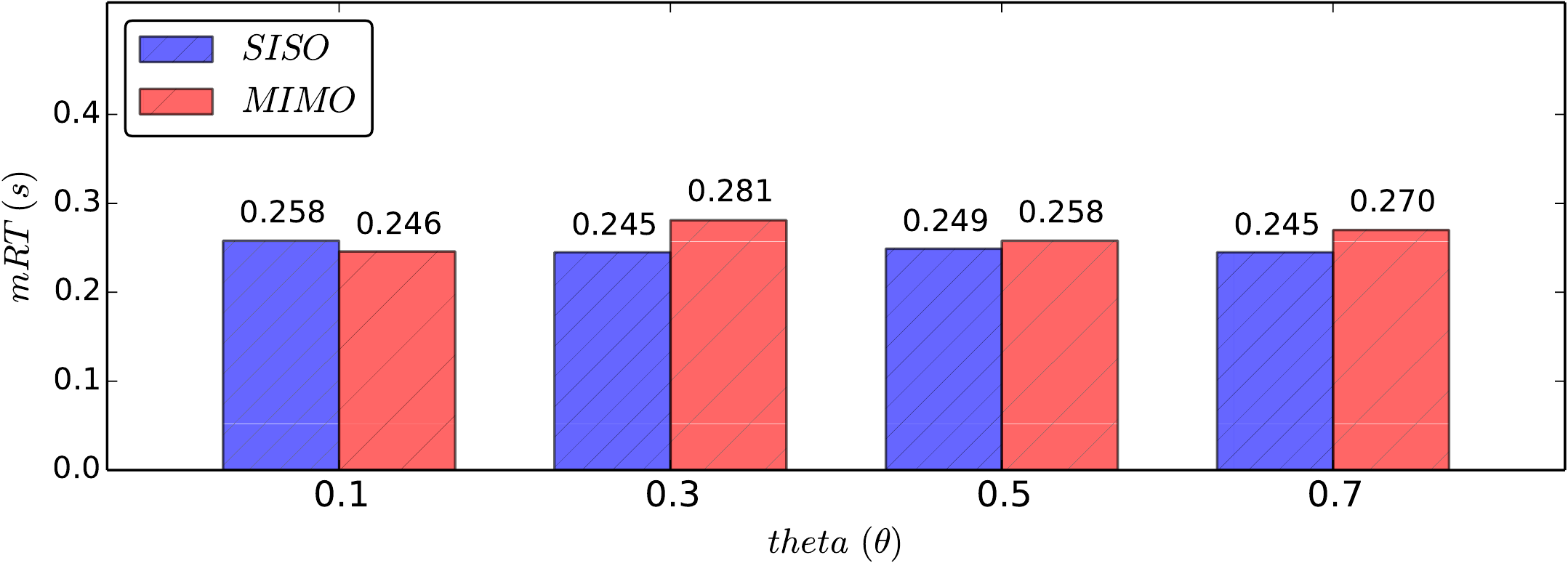}}
\endminipage\hfill
\minipage{1\textwidth}
\subfigure[\label{subfig_sigma}]{\includegraphics[width=.48\textwidth]{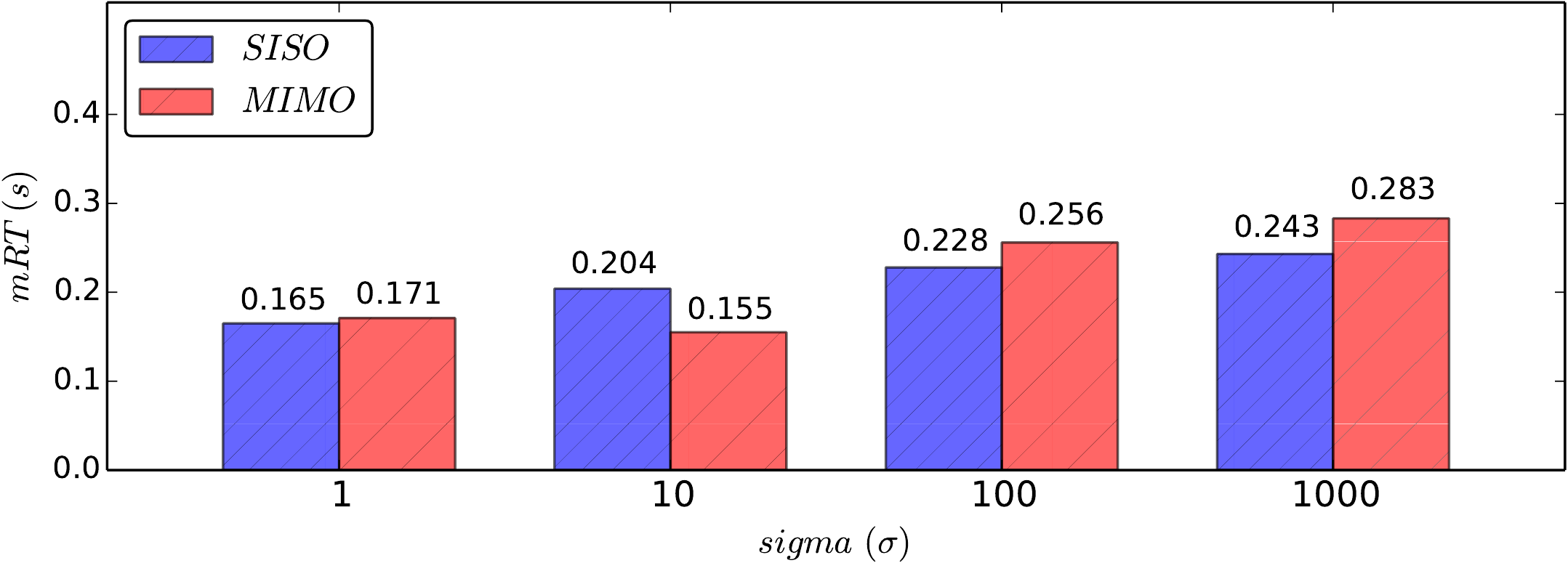}}
\endminipage
\caption{$\mRT$ with respect to theta ($\theta$) and sigma ($\sigma$) for \hinf (Fig~\ref{subfig_theta}) and MCC-KF (Fig.~\ref{subfig_sigma}) respectively.}\label{fig:theta_sigma}
\end{figure}

Apart from Kalman filter, \hinf and MCC-KF need tuning parameters $\theta$ and $\sigma$, respectively. In order to take a decision for the values of these parameters, we setup experiments with the workload WL1 described in Section~\ref{subsection:workload}. Note that the experiments have run for both SISO and MIMO models. Fig.~\ref{fig:theta_sigma} present the average $\mRT$ for the total duration of each experiment with respect to $\theta$ and $\sigma$ value for \hinf and MCC-kF (SISO and MIMO models) respectively.

Fig.~\ref{subfig_theta}~illustrates the different average $\mRT$ values while the \hinf controller is applied to track the CPU utilization of each component. For the SISO case, there is a small difference on the $\mRT$ with different values of $\theta$, however for small values (i.e., $0.1$, $0.3$) the controller fails to track the CPU utilization accurately because of the over/under estimated values of the controller gain. When $\theta$ is $0.5$ or $0.7$ the CPU utilization tracking is better and thus we select $\theta$ to be $0.7$ whereupon the \hinf SISO controller performs better. For the MIMO case it is clear that when $\theta$ is at low values (e.g., $0.1$ and $0.3$) the average $\mRT$ is relatively high ($0.278$s and $0.275$s respectively). When $\theta$ is $0.5$ or $0.7$, the average $\mRT$ stays at lower values and specifically $0.223$s and $0.239$s respectively. However, the gain increases when $\theta$ also increases which makes the CPU usage tracking more aggressive and thus the allocations are kept for many samples at the component's CPU usage upper bound ($100\%$). At this situation there are many CPU resources unutilized and this is the reason for the low $\mRT$ values. To keep the CPU usage tracking using the \hinf filter non-aggressive, we choose for the experiments, $\theta$ to be $0.1$.

As in the \hinf case, MCC-KF needs a parameter tuning in order to accurately predict the CPU allocations based on the past utilizations. For this reason, we setup experiments with the same context as the \hinf. Fig.~\ref{subfig_sigma} presents the average $\mRT$ for the total duration of each experiment with respect to $\sigma$ value for both SISO and MIMO. As we can see, both SISO and MIMO models, perform with almost the same trend. For the SISO case we can see that as the value of $\sigma$ increases, the average $\mRT$ also increases. The average $\mRT$ is $0.165$s, $0.204$s, $0.228$s and $0.243$s when $\sigma$ is $1$, $10$, $100$ and $1000$ respectively. For the MIMO case it can be observed that when $\sigma$ is $1$ or $10$) the average $\mRT$ is $0.142$s and $0.130$s respectively. When $\sigma$ is $100$, the average $\mRT$ is $0.262$s and when it is $1000$ the average $\mRT$ is $0.342$s. For both SISO and MIMO model, low values of $\sigma$ such as $1$ and $10$ make the MCC-KF gain to be near $0$ and thus the allocations are not updated accordingly while keeping the initial CPU allocation for each component ($100\%$). When the value of $\sigma$ is $100$ or $1000$, the allocations are updated normally as the CPU usage is being tracked. Thus, to keep the CPU allocations updating using the MCC-KF filter (SISO and MIMO model), we select to keep $\sigma$ at $100$ for the next experiments.

% ===============================================
% Kalman filter SISO
% ===============================================
\subsection{Kalman filter  - SISO}
\begin{figure*}[!ht]
\centering
\mbox{
\subfigure[\label{subfig7a}]{\includegraphics[width=.3\textwidth]{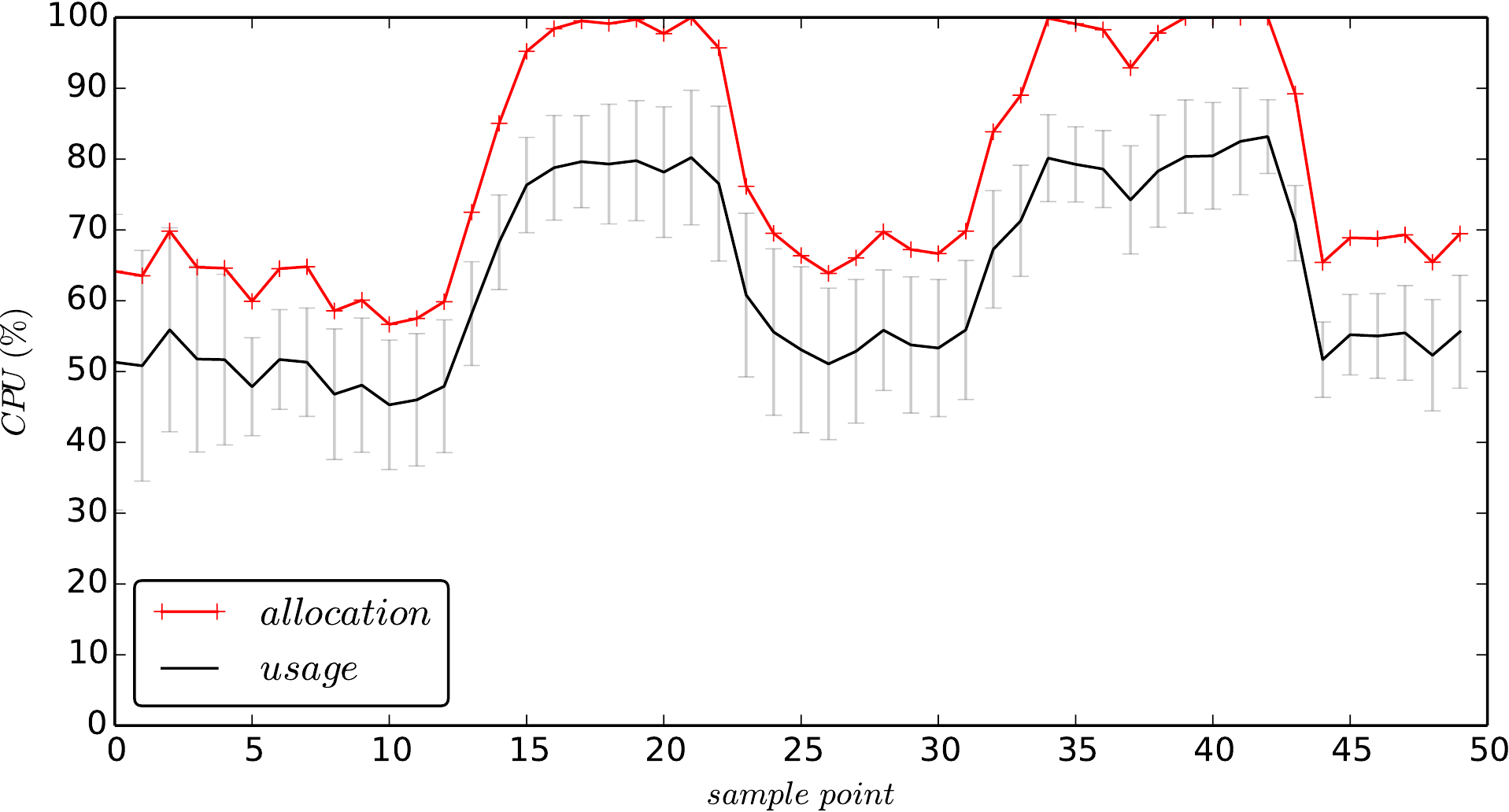}}\quad
\subfigure[\label{subfig7b}]{\includegraphics[width=.3\textwidth]{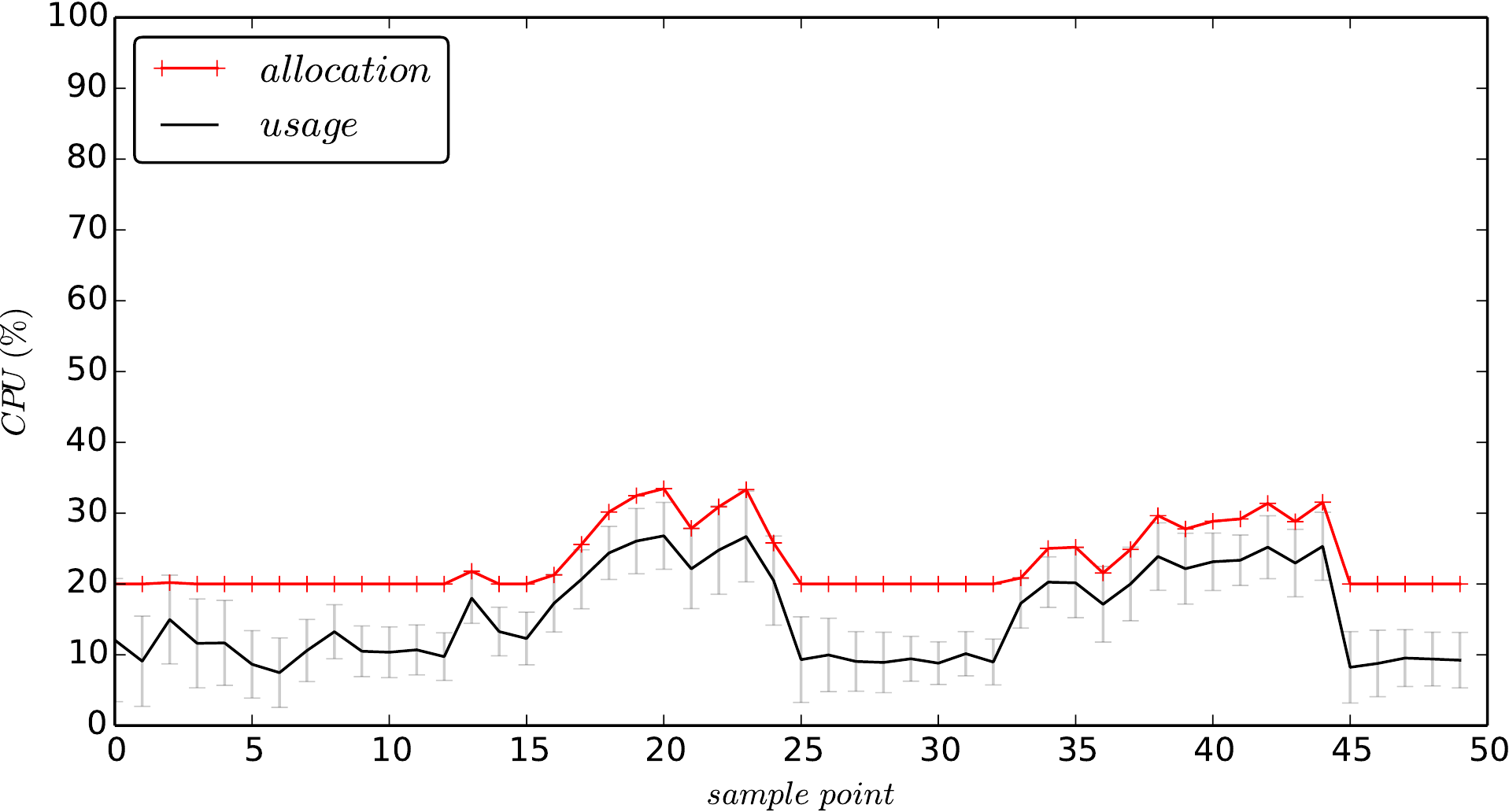}}\quad
\subfigure[\label{subfig7c}]{\includegraphics[width=.3\textwidth]{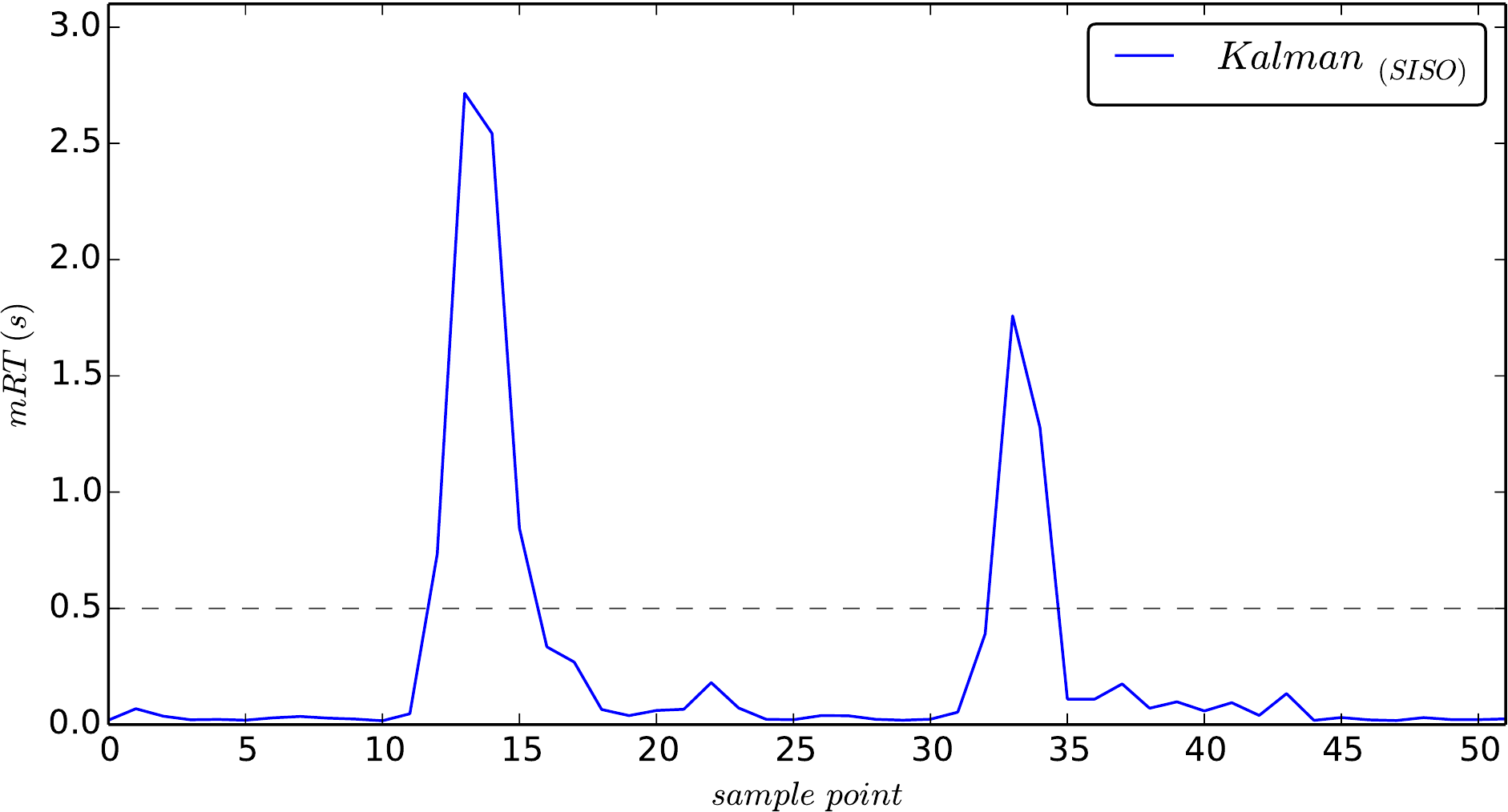}}
}\caption{Kalman - SISO filter. Fig.~\ref{subfig7a}: CPU usage and allocation of the web server component. Fig.~\ref{subfig7b}: CPU usage and allocation of the database server component. Fig.~\ref{subfig7c}: $\mRT$ with respect to time for RUBiS application.}\label{figure7}
\end{figure*}

Kalman filters form the current state of the art control approach for the CPU resource provisioning problem. The SISO Kalman filter model was first implemented into the ViResA project and then installed in both components in order to evaluate its performance. It is worth mentioning that Kalman filters require a less complex setup (compared to the MCC-KF and \hinf ones). Fig.~\ref{subfig7a} and Fig.~\ref{subfig7b} show the CPU usages-allocations of both the web server and database server components for the Kalman filter. Fig.~\ref{subfig7c} shows the $\mRT$ of the RUBiS application requests over time. The Kalman filter predicts and adjust the CPU allocations for a duration of $50$ samples on both components separately. In Fig.~\ref{subfig7a} and Fig.~\ref{subfig7b}, the predicted and adjusted Kalman allocations, properly follow the CPU usage. However, with abrupt CPU changes the $\mRT$ of the RUBiS application exhibits high increase.

% ===============================================
% Hinf filter SISO
% ===============================================
\subsection{\hinf filter  - SISO}
The \hinf SISO controllers are implemented on both RUBiS application components (i.e., web and database servers) as described in Section~\ref{siso_tier_control}. It is worth mentioning that as $\theta\rightarrow 0$ the \hinf gain approaches that of the Kalman gain. For our workload profile and experimental testbed, we set $\theta = 0.7$ for the SISO controller and $c=0.8$.

\begin{figure*}[!ht]
\centering
\mbox{
\subfigure[\label{subfig8a}]{\includegraphics[width=.3\textwidth]{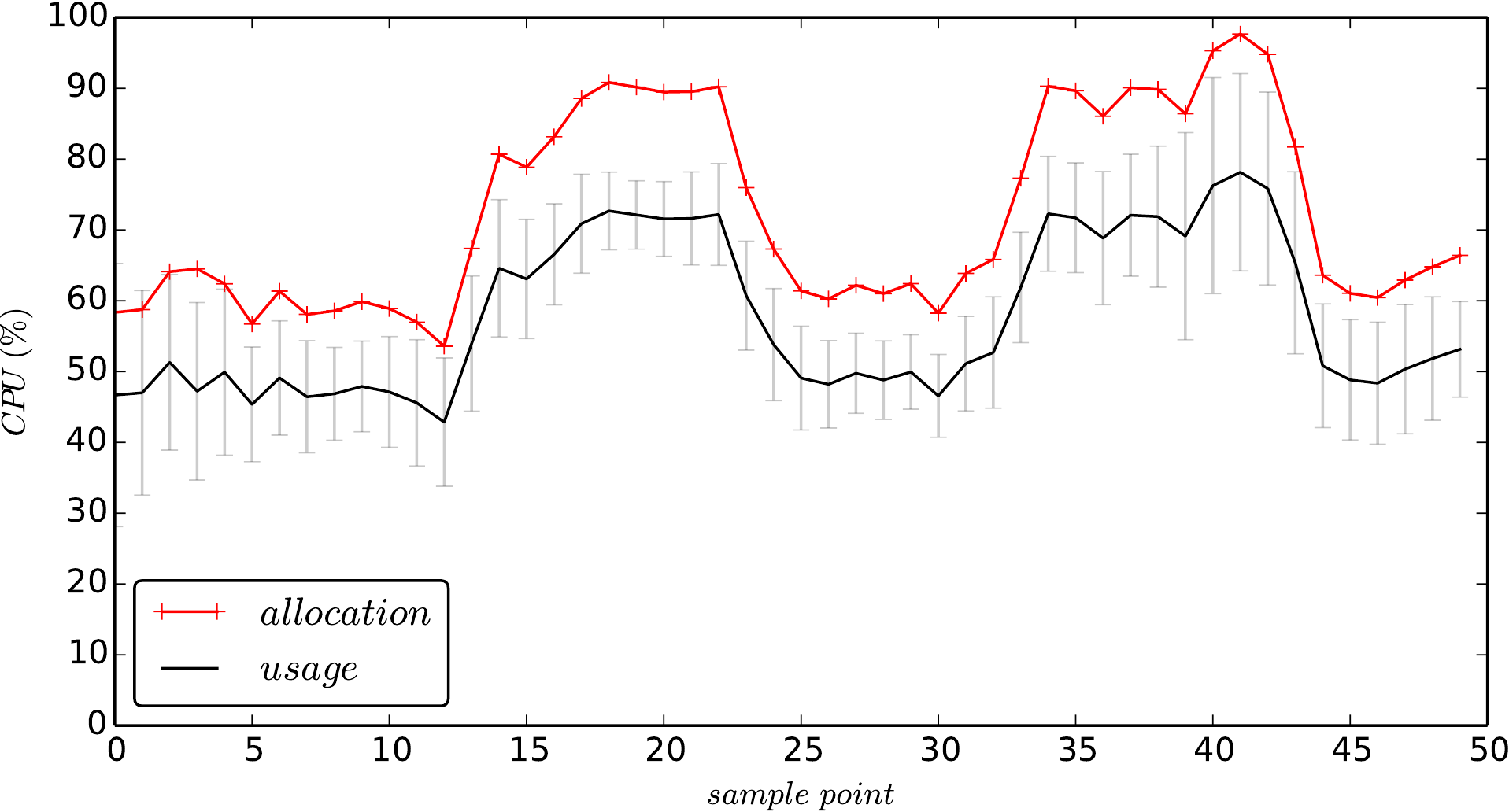}}\quad
\subfigure[\label{subfig8b}]{\includegraphics[width=.3\textwidth]{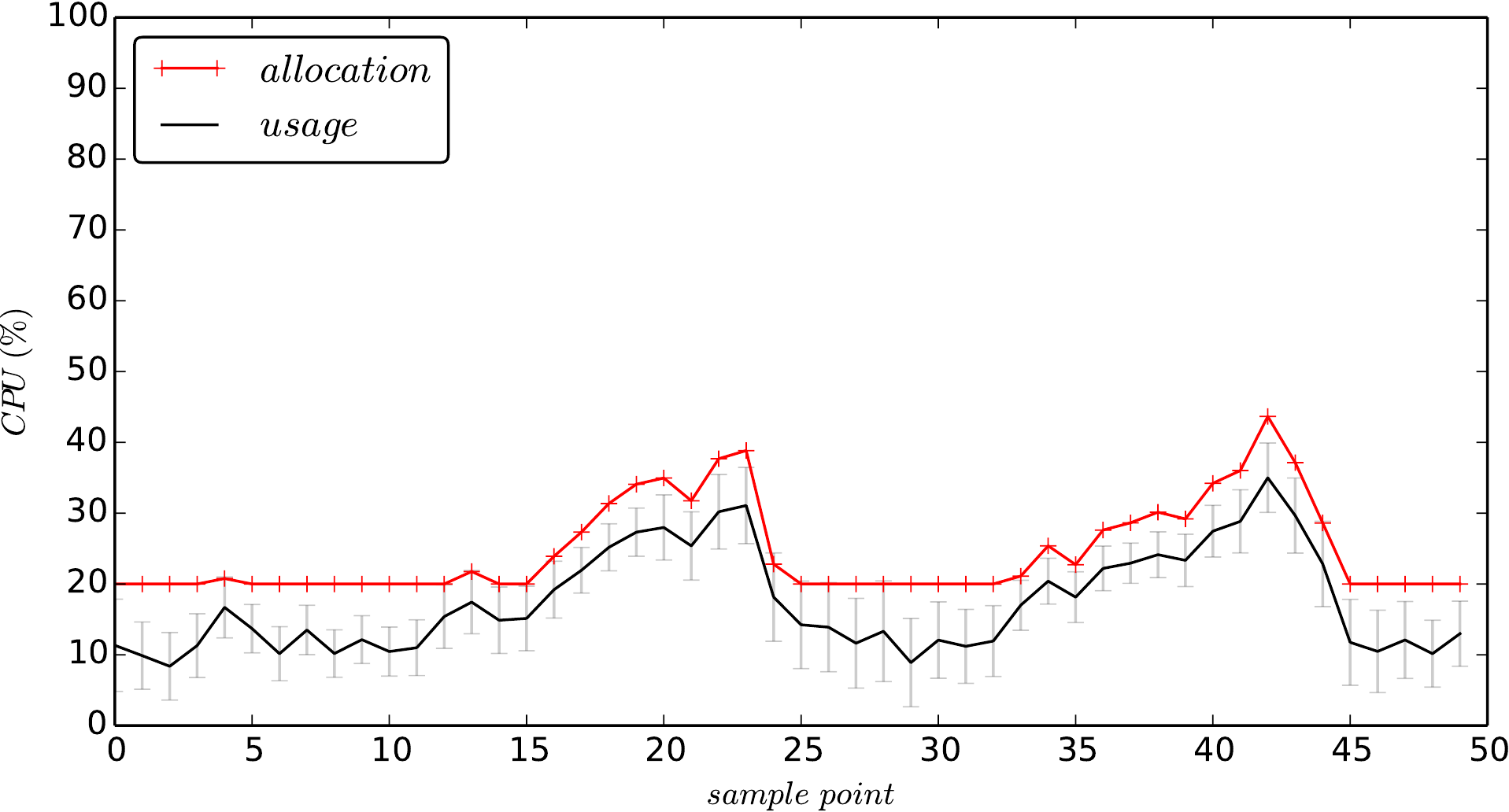}}\quad
\subfigure[\label{subfig8c}]{\includegraphics[width=.3\textwidth]{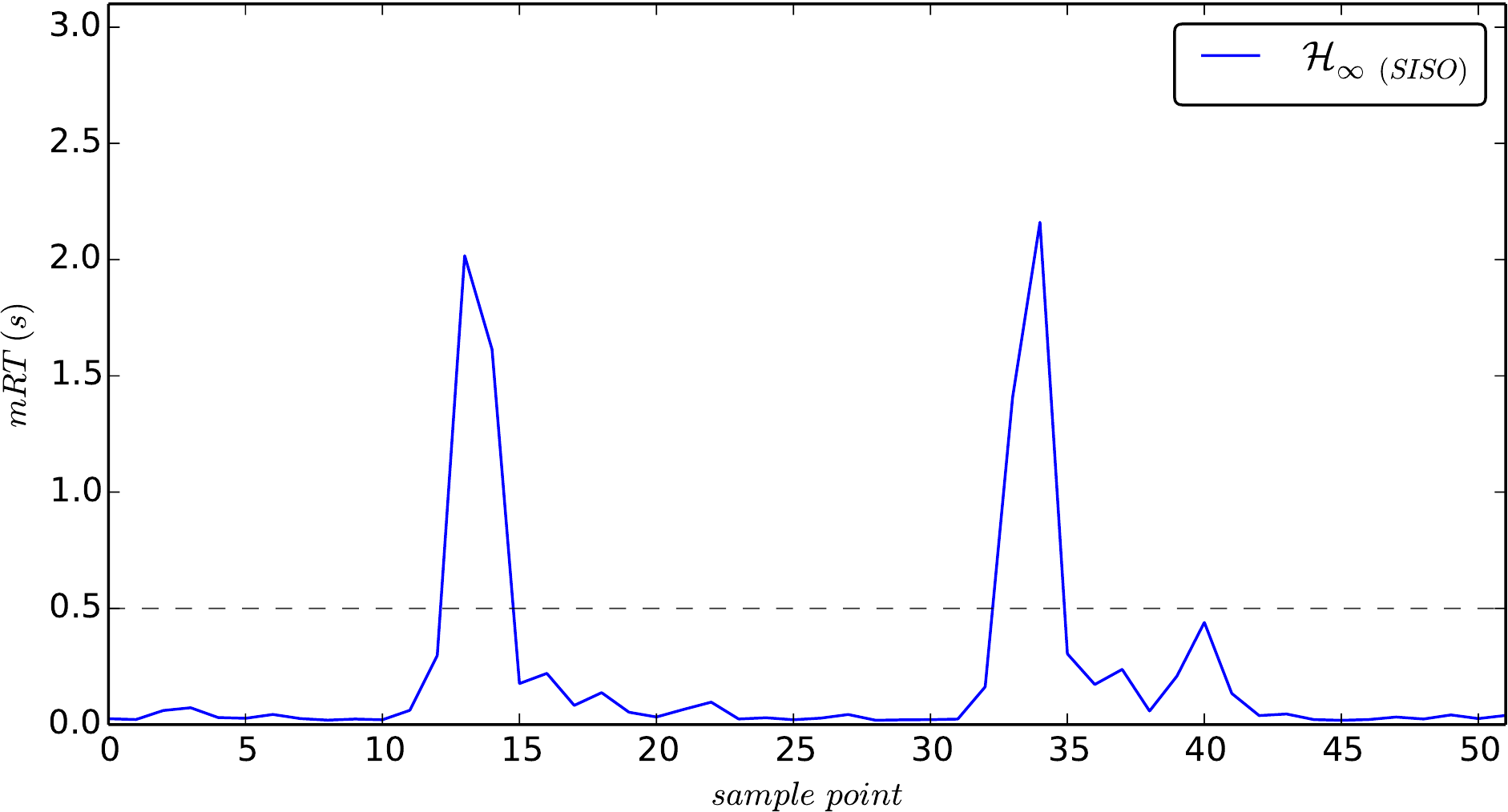}}
}\caption{\hinf - SISO filter. Fig.~\ref{subfig8a}: CPU usage and allocation of the web server component. Fig.~\ref{subfig8b}: CPU usage and allocation of the database server component. Fig.~\ref{subfig8c}: $\mRT$ with respect to time for RUBiS application.}\label{figure8}
\end{figure*}

Fig.~\ref{subfig8a} and Fig.~\ref{subfig8b} show the CPU usages-allocations of both the web server and database server components using the \hinf filter. Fig.~\ref{subfig8c} presents the $\mRT$ of the RUBiS application requests over time. In this experiment, the \hinf SISO filter predicts and adjusts the CPU allocations for a duration of $50$ samples on both components separately. Evaluation is based on the same workload presented in Section~\ref{subsection:workload}. Note that, the \hinf filter proposed in \cite{2010:CDC_themisEva} was evaluated only via simulation using synthetic data. Also, in previous authors' work \cite{Makridis:2017} the \hinf filter was applied only on the web server component while the database component was statically fully allocated. Hence, different to the aforementioned works, the \hinf filter presented here is the first time that is evaluated via a real testbed. Overall, the \hinf SISO filter performs well during the task of allocating resources since it provides extra resources to the application in order to keep the $\mRT$ low.

\begin{figure*}[!ht]
\centering
\mbox{
\subfigure[\label{subfig9a}]{\includegraphics[width=.3\textwidth]{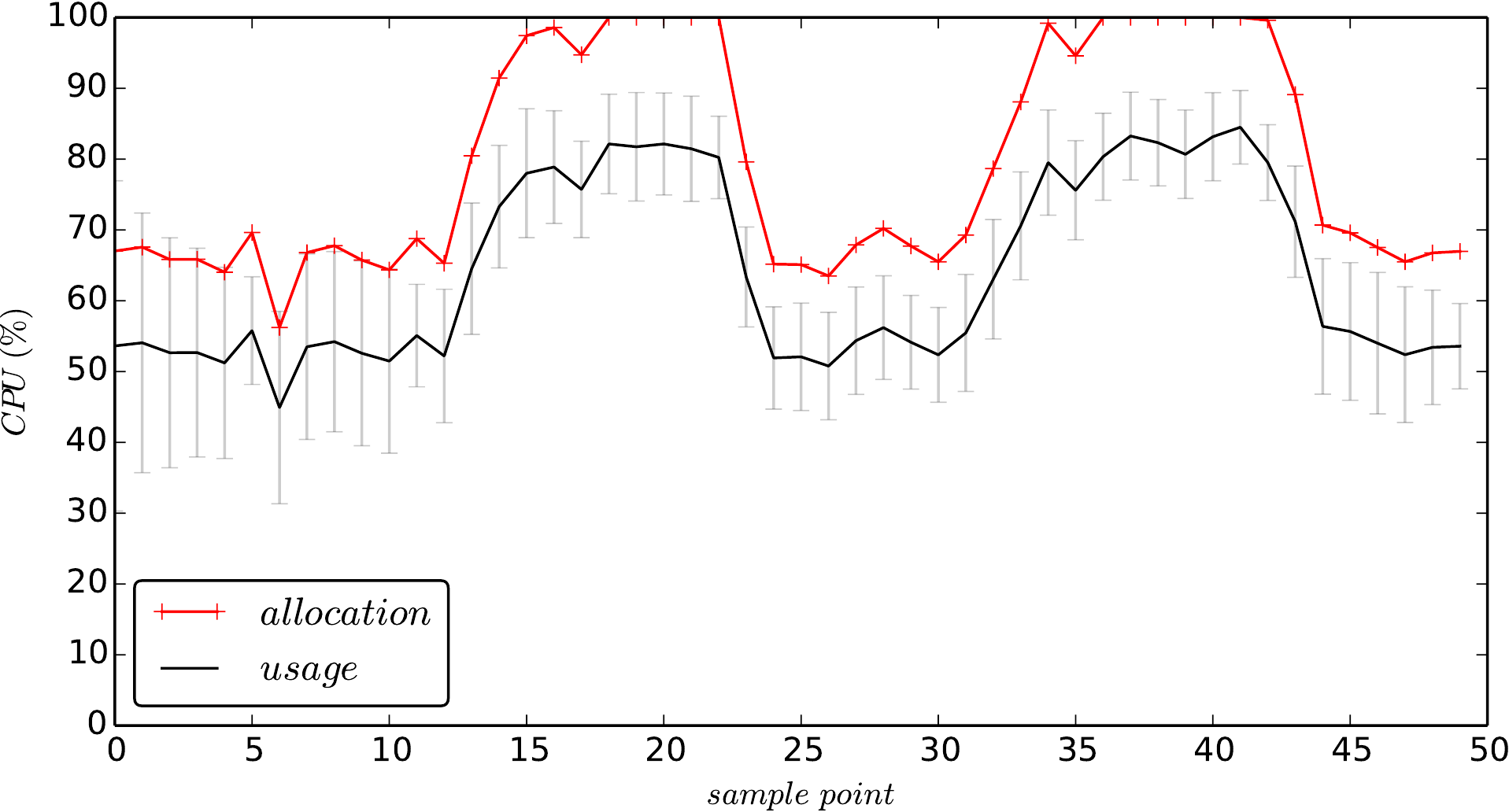}}\quad
\subfigure[\label{subfig9b}]{\includegraphics[width=.3\textwidth]{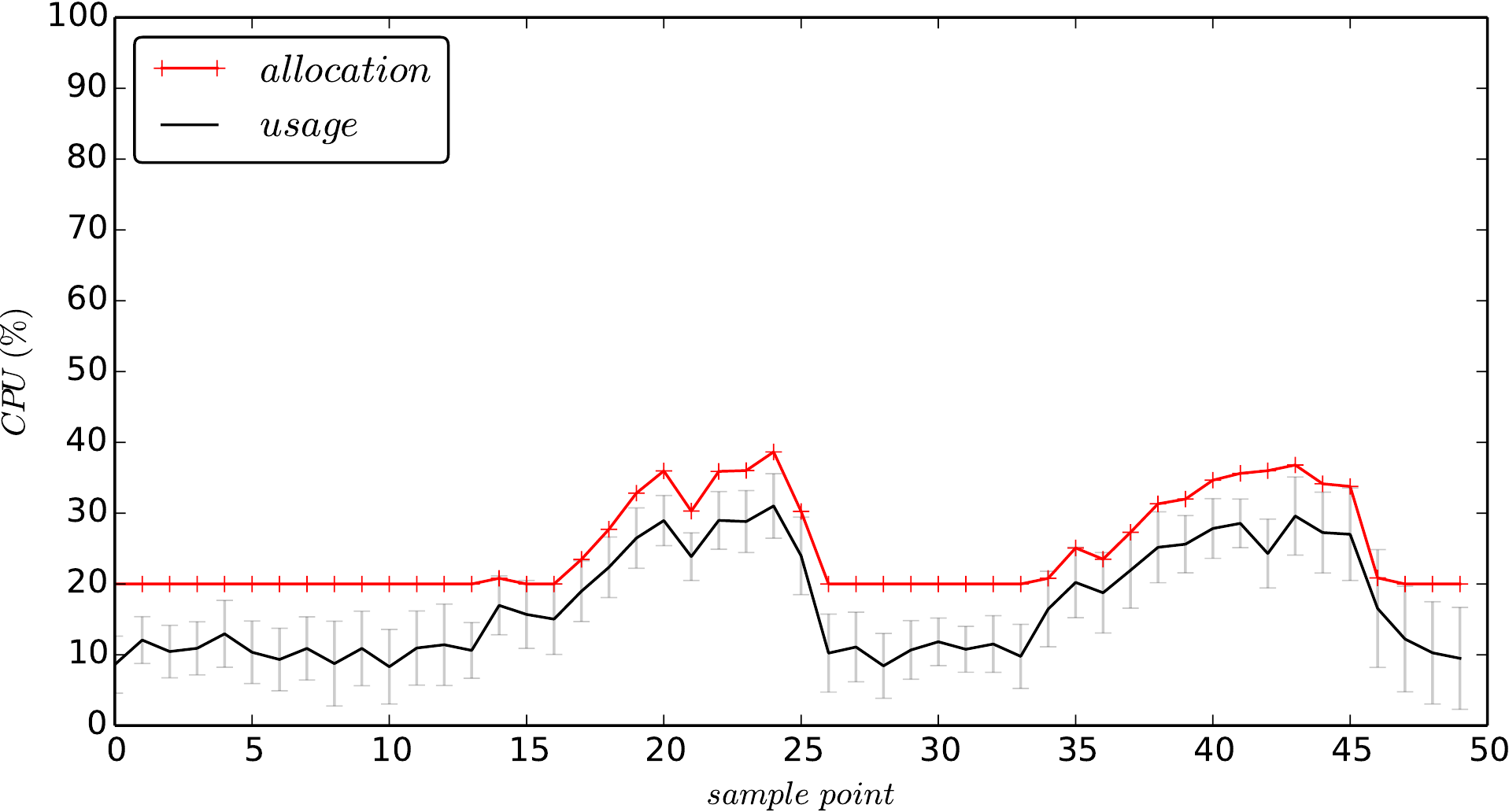}}\quad
\subfigure[\label{subfig9c}]{\includegraphics[width=.3\textwidth]{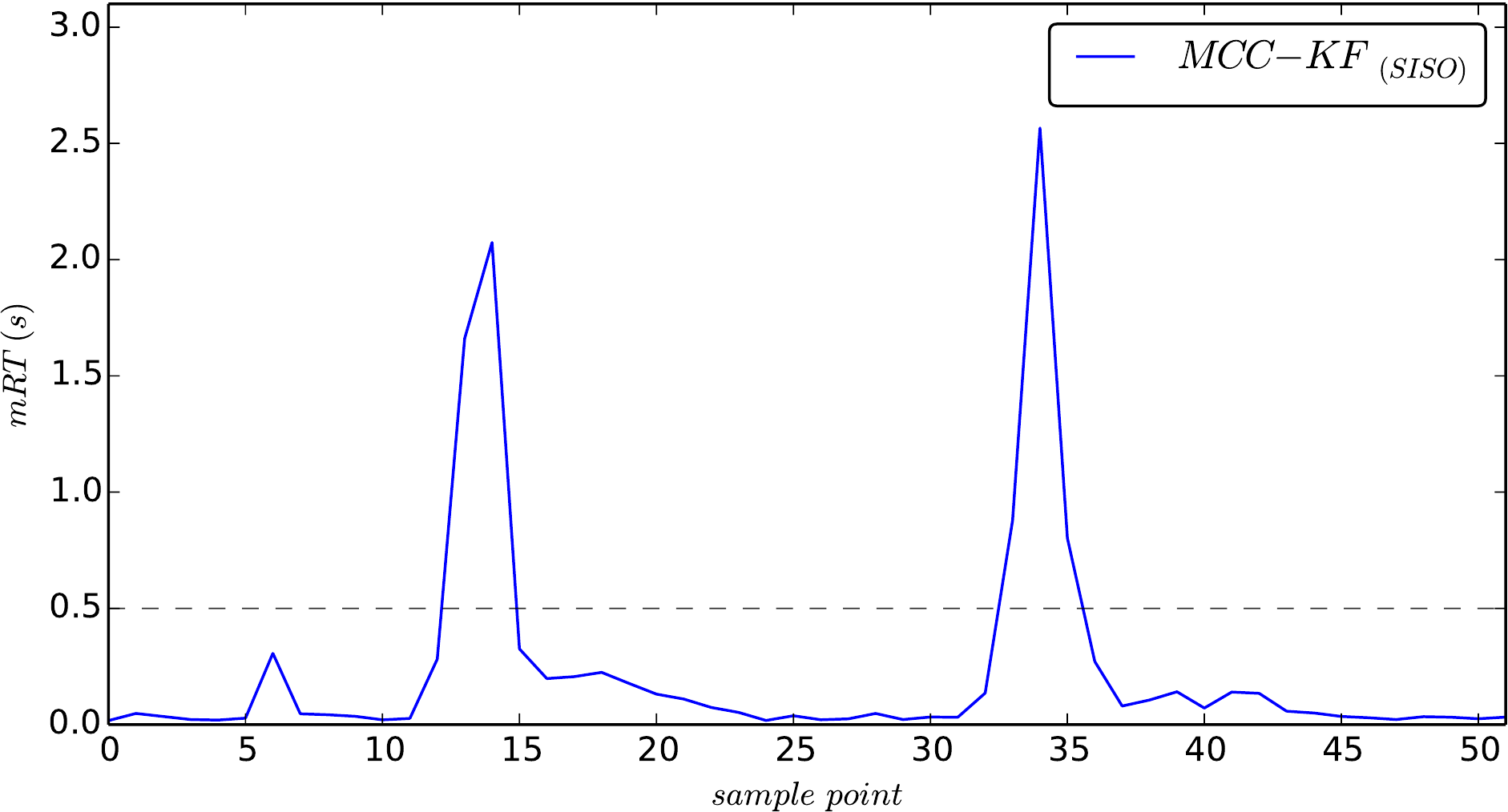}}
}\caption{MCC-KF - SISO filter. Fig.~\ref{subfig9a}: CPU usage and allocation of the web server component. Fig.~\ref{subfig9b}: CPU usage and allocation of the database server component. Fig.~\ref{subfig9c}: $\mRT$ with respect to time for RUBiS application.}\label{figure9}
\end{figure*}

% ===============================================
% MMC-KF filter SISO
% ===============================================
\subsection{MCC-Kalman filter (MCC-KF) - SISO}
Fig.~\ref{subfig9a} and Fig.~\ref{subfig9b} depict the graphs of the CPU usages-allocations for both server components (i.e., web server and database server) respectively while the MCC-KF filter is applied. The Fig.~\ref{subfig9c} shows the $\mRT$ of the RUBiS application requests over time.
Similar to the previous experiment with \hinf, the MCC-KF operates well enough at sudden CPU usage changes, while the $\mRT$ is kept at relatively low values except from the periods of the workload injection. During all other regions, the $\mRT$ stays at low levels for the reason that the remaining amount of CPU resources is sufficient for serving the stable workload.

% ===============================================
% SISO controllers comparison
% ===============================================
\subsection{Comparison - SISO controllers}
Table~\ref{table:siso} presents application performance metrics for the SISO controllers. For both workloads (WL1 and WL2) the \hinf filter and the equivalent MCC-KF perform well (with the \hinf performing better overall) and both superior compared to the Kalman filter. This is clearly illustrated in the experiment whereby for WL1 the Kalman filter offers an average $\mRT$ of $0.260$s, while the \hinf filter and MCC-KF filter achieve an average $\mRT$ of $0.224$s and $0.246$s, respectively. In addition, the Kalman and MCC-KF filters have similar SLO obedience at $89.0\%$ and $88.9\%$ respectively while the \hinf filter performs slightly better with $90.0\%$ overall SLO obedience. The WL2 follows almost the same pattern regarding the average $\mRT$. The \hinf and MCC-KF filters perform both superior compared to the Kalman filter. Specifically, the \hinf and MCC-KF filters manage to track the CPU usage with average $\mRT$ $0.789$s and $0.792$s respectively while Kalman remains at $0.908$s.

\begin{center}
\vspace{0.3cm}
\begin{tabular}{ |p{2.9cm}|p{1.35cm}|p{1.35cm}|p{1.35cm}| }
\hline
\multicolumn{4}{|c|}{Application Evaluation Metrics (SISO) - WL} \\
\hline
Controller & Kalman & \hinf & MCC-KF \\
\hline
\hline
Workload & \multicolumn{3}{c|}{WL1} \\
\hline
 Compl. requests & 34044 & 34061 & 34152 \\
 Avg. {\begin{small}VM1 CPU\%\end{small}} & 62.8 &  57.6 & 63.1 \\
 Avg. {\begin{small}VM2 CPU\%\end{small}} & 17.3 & 17.4 & 17.4 \\
 Avg. $\mRT$ (s) & 0.260 & 0.224 & 0.246 \\
 SLO obedience & 89.0\% & 90.0\% & 88.9\% \\
\hline
\hline
Workload & \multicolumn{3}{c|}{WL2} \\
\hline
 Compl. requests & 36020 & 36561 & 36498 \\
 Avg. {\begin{small}VM1 CPU\%\end{small}} & 64.1 &  64.8 & 65.4 \\
 Avg. {\begin{small}VM2 CPU\%\end{small}} & 18.6 & 18.8 & 18.6 \\
 Avg. $\mRT$ (s) & 0.908 & 0.789 & 0.792 \\
 SLO obedience & 58.6\% & 59.6\% & 59.6\% \\
\hline
\end{tabular}
\captionof{table}{SISO-based controller performance under the workloads WL1 and WL2.}\label{table:siso}
\end{center}

\begin{figure*}[ht]
\centering
\mbox{
\subfigure[\label{subfig10a}]{\includegraphics[width=.32\textwidth]{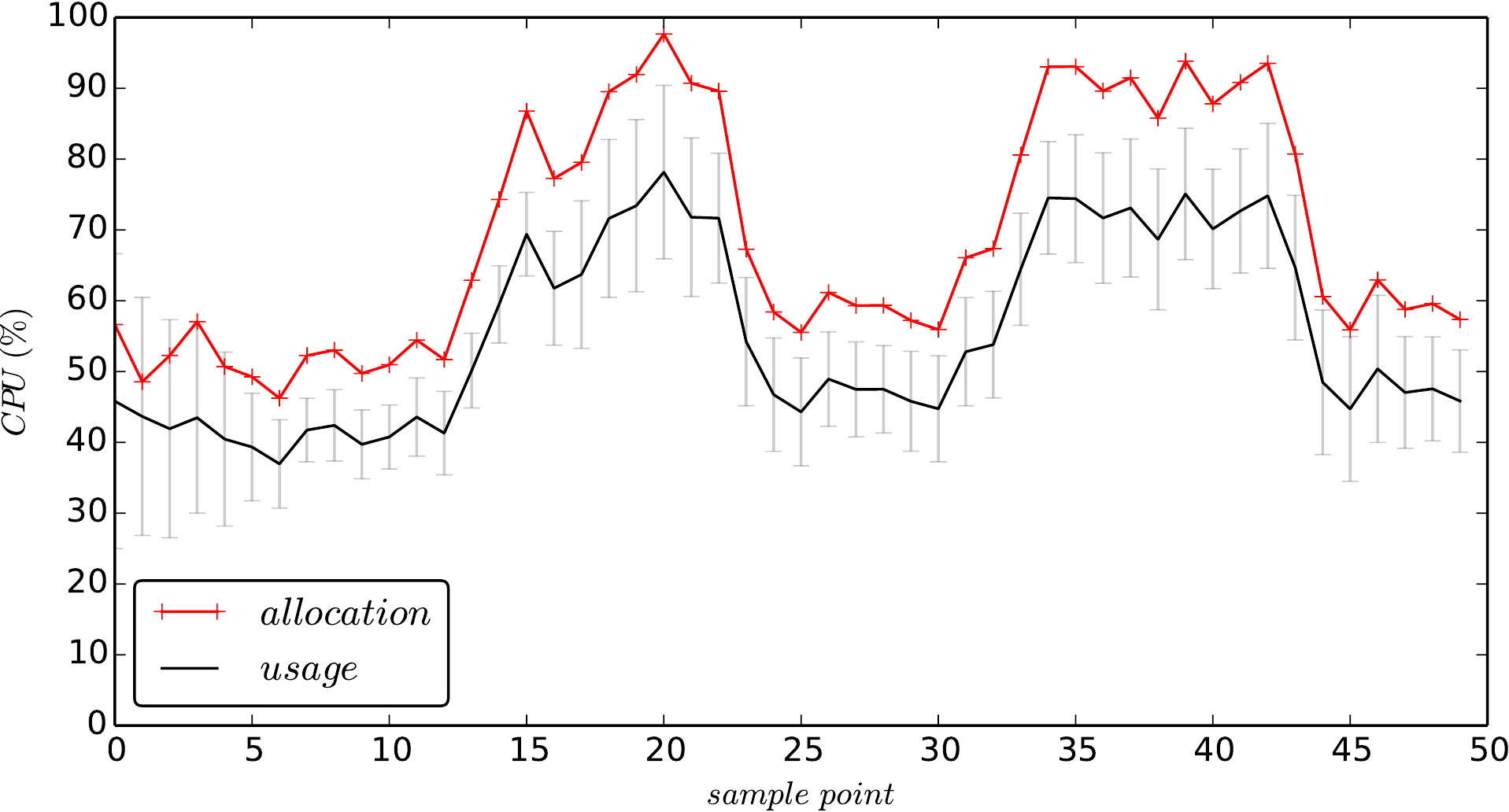}}\quad
\subfigure[\label{subfig10b}]{\includegraphics[width=.32\textwidth]{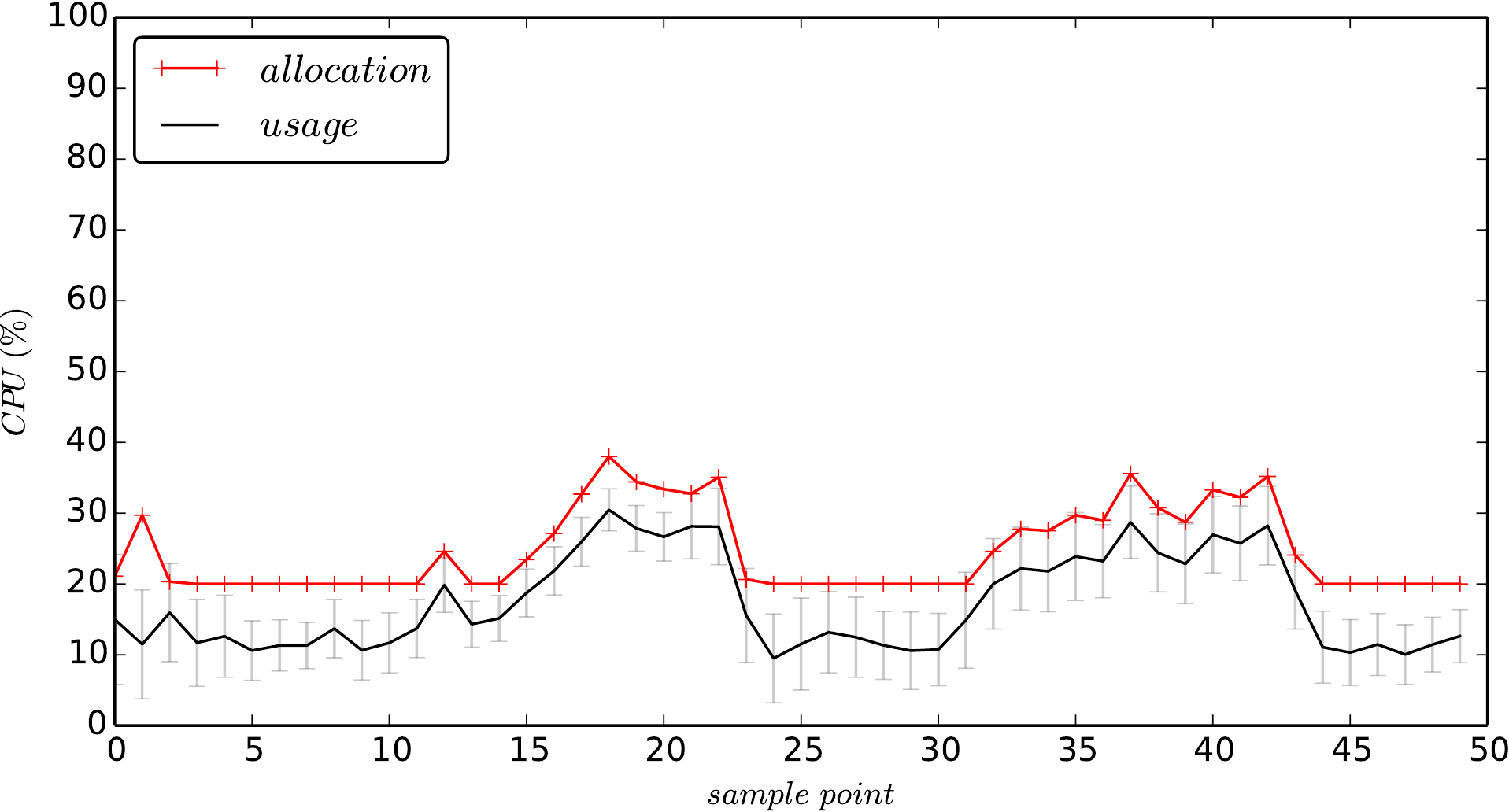}}\quad
\subfigure[\label{subfig10c}]{\includegraphics[width=.32\textwidth]{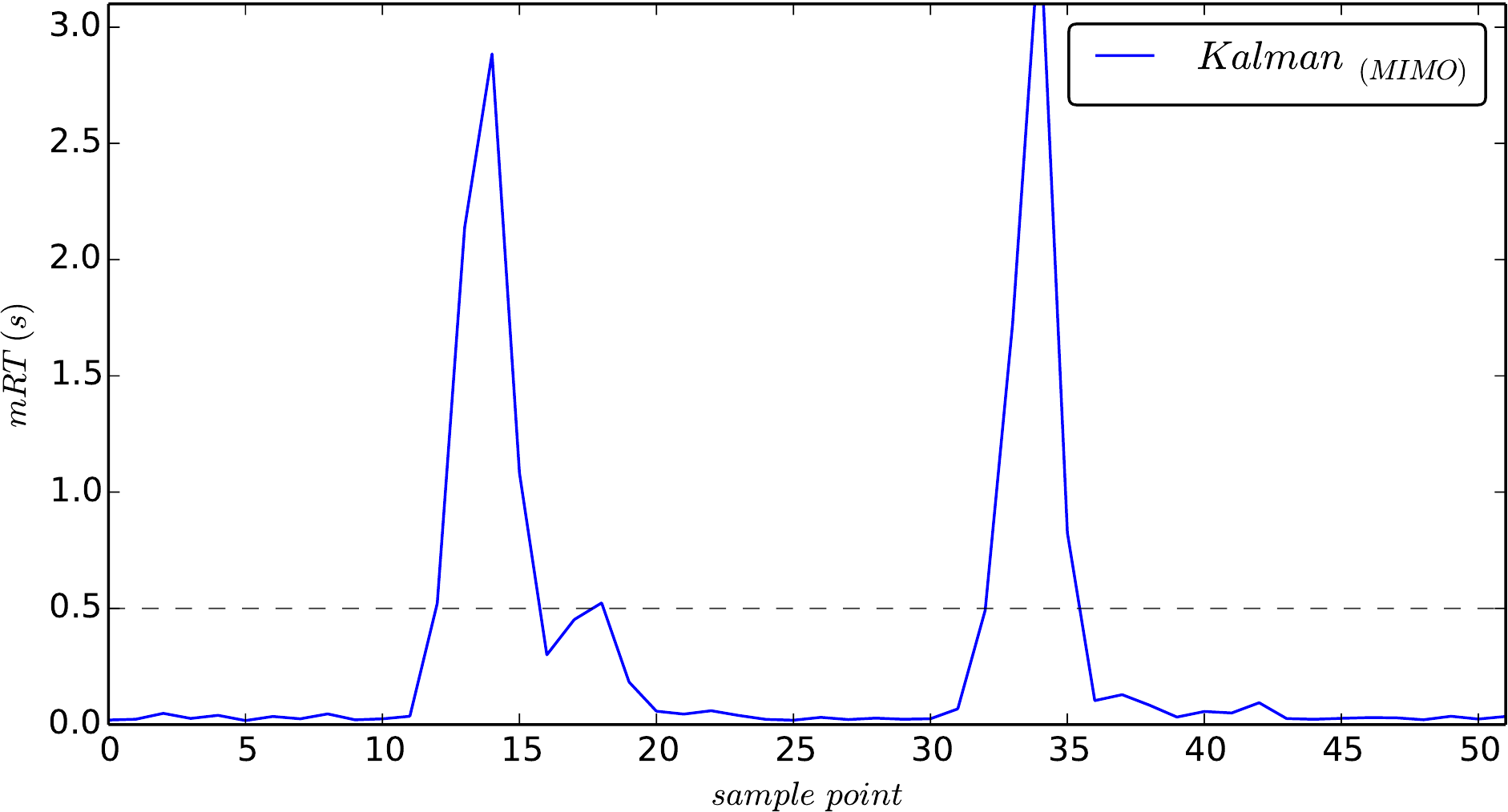}}
}\caption{Kalman - MIMO filter. Fig.~\ref{subfig10a}: CPU usage and allocation of the web server component. Fig.~\ref{subfig10b}: CPU usage and allocation of the database server component. Fig.~\ref{subfig10c}: $\mRT$ with respect to time for RUBiS application.}\label{figure10}
\end{figure*}

\begin{figure*}[ht]
\centering
\mbox{
\subfigure[\label{subfig11a}]{\includegraphics[width=.32\textwidth]{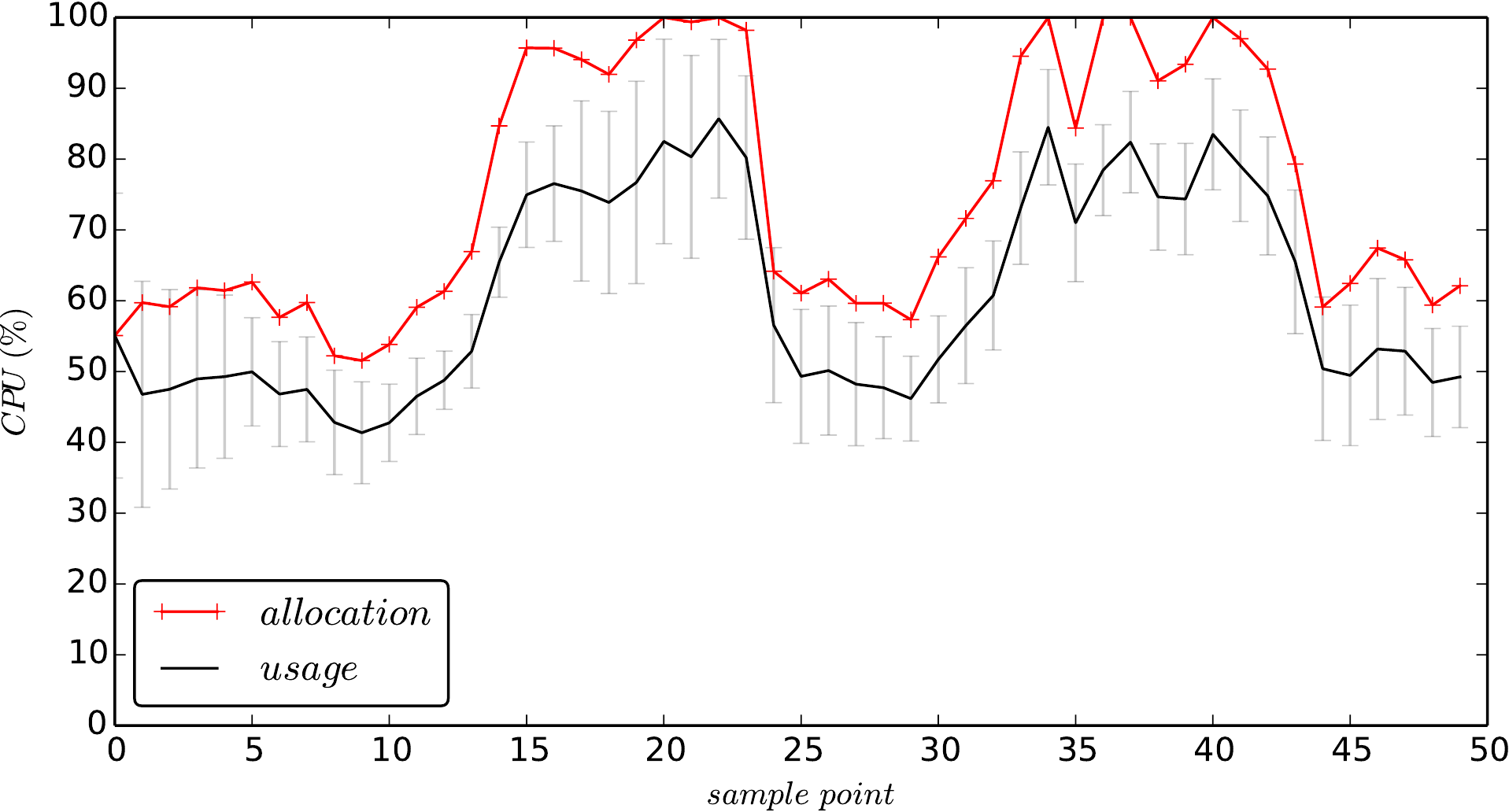}}\quad
\subfigure[\label{subfig11b}]{\includegraphics[width=.32\textwidth]{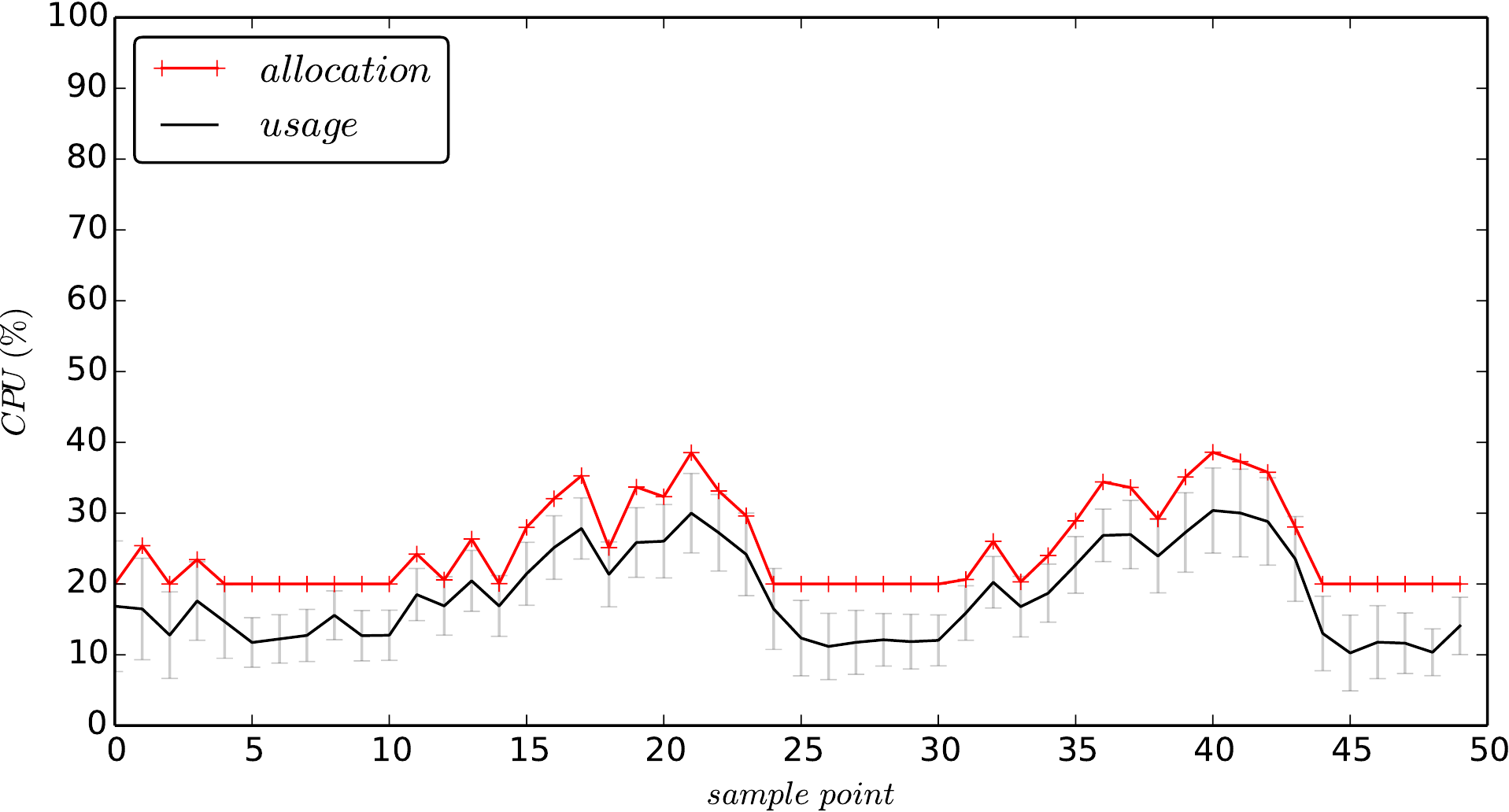}}\quad
\subfigure[\label{subfig11c}]{\includegraphics[width=.32\textwidth]{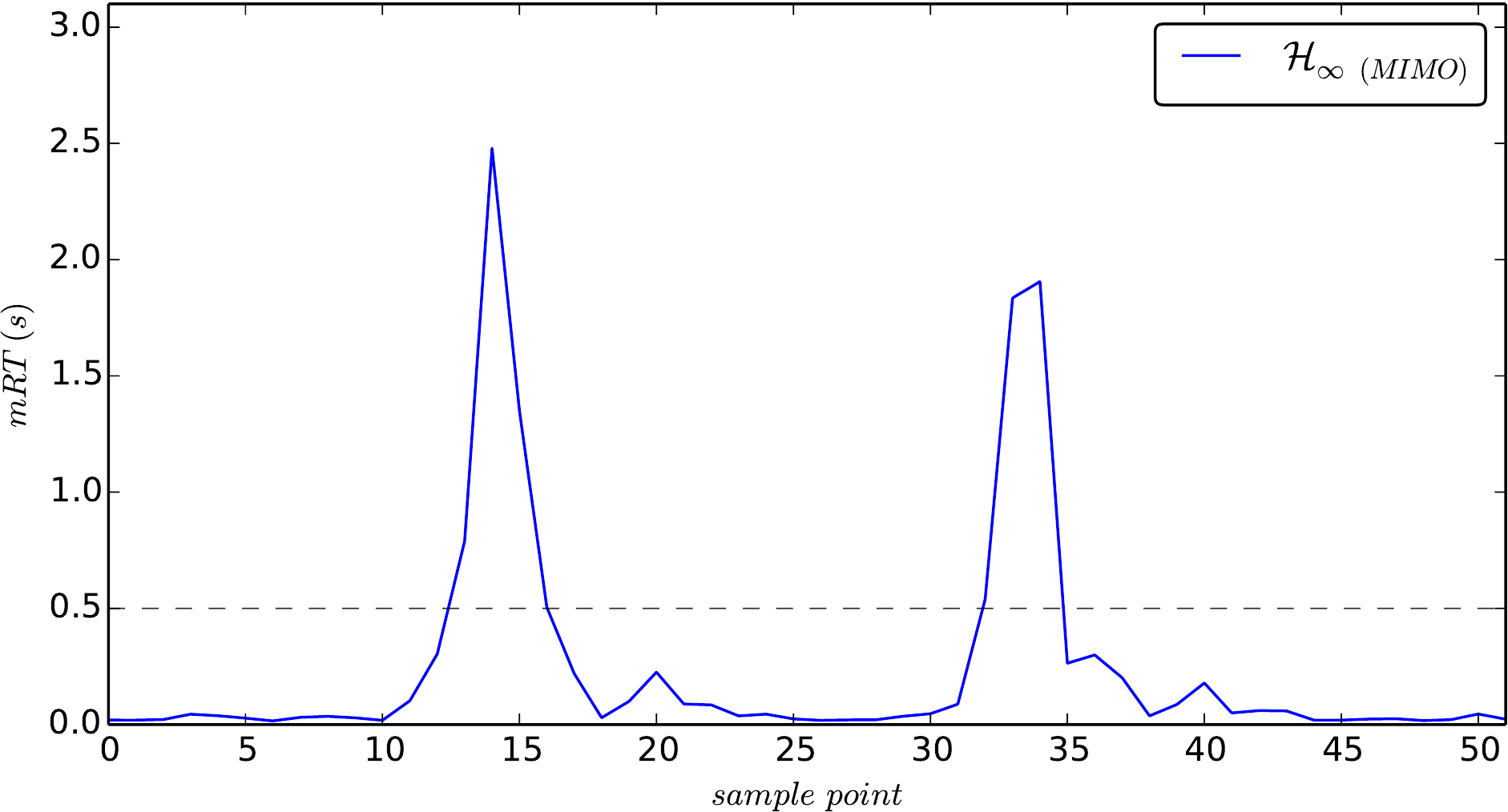}}
}\caption{\hinf - MIMO filter. Fig.~\ref{subfig11a}: CPU usage and allocation of the web server component. Fig.~\ref{subfig11b}: CPU usage and allocation of the database server component. Fig.~\ref{subfig11c}: $\mRT$ with respect to time for RUBiS application.}\label{figure11}
\end{figure*}

\begin{figure*}[ht]
\centering
\mbox{
\subfigure[\label{subfig12a}]{\includegraphics[width=.32\textwidth]{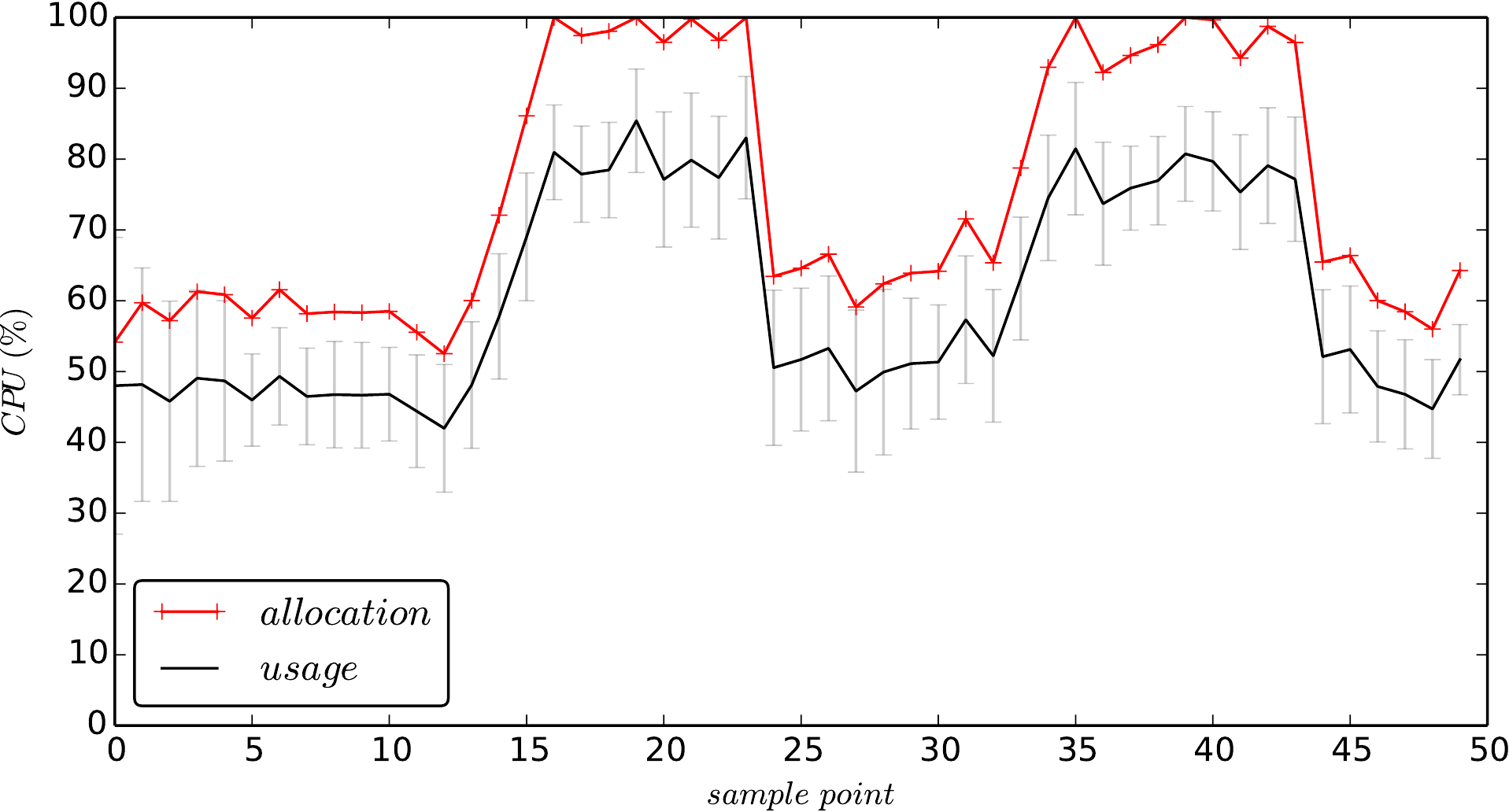}}\quad
\subfigure[\label{subfig12b}]{\includegraphics[width=.32\textwidth]{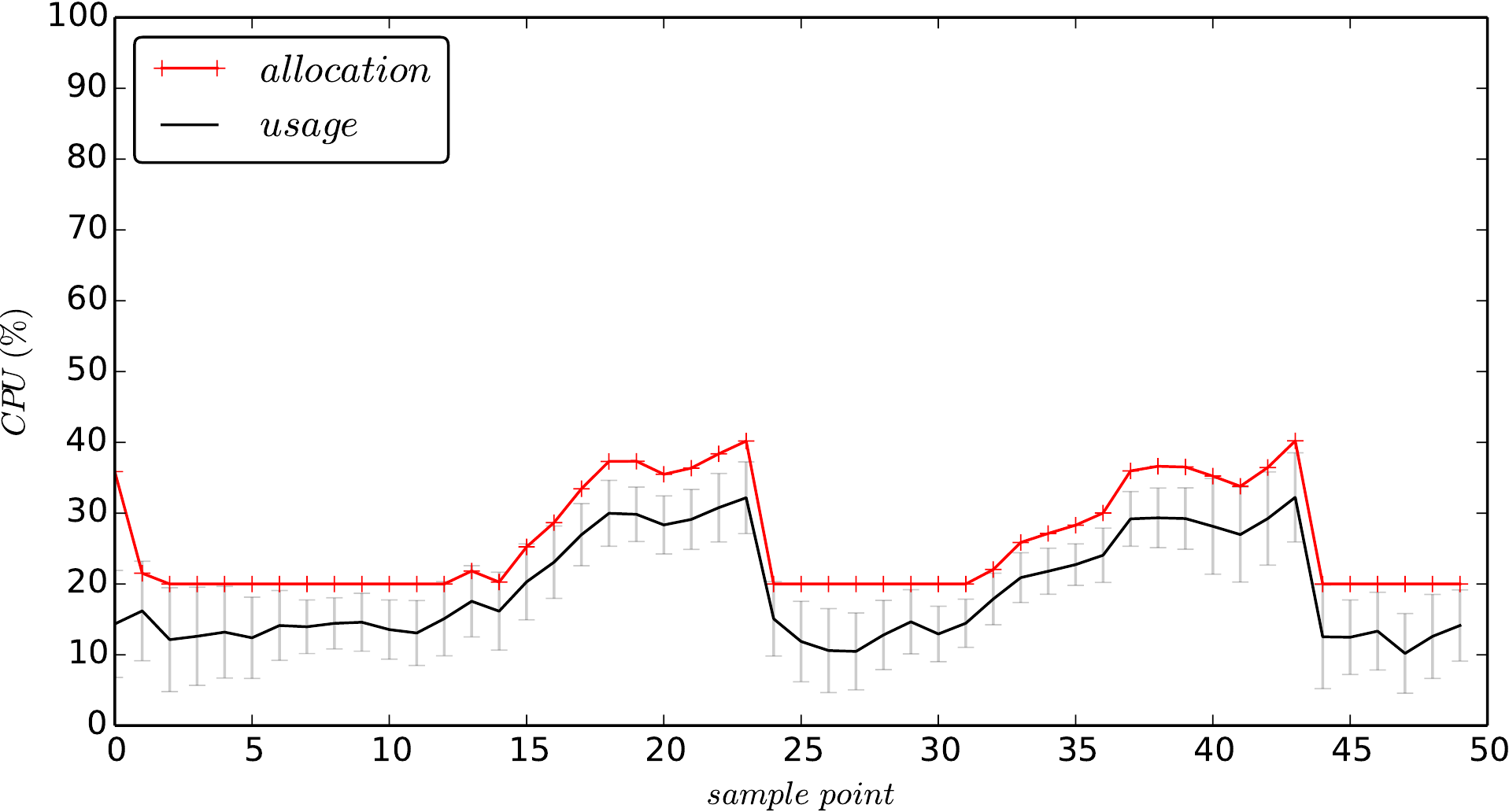}}\quad
\subfigure[\label{subfig12c}]{\includegraphics[width=.32\textwidth]{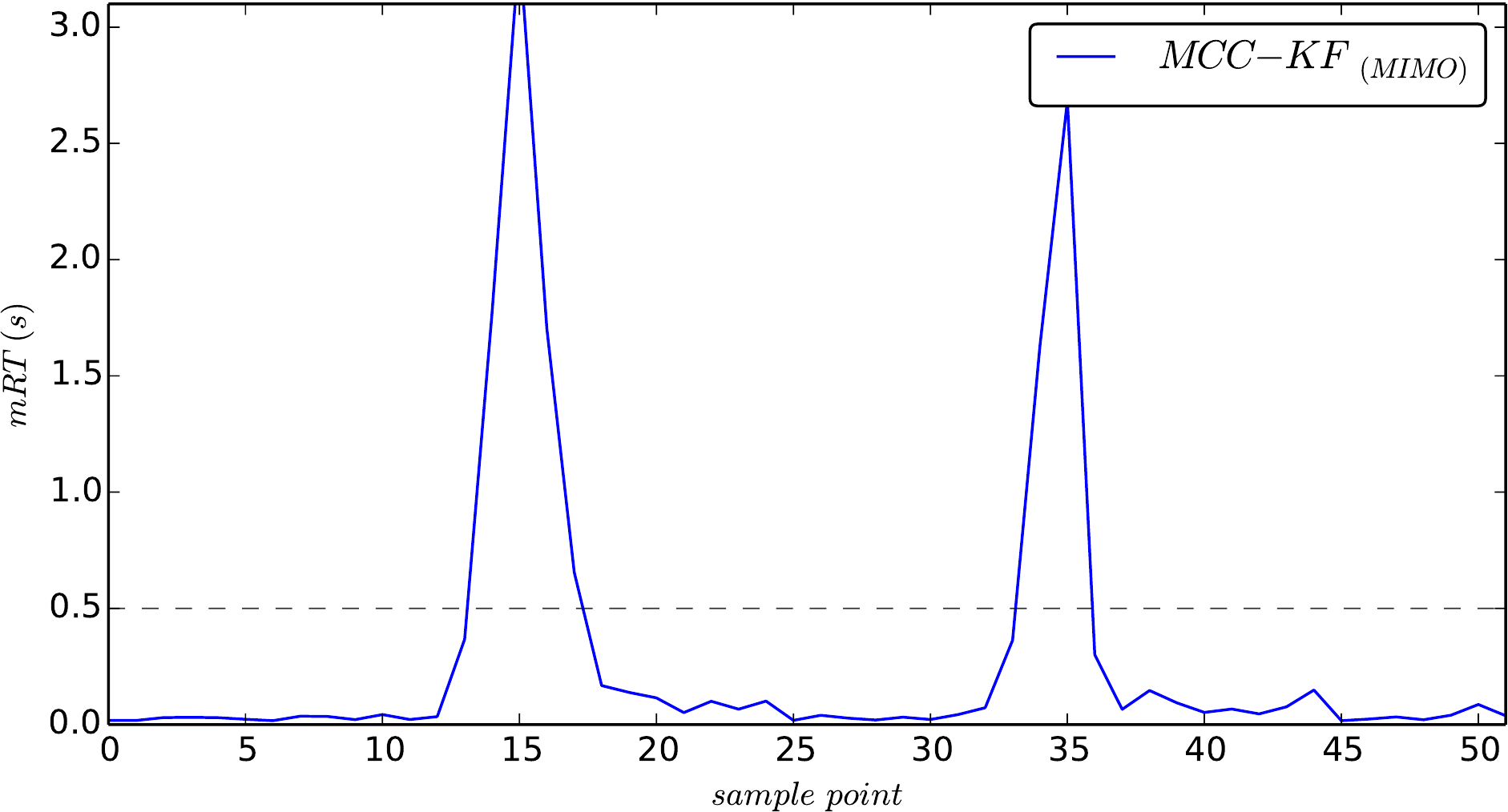}}
}\caption{MCC-KF - MIMO filter. Fig.~\ref{subfig12a}: CPU usage and allocation of the web server component. Fig.~\ref{subfig12b}: CPU usage and allocation of the database server component. Fig.~\ref{subfig12c}: $\mRT$ with respect to time for RUBiS application.}\label{figure12}
\end{figure*}

\subsection{Kalman filter - MIMO}
The Kalman MIMO controller has been previously evaluated in \cite{selfAdaptive} and in \cite{Kalyvianaki:2014}. However, in this experiment the Kalman MIMO controller is compared with the other two MIMO controllers that are designed and implemented in the ViResA project. The graphs in Fig.~\ref{subfig10a} and Fig.~\ref{subfig10b} present the CPU usages-allocations for both components during Kalman MIMO controller prediction and allocation of CPU resources. Fig.~\ref{subfig10c} shows the $\mRT$ of the RUBiS application over time after the allocations that the Kalman filter adapts. As can be seen in Fig.~\ref{subfig10a} and Fig.~\ref{subfig10b} at sample point $10$, the Web Server utilizations affects directly the Database utilizations which show the correlation and the inter-component resource couplings between the VMs. An abrupt Web Server utilization change causes a direct Database utilization change and thus the $\mRT$ is affected. During the stable workload regions, the Kalman manages to optimally allocate resources leaving the $\mRT$ unaffected.

\subsection{\hinf filter - MIMO}\label{hinf_mimo}
Fig.~\ref{subfig11a} and Fig.~\ref{subfig11b} present the graphs of the CPU usages-allocations for both components during \hinf MIMO controller prediction and allocation of CPU resources. The \hinf MIMO was installed on a remote physical machine in order to predict and control the states of the system for both components as shown in Fig.~\ref{mimo-system}. Fig.~\ref{subfig11c} shows the $\mRT$ of the RUBiS application over time.

\subsection{MCC-Kalman filter (MCC-KF) - MIMO}
Experimental results for the MCC-KF MIMO controller are shown on Fig.~\ref{subfig12a}, Fig.~\ref{subfig12b} and Fig.~\ref{subfig12c}. The MCC-KF computed and adjusted allocations with respect to the CPU usages of both components are evident in Fig.~\ref{subfig12a} and Fig.~\ref{subfig12b} for the Web Server and Database Server components respectively. Kernel size $\sigma$ of the correntropy criterion is set to $100$ to provide sufficient weight in the second and higher-order statistics of the MCC-KF.
As in the \hinf MIMO controller, the MCC-KF MIMO controller is also installed on a remote physical machine in order to estimate the states and control the allocation of the system for both components (see Fig.~\ref{mimo-system}). The $\mRT$ of the RUBiS application over time during MCC-KF MIMO control resource allocation is presented in Fig.~\ref{subfig12c}.

% ===============================================
% MIMO controllers comparison
% ===============================================
\subsection{Comparison - MIMO controllers}
Table~\ref{table:mimo_wl} presents the performance results for MIMO controllers under the workload WL1 and WL2. Clearly a similar trend, as in the SISO cases, follows. In this context, for both workloads, the MCC-KF and equivalent \hinf filters offer better performance (with the \hinf slightly better overall) compared to the Kalman filter. It is worth mentioning that the Kalman filter is not able to predict the next state of the system in abrupt workload changes and thus its SLO obedience is the lowest and its average $\mRT$ is the highest for both workloads. The \hinf has the lower average $\mRT$ with $0.270$s and $0.834$s under the WL1 and WL2. Slightly higher values of average $\mRT$ has the MCC-KF filter with $0.290$s and $0.835$s while using Kalman filter the average $\mRT$ is at $0.309$s and $0.896$s under workload WL1 and WL2 respectively.

\begin{center}
\vspace{0.1ex}
\begin{tabular}{ |p{2.9cm}|p{1.35cm}|p{1.35cm}|p{1.35cm}| }
\hline
\multicolumn{4}{|c|}{Application Evaluation Metrics (MIMO) - WL} \\
\hline
Controller & Kalman & \hinf & MCC-KF \\
\hline
\hline
Workload & \multicolumn{3}{c|}{WL1} \\
\hline
 Cmpl. requests & 34053 & 34402 & 34332 \\
 Avg. {\begin{small}VM1 CPU\%\end{small}} & 58.4 &  61.3 & 61.1 \\
 Avg. {\begin{small}VM2 CPU\%\end{small}} & 17.6 & 18.0 & 17.8 \\
 Avg. $\mRT$ (s) & 0.309 & 0.270 & 0.290 \\
 SLO obedience & 87.2\% & 88.1\% & 88.1\% \\
\hline
\hline
Workload & \multicolumn{3}{c|}{WL2} \\
\hline
 Compl. requests & 36140 & 36523 & 36441 \\
 Avg. {\begin{small}VM1 CPU\%\end{small}} & 64.1 &  59.9 & 62.0 \\
 Avg. {\begin{small}VM2 CPU\%\end{small}} & 18.4 & 18.5 & 18.8 \\
 Avg. $\mRT$ (s) & 0.896 & 0.834 & 0.835 \\
 SLO obedience & 59.2\% & 61.7\% & 61.0\% \\
\hline
\end{tabular}
\captionof{table}{MIMO-based controller performance under workloads WL1 and WL2.}\label{table:mimo_wl}
\end{center}

\begin{figure}[!h]
\centering
\minipage{1\textwidth}
\subfigure[\label{subfig13a}]{\includegraphics[width=.48\textwidth]{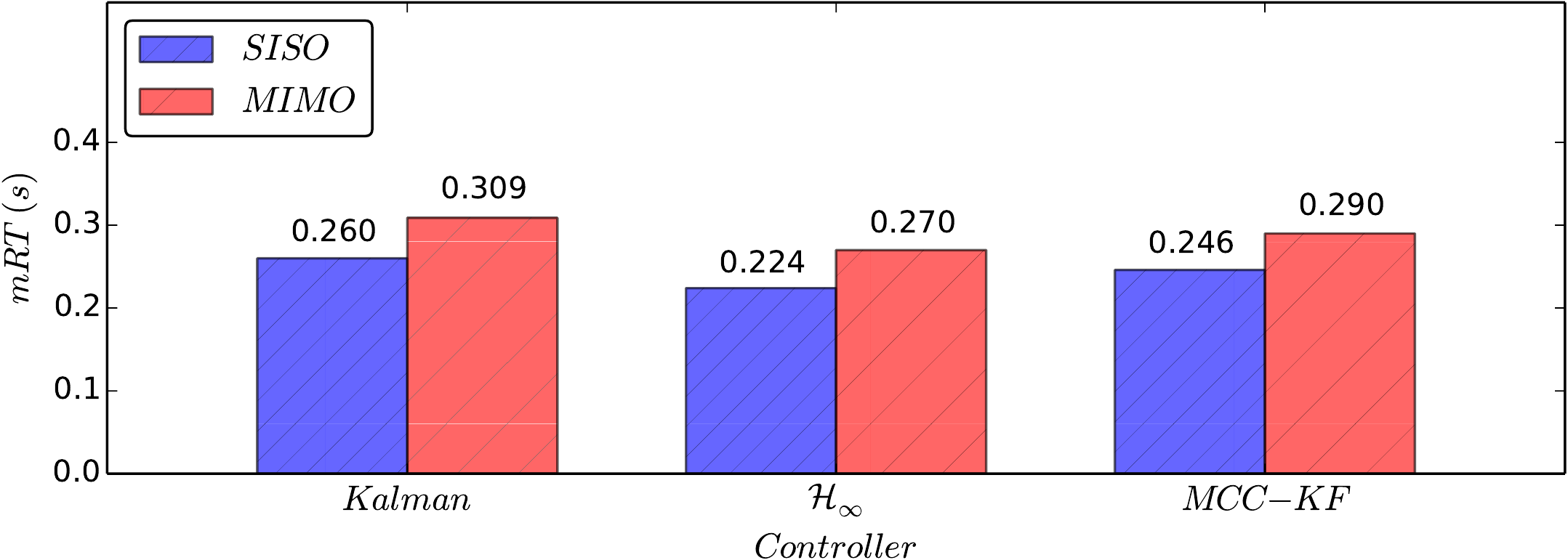}}
\endminipage\hfill
\minipage{1\textwidth}
\subfigure[\label{subfig13b}]{\includegraphics[width=.48\textwidth]{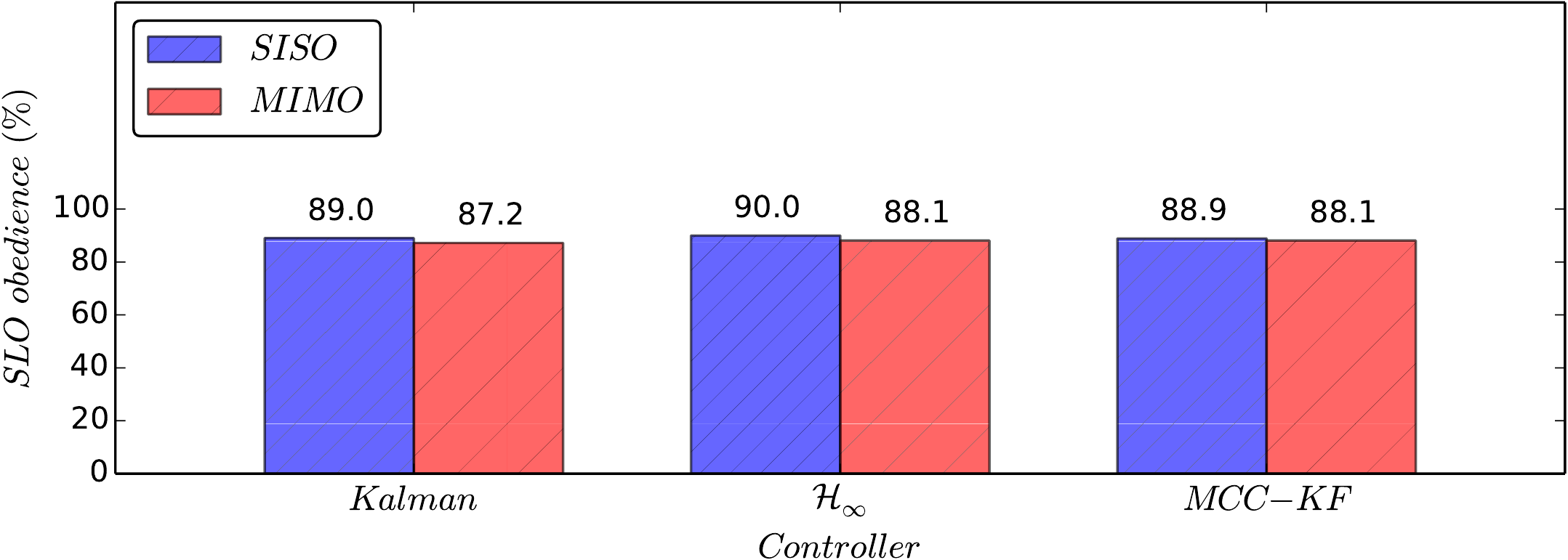}}
\endminipage
\vspace{-10px}
\caption{Controller comparison and evaluation. Fig.~\ref{subfig13a}: $\mRT$ for each controller SISO and MIMO.  Fig.~\ref{subfig13b}: SLO obedience for each controller SISO and MIMO.}\label{figure13}
\end{figure}

Fig.~\ref{figure13} shows the performance evaluation of both SISO and MIMO models. Overall, SISO controllers have lower average $\mRT$ and higher SLO obedience than MIMO controllers. However, the MIMO model offers central control of the CPU resources for cloud applications, while the SISO model can only be applied on individual virtualized servers. SISO controllers do not consider the correlation between the application's components, while MIMO controllers do consider such correlation via the covariance calculations. MIMO controllers are installed on a remote physical machine which may adds extra delays in the control signals thus affecting $\mRT$ accordingly.

The \hinf controller offers the best performance among the SISO and MIMO models as it achieves the lower average $\mRT$ and the higher SLO obedience. MCC-KF controller follows with a slightly higher average $\mRT$ in both SISO and MIMO models than \hinf. %The SLO obedience of the MCC-KF is the same for the MIMO model and about $1\%$ lower than \hinf for the SISO model. 
Lastly, the Kalman filter offers the lower performance with the highest values of average $\mRT$ and the lowest values of the total SLO obedience.

\subsection{Remarks}
The following remarks are highlighted for the SISO and MIMO approaches:

\begin{remark}
SISO controllers do not consider inter-component resource couplings due to their independent action on each physical machine (installed on the Domain-0 of each physical machine) of a cloud application. Since they do not need to exchange any messages with other physical machines, SISO controllers do not experience any delays.
\end{remark}

\begin{remark}
Another benefit of having SISO controllers is that each component of a cloud application can host a different SISO controller depending on the dynamics of workload variation (e.g., Kalman filter can be used for smoother CPU usage components, while \hinf or MCC-KF for abrupt ones).
\end{remark}

\begin{remark}
MIMO controllers encompass inter-component resource couplings (cross-coupling) via the covariance calculation process.
\end{remark}

\begin{remark}
MIMO controllers are installed on remote servers which may experience network delays resulting in possible application performance degradation. Nevertheless, as modern cloud applications are hosted on multi-tier applications, a centralized MIMO controller is preferred over SISO controllers in order to centrally estimate needed the allocations of each component, while taking into account the resource couplings.
\end{remark}

% ===============================================
%
%
% Related work
%
%
% ===============================================
\section{Related work}\label{sec:relatedwork}

In~\cite{utilizationAndSLO} and~\cite{utilityDriven}, the authors directly control application response times through runtime resource CPU allocation using an \emph{offline} system identification approach with which they tried to model the relationship between the response times and the CPU allocations in regions where it is measured to be linear. However, as this relationship is application-specific and relies on offline identification performance models, it cannot be applied when multiple applications are running concurrently and it is not possible/easy to adjust to new conditions. This triggered the need for searching for other approaches.

The authors in \cite{utilityDriven} and \cite{autoParam} were among the first to connect the control of the application CPU utilization within the VM with the response times. The use of control-based techniques has emerged as a natural approach for resource provisioning in a virtualized environment. Control-based approaches have designed controllers to continuously update the maximum CPU allocated to each VM based on CPU utilization measurements. For example, Padala \emph{et al.}~\cite{adaptiveControlV} present a two-layer non-linear controller to regulate the utilization of the virtualized components of multi-tier applications.

Multi-Input-Multi-Output (MIMO) feedback controllers have also been considered; see, for example, \cite{resourceProvisioning} and~\cite{selfAdaptive}. These controllers make global decisions by coupling the resource usage of all components of multi-tier server applications. In addition, the resource allocation problem across consolidated virtualized applications under conditions of contention have been considered in~\cite{adaptiveControlV, automatedControl}: when some applications demand more resources than physically available, then the controllers share the resources among them, while respecting the user-given priorities.
Kalyvianaki \emph{et al.}~\cite{selfAdaptive,Kalyvianaki:2014} were the first to formulate the CPU allocation problem as a state prediction one and propose adaptive Kalman-based controllers to predict the CPU utilization and maintain the CPU allocation to a user-defined threshold. Even though the standard Kalman filter provides an optimal estimate when the noise is Gaussian, it may perform poorly if the noise characteristics are non-Gaussian.

To account for uncertainties in the system model and noise statistics, in this work we propose the use of two controllers for the state estimation of the CPU resources: (a) an \hinf filter in order to minimize the maximum error caused by the uncertainties in the model and (b) the Maximum Correntropy Criterion Kalman Filter which measures the similarity of two random variables using information of high-order signal statistics. This type of controllers showed improved performance in saturation periods and sudden workload changes. The system model in this paper is represented in the form of a random walk (first-order autoregressive model), while  a higher-order autoregressive model can be adopted if increased modeling complexity is required. For such a linear system, robust state estimation techniques have been proposed. Other advanced filtering techniques, such as mixed Kalman/\hinf filtering \cite{1996:RNC235} and Robust Student's t-Based Kalman Filter \cite{2017:robustStudent-t} can also be used. Moreover, due to the lower and upper bounds on the values of the states, one can consider filtering for nonlinear systems. Nonlinear filtering based on the Kalman filter, includes the extended and unscented Kalman filters. Additionally, particle filtering (or sequential Monte Carlo methods) is a set of genetic-type particle Monte Carlo methodologies that provide a very general solution to the nonlinear filtering problem \cite{2006:hinf}. Overall, there is room for developing and testing advanced estimation techniques for this problem and this paper provides a springboard for researchers to make enhanced contributions towards this direction.

Researchers have also used neuro-fuzzy control for controlling CPU utilization decisions of virtual machines level controllers. For example, Sithu \emph{et al.}~\cite{sithu_resource_2011} present CPU load profiles to train the neuro-fuzzy controller to predict usage with 100\% allocation but not considering CPU allocation in the training. Deliparaschos \emph{et al.}~\cite{ecc:2016}, on the other hand,  presented the training of neuro-fuzzy controller via  extensive use of data from established controllers, such as, \cite{Kalyvianaki:2014}.

% ===============================================
%
%
% CONCLUSIONS AND FUTURE DIRECTIONS
%
%
% ===============================================
\section{Conclusions and Future Directions}\label{conclusions}

% ===============================================
% Conclusions
% ===============================================
\subsection{Conclusions}

In this paper, we propose and present a rigorous study of SISO and MIMO models comprising adaptive robust controllers (i.e., \hinf filter, MCC-K filter) for the CPU resource allocation problem of VMs, while satisfying certain quality of service requirements. The aim of the controllers is to adjust the CPU resources based on observations of previous CPU utilizations. For comparison purposes, tests were performed on an experimental setup implementing a two-tier cloud application. Both proposed robust controllers offer improved performance under abrupt and random workload changes in comparison to the current state-of-the-art. Our experimental evaluation results show that (a) SISO controllers perform better than the MIMO ones; (b) \hinf and MCC-KF have improved performance than the Kalman filter for abrupt workloads; (c) \hinf and MCC-KF have approximately equal performance for both SISO and MIMO models. In particular, the proposed robust controllers manage to reduce the average $\mRT$ while keeping the SLO violations low.

% ===============================================
% Future Directions
% ===============================================
\subsection{Future Directions}

The system considered in this paper addresses only CPU capacity, and resource needs are coupled across multiple dimensions (i.e., compute, storage, and network bandwidth). Therefore, workload consolidation should be performed while catering for resource coupling in multi-tier virtualized applications in order to provide timely allocations during abrupt workload changes. In this context, part of ongoing research considers use of system identification/learning to extract coupling information between resource needs for workload consolidation while meeting the SLOs.

The heterogeneity of cloud workloads could be challenging for choosing a suitable headroom online for minimizing the resources provided while meeting the SLO. Thus, it is important to evaluate the system performance under different types of workload as assessed in \cite{ardagna2017generalized}. Therefore, while CPU usage headroom varies with different workload patterns and frequencies, adapting  the headroom could be beneficial in terms of CPU resource savings while meeting the SLOs.

Data centers distributed in different geo-location can work collaboratively, and these are normally connected via high-speed Internet or dedicated high-bandwidth communication links. VMs can be migrated, if necessary, within a data center or even across data centers \cite{Endo:2011}. The combination of dynamically changing VMs sizes and their migration is very challenging and an open problem. Current clouds (e.g., Google Compute Engine, Microsoft Azure) provide limited support for real-time (RT) performance to VMs; in other words, they cannot guarantee a SLO on low-latency \cite{xi_rt-open_2015}. However, cloud services benefit from RT applications since they can host computation-intensive RT tasks (e.g., gaming consoles). An RT cloud, ensures latency guarantee for tasks running in VMs, provides RT performance isolation between VMs, and allows resource sharing between RT and non-RT VMs. 

Moreover, we are investigating the possibility of accelerating workload optimization through rapid adaptation to changing throughput and latency requirements via programmable logic technology (i.e., FPGA) over CPU; and GPU-based alternatives for high performance virtualized applications. Even though FPGAs run on lower speeds relative to CPUs, their inherent parallelism allows process multi-tasking resulting to dramatic speed enhancement and enabling hardware consolidation due to the ability of a single FPGA simultaneously performing the tasks of multiple servers.

% ===============================================
% Acknowledgements
% ===============================================
\section*{Acknowledgements}

The authors would like to thank Ms. Kika Christou, from the Information Systems \& Technology Service of Cyprus University of Technology, for her continued support on the computing and networking infrastructure aspects.

% ===============================================
% Bibliography
% ===============================================
\bibliographystyle{IEEEtran}
\bibliography{paper.bib}

\end{document}